\documentclass{aa}
\usepackage{graphicx}
\usepackage[varg]{txfonts}
\usepackage{natbib,twoopt}
\usepackage[breaklinks=true]{hyperref} 
\bibpunct{(}{)}{;}{a}{}{,} 
\makeatletter
\newcommandtwoopt{\citeads}[3][][]{\href{http://adsabs.harvard.edu/abs/#3}%
{\def\hyper@linkstart##1##2{}%
\let\hyper@linkend\@empty\citealp[#1][#2]{#3}}}
\newcommandtwoopt{\citepads}[3][][]{\href{http://adsabs.harvard.edu/abs/#3}%
{\def\hyper@linkstart##1##2{}%
\let\hyper@linkend\@empty\citep[#1][#2]{#3}}}
\newcommandtwoopt{\citetads}[3][][]{\href{http://adsabs.harvard.edu/abs/#3}%
{\def\hyper@linkstart##1##2{}%
\let\hyper@linkend\@empty\citet[#1][#2]{#3}}}
\newcommandtwoopt{\citeyearads}[3][][]%
{\href{http://adsabs.harvard.edu/abs/#3}
{\def\hyper@linkstart##1##2{}%
\let\hyper@linkend\@empty\citeyear[#1][#2]{#3}}}
\makeatother

\usepackage{lscape}
\newcommand{\vsin}{\hbox{$v \sin i$}} 
\usepackage{xcolor}
\usepackage[export]{adjustbox}

\begin{document}

   \title{The CARMENES search for exoplanets around M dwarfs}
   \subtitle{Benchmarking the impact of activity in \\ high-precision radial velocity measurements}

\titlerunning{Benchmarking impact of activity on high-precision RVs}

   \author{S.\,V.~Jeffers\inst{1}, J.\,R.~Barnes\inst{2}, P.~Sch\"ofer\inst{3},  A.~Quirrenbach\inst{4}, M.~Zechmeister \inst{3}, P.\,J.~Amado \inst{5},  J.\,A.~Caballero\inst{6}, M.~Fern\'andez \inst{5}, E.~Rodr\'iguez\inst{5}, I.~Ribas \inst{7,8}, A.~Reiners \inst{3}, C.~Cardona~Guill\'en \inst{9,10}, C.~Cifuentes \inst{6}, S.~Czesla \inst{11}, A.\,P.~Hatzes \inst{12}, M.~K\"urster \inst{13}, D.~Montes\inst{14}, J.\,C.~Morales \inst{7,8},  S.~Pedraz\inst{15},
   S.~Sadegi \inst{4,12}}

\authorrunning{Jeffers et al.}

   \institute{$^{1}$ Max Planck Institute for Solar System Research, Justus-von-Liebig-Weg 3, D-37077 G\"ottingen, Germany\\
    \email{sandrajeffers.astro@gmail.com}\\
	 $^{2}$ School of Physical Sciences, The Open University, Walton Hall, MK7 6AA, Milton Keynes, United Kingdom\\   
	 $^{3}$ Institut f\"ur Astrophysik, Georg-August-Universit\"at, Friedrich-Hund-Platz 1, D-37077 G\"ottingen, Germany  \\
	 $^{4}$ Landessternwarte, Zentrum für Astronomie der Universit\"at Heidelberg, K\"onigstuhl 12, D-69117 Heidelberg, Germany \\
	 $^{5}$ Instituto de Astrof\'isica de Andaluc\'ia (CSIC), Glorieta de la Astronomía s/n, E-18008 Granada, Spain\\
	 $^{6}$ Centro de Astrobiolog\'ia (CSIC-INTA), ESAC, Camino Bajo del Castillo s/n, E-28692 Villanueva de la Cañada, Madrid, Spain\\
	 $^{7}$ Institut de Ci\`encies de l’Espai (CSIC), Campus UAB, c/ de Can Magrans s/n, E-08193 Bellaterra, Barcelona, Spain \\
     $^{8}$ Institut d’Estudis Espacials de Catalunya, E-08034 Barcelona, Spain\\
	 $^{9}$  Instituto de Astrof\'isica de Canarias, c/ V\'a L\'actea s/n, E-38205 La Laguna, Tenerife, Spain\\
     $^{10}$ Departamento de Astrof\'isica, Universidad de La Laguna, E-38206 Tenerife, Spain\\
     $^{11}$ Hamburger Sternwarte, Universit\"at Hamburg, Gojenbergsweg 112, D-21029 Hamburg, Germany  \\
     $^{12}$ Thüringer Landessternwarte Tautenburg, Sternwarte 5, D-07778 Tautenburg, Germany \\
     $^{13}$ Max-Planck-Institut f\"ur Astronomie, K\"onigstuhl 17, D-69117 Heidelberg, Germany\\
	 $^{14}$ Facultad de Ciencias F\'isicas, Departamento de F\'isica de la Tierra y Astrof\'isica \& IPARCOS-UCM (Instituto de F\'isica de Part\'iculas y del Cosmos de la UCM), Universidad Complutense de Madrid, E-28040 Madrid, Spain\\
 $^{15}$ Centro Astron\'omico Hispano-Alem\'an (CSIC-Junta de Andaluc\'ia), Observatorio Astron\'omico de Calar Alto, Sierra de los Filabres, E-04550 G\'ergal, Almer\'ia, Spain  }


   \date{Received dd July 2021 / Accepted dd Month 2021} 
 
  \abstract
   {Current exoplanet surveys using the radial velocity (RV) technique are targeting M dwarfs because any habitable zone terrestrial-mass planets will induce a high RV and orbit on shorter periods than for more massive stars.  One of the main caveats is that M dwarfs show a wide range of activity levels from inactive to very active, which can induce an asymmetry in the line profiles and, consequently, a spurious RV measurement.  
     }
   {We aim to benchmark the impact of stellar activity on high-precision RV measurements using regular-cadence CARMENES visible and near-infrared observations of the active M3.5 dwarf \object{EV~Lac}.}
   {We used the newly developed technique of low-resolution Doppler imaging to determine the centre-of-light, or spot-induced RV component, for eight observational epochs.   }
   {We confirm a statistically significant and strong correlation between the independently measured centre-of-light and the chromatic index, which is a measure of the amplitude variation with wavelength of the RVs.  We also find circular ``closed-loop'' relations of several activity indices with RV for a subset of data that covers only several rotation periods. We also investigate the implications of large phase gaps in the periodograms of activity indicators. Finally, by removing the spot-induced RV component we improve the planet-mass sensitivity by a factor of at least three.}
  {We conclude that for active M stars, a regular-cadence observing strategy is the most efficient way to identify and eliminate sources of correlated noise.  }


   \maketitle
%

\section{Introduction}

One of the main drivers of exoplanet research is the detection of  Earth-mass planets orbiting in the habitable zones of their host stars.
Low-mass M dwarf stars are currently the focus of high-precision radial velocity (RV) exoplanet surveys because the relative amplitude of the RV induced by an Earth-mass planet is larger for a lower-mass star than that of the higher-mass G dwarfs. One of the first exoplanet surveys at near-infared (NIR) wavelengths is the CARMENES survey (Calar Alto high-Resolution search for M dwarfs with Exo-earths with Near-infrared and optical 
\'{E}chelle Spectrographs; \citealt{Quirrenbach2018SPIE10702E..0WQ}). The guaranteed time observations (GTO) program comprises over 350 M dwarfs that were selected as the brightest $J$-band stars in each spectral type bin in addition to being observable from Calar Alto and, without companions at less than 5\,arcsec.
Consequently, the GTO sample also includes several very active stars.    The long time-span of the CARMENES GTO observations of active stars makes them ideal targets for investigating their stellar activity using high-precision RV observations.  A summary of recent results was given by \cite{Quirrenbach2020SPIE11447E..3CQ}.

When stellar activity features such as starspots are present, they induce an asymmetry in the shape of spectral lines. The magnitude of the effect depends on the strength and location of the activity regions. The asymmetries induce a shift of the line centre, and consequently contribute an additional activity-induced RV shift. The stellar activity of M dwarfs has been extensively investigated in the literature; these dwarfs can be very inactive, but some are the most active stars \citep[e.g.][]{Browning2010AJ....139..504B, Reiners2012AJ....143...93R, West2015ApJ...812....3W, Newton2017ApJ...834...85N, Jeffers2018A&A...614A..76J, Schoefer2019A&A...623A..44S}.  
The variation amplitudes can range from cm\,s$^{-1}$ on timescales much shorter than the stellar rotation period in the case of convective motions and predicted oscillations to m\,s$^{-1}$ to timescales of several rotation periods in the case of starspots.  On timescales of several years, active stars can exhibit stellar activity cycles.   The impact of stellar activity on detecting any orbiting exoplanets was simulated for M dwarfs by \cite{Barnes2011MNRAS.412.1599B} using realistic starspot patterns extrapolated from the Sun  \citep{Jeffers2005MNRAS.359..729J}.  \cite{Barnes2011MNRAS.412.1599B} reported spot-induced variations ranging from 1\,m\,s$^{-1}$ to 1\,km\,s$^{-1}$ for different levels of stellar activity.  The precision of current instrumentation reaches RV precisions of 1\,$<$\,m\,s$^{-1}$ \citep[e.g. the ESPRESSO, EXPRES and MAROON-X instruments;][]{Pepe2013Msngr.153....6P, Jurgenson2016SPIE.9908E..6TJ,Bean2020AAS...23522504B}, which means that it is not instrumental precision but intrinsic stellar variability that effectively limits the detection of exoplanets.

\begin{table}[]
    \caption{\label{tab:stellarparam}Fundamental stellar parameters of EV~Lac}
    \centering
    \begin{tabular}{lcl}
     \noalign{\smallskip}
        \hline
        \hline
        Parameter & Value & Reference \\
        \hline
Name & EV\,Lac &  \\
BD & +43 4305 & Arg03 \\
GJ & 873 &  Gli57\\
Karmn & J22468+443  & Cab16 \\
$\alpha$ (J2000) & 22:46:49.73 & {\it Gaia} EDR3 \\
$\delta$ (J2000) & +44:20:02.4 & {\it Gaia} EDR3 \\
$d$ [pc] & 5.05 $\pm$ 0.01 & {\it Gaia} EDR3\\
$G$ [mag] &$9.005\pm0.003$ & {\it Gaia} EDR3 \\
$J$ [mag] & $6.106\pm0.030$& 2MASS \\
Sp. Type & M3.5\,V & PMSU \\
$T_{\rm eff}$ [K] & 3291$\pm$51 & Pass19 \\
$\log{g}$ [cgs] & 5.11$\pm$0.07 &  Pass19 \\
$L_\star$ [$L_{\odot}$] & 0.01288$\pm$0.00014 &  Cif20 \\
$R_\star$ [$R_{\odot}$] & 0.349 $\pm$ 0.010 & Sch19 \\
$M_\star$ [$M_{\odot}$] & 0.344 $\pm$ 0.015 & Sch19 \\
$[{\rm Fe/H}]$ & --0.19 $\pm$ 0.16 &  Pass19\\
pEW(H$\alpha$) [\AA] & $-3.62$ to $-4.74$ & Jef18 \\ 
$v\sin{i}$ [km\,s$^{-1}$] & $ 3.5\pm$1.5 & Rei18 \\
$P_{\rm rot}$ [d] & 4.349$\pm$0.004 & Rev20 / this work \\
$\log{R'_{\rm HK}}$&  --3.75  & BSa18 \\ 
L$_{\rm X}$/L$_{\rm bol}$ & --3.10 & Wri11 \\ 
Kinematic group & Ursa Major association & Cor21 \\
Age [Myr] & 300 & Kin03 \\
Inclination [deg] &  60  & Mor08 \\
        \hline
    \end{tabular}
    \tablebib{{All stellar parameters for the over 350 stars in the CARMENES GTO sample have been derived in a uniform manner.  The references to the quantities are as follows:
        2MASS: \cite{Skrutskie2006AJ....131.1163S};
        Arg03: \cite{Argelander1903BD....C......0A}
        BSa18: \cite{BoroSaikia2018A&A...616A.108B};
        Cab16: \cite{Caballero2016csss.confE.148C};
        {\it Gaia} EDR3: \cite{Riello2021A&A...649A...3R};
        Gli57: \cite{Gliese1957MiABA...8....1G};
        Cif20: \cite{Cif2020A&A...642A.115C};
        Cor21: Cort\'es-Contreras et al. (in preparation);
        Jef18: \cite{Jeffers2018A&A...614A..76J};
        Kin03: \cite{King2003AJ....125.1980K};
        Mor08: \cite{Morin2008MNRAS.390..567M};
        Pass19: \cite{Passegger2019A&A...627A.161P};
        Rei18: \cite{Reiners2018A&A...612A..49R};
        Rev20: \cite{Revilla:Thesis:2020};
        PMSU: \cite{Hawley1996AJ....112.2799H};
        Sch19: \cite{Schweitzer2019A&A...625A..68S};
        Wri11: \cite{wright2011ApJ...743...48W}.
    }}
\end{table}

\begin{figure}
\centering
\includegraphics[angle=270,width=0.45\textwidth]{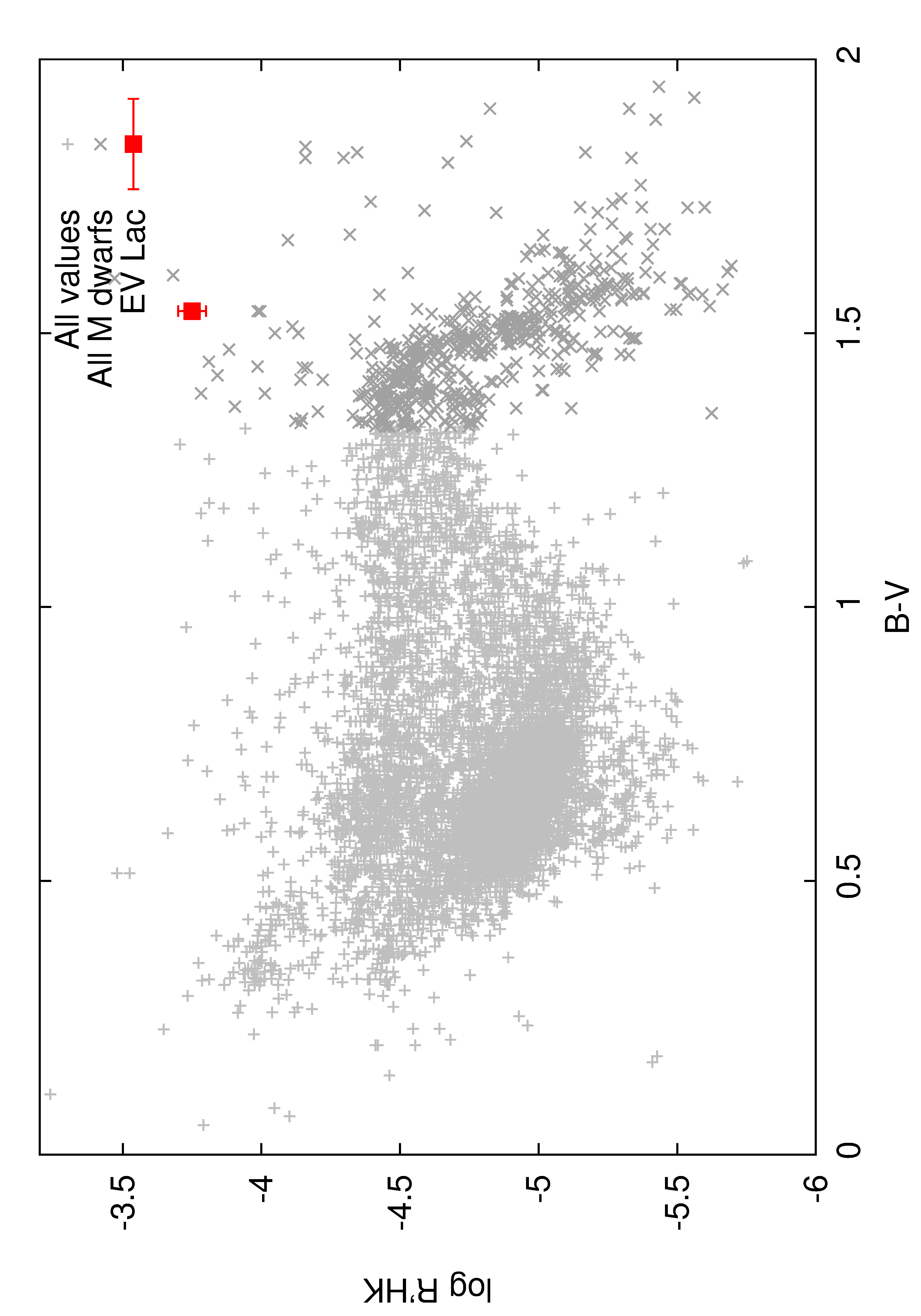}
\caption{$\log{R'_{\rm HK}}$of EV~Lac in relation to the $\log{R'_{\rm HK}}$values of more than 4000 main-sequence stars ranging in spectral types from late F to M, from \protect\cite{BoroSaikia2018A&A...616A.108B}.}
\label{fig:RHK_EVLac}
\end{figure}

\begin{figure*}
    \centering
    \includegraphics[width=0.75\textwidth]{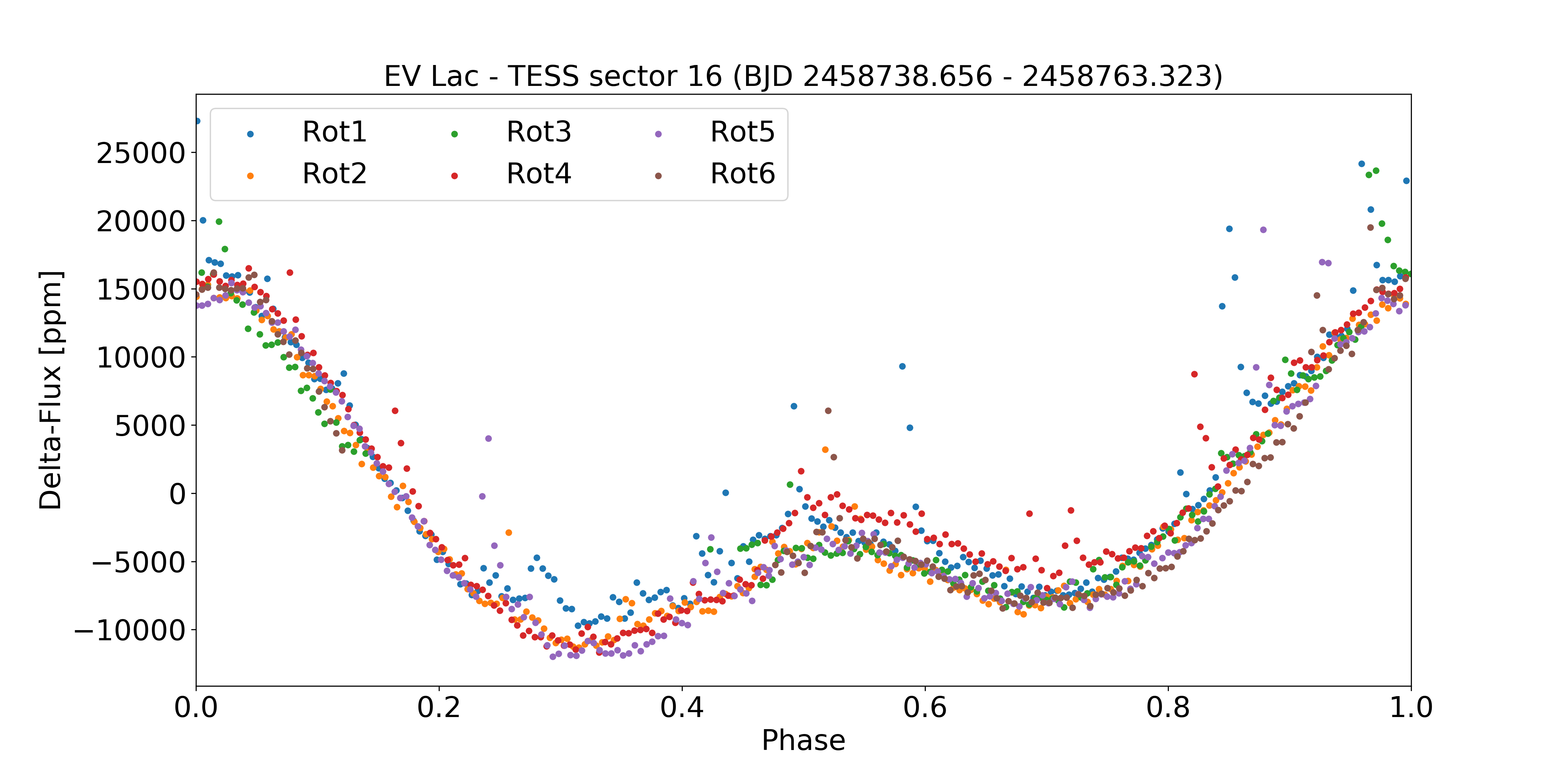}
    \caption{Phase-folded {\em TESS} light curve of EV~Lac from sector 16 covering six rotational periods.  The computed rotational period is 4.349$\pm$0.004\,d. }
    \label{fig:TESSlc}
\end{figure*}

The H$\alpha$ line is an accessible diagnostic to indicate chromospheric activity.  Many studies \citep[e.g.][among many others]{Browning2010AJ....139..504B,  West2015ApJ...812....3W, Newton2017ApJ...834...85N, Jeffers2018A&A...614A..76J} 
have consistently shown that approximately 10\,\% of the early-M dwarfs are H$\alpha$ active, while at least 70\,\% of the lower-mass late-M dwarfs are H$\alpha$ active.  The reason is that M dwarfs have much longer rotational braking times.

Reconstructions of photospheric starspots on rapidly rotating M dwarfs typically show that cool spots can be found anywhere on the surface (\citealt{barnes01mdwarfs, barnes04hkaqr}). For fully convective M dwarfs, this scenario persists (\citealt{barnes15mdwarfs, Barnes2017MNRAS.466.1733B}), often with significant high latitude structure. This finding is consistent with the large-scale dipolar field reconstructions for more moderately rotating, fully convective stars \citep{Morin2008MNRAS.384...77M, Morin2008MNRAS.390..567M, Phan-Bao2009ApJ...704.1721P} and also for the complex structure of smaller features derived using molecular lines \citep[e.g.][]{Afram2019A&A...629A..83A}. 

\begin{figure}
\centering
\includegraphics[angle=270,width=0.45\textwidth]{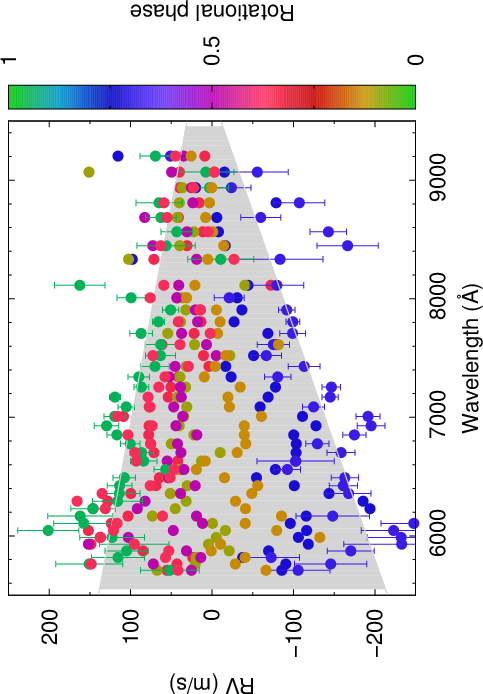}
\caption{Variation of RV as a function of order number for CARMENES visible channel.   The slope of the RV-log($\lambda$) correlation is the CRX value.  Eight CARMENES spectra are shown and colour-coded based on stellar rotational phase.  The upper and lower borders of the shaded grey region are the slopes of the spectra with the highest and lowest RV variations, as indicated by the green and blue points respectively.  An important point to note is that the two outer extremes of the shaded grey area will not converge at redder wavelengths. The same eight spectra are used to reconstruct the first low-resolution spot map in Figure 6. }
\label{fig:RV-per-order}
\end{figure}

The advantage of the CARMENES spectrograph is its large wavelength coverage which ranges from visible to NIR wavelengths (see below).  This is important to identify the features of stellar activity because they induce RV variations that are wavelength dependent, whereas for a planetary companion, the RV variations are wavelength independent. With the large wavelength coverage of CARMENES, an additional diagnostic is the chromatic index (CRX), which is the slope of the RV-log($\lambda$) correlation per spectral order.  The CRX quantifies the wavelength dependence  of the RV and was first presented by  \cite{Zechmeister2018A&A...609A..12Z}.  Examples of the dependence of the RVs as a function of spectral order and stellar rotational phase are shown in Figure~\ref{fig:RV-per-order} for the CARMENES visible channel.  The errors are too high in the NIR channel to demonstrate the decrease in RVs at redder wavelengths. This means that although CRX can be measured in the NIR channel, the uncertainties are greater than the visible channel, as we show below. We do not expect the slopes in the CRX values to converge at redder wavelengths, as would be suggested by the linear outer envelopes of the shaded region in Figure~\ref{fig:RV-per-order}. Instead, from a first-order black-body scaling of the spot and photospheric flux ratios, the scatter in RVs should remain similar to the scatter seen at approximately 9500\AA. This behaviour is also predicted by examining the wavelength-dependent ratio of model PHOENIX spectra appropriate for the spot and photosphere temperatures \citep{husser13atlas}.  In Figure~\ref{fig:RV-per-order}, the extremes of the RVs are coloured green and blue which results from that EV~Lac shows a double-dip pattern in CRX and RV.

In an analysis of stellar activity, \cite{TalOR2018A&A...614A.122T} computed the CRX for a subsample of the CARMENES GTO sample. For some of their stars, they reported a correlation between the standard deviation of the RVs and stellar \vsin\, and an RV-CRX anti-correlation, which they assumed to be caused by the presence of dark starspots on the stellar surface.  Additionally, \cite{Baroch2020A&A...641A..69B} analysed the CRX of the mid-M dwarf YZ CMi in combination with photometric observations.  Using the CRX, \cite{Baroch2020A&A...641A..69B} were able to constrain the spot-filling factor of the stellar surface, the temperature contrast between the spots and the unspotted photosphere, and the location of spots.   The results of \cite{Baroch2020A&A...641A..69B} also indicated that the convective blueshift decreases for M dwarfs compared to earlier solar-type stars, which is consistent with the results of decreasing convective blueshift with decreasing mass which has been reported by \cite{Meunier2017A&A...607A.124M} for G and K dwarfs and more recently by \cite{Liebing2021A&A...654A.168L} for late-F to early-M dwarfs.  

With the advent of many long-term high-precision RV surveys for exoplanets, we are now entering an era of understanding the long-term intrinsic variability of the target stars.  As part of the CARMENES GTO survey, we observed the active M3.5 dwarf EV~Lac with observations that were secured with a regular cadence spanning a total period of just over one year.  The regular cadence, where observations span not more than a few stellar rotation cycles, is  important for capturing the evolution of the stellar activity features, which can be a significant source of correlated noise in RV measurements. EV~Lac is a particularly suitable target as it has very high levels of stellar activity for its relatively low \vsin\ value. As previously noted by \cite{Barnes2011MNRAS.412.1599B}, for a fixed spot pattern, doubling \vsin\ doubles activity-induced jitter. 

We perform an in-depth investigation of the stellar activity of EV~Lac using high-precision RVs from  CARMENES and observations from the {\em TESS} satellite.   In Sect.~2 we summarise the stellar activity of EV~Lac and present the new CARMENES and {\em TESS} observations in Sect.~3.  In Sect.~4, we present the wavelength dependence of the RVs and in Section 5 we apply the recently developed technique of low-resolution Doppler imaging to identify the apparent spot-induced RV component.  In Sect.~6 we use the results to investigate the evolution of the stellar activity indices as a function of rotation phase, and in Sect.~7 we quantify the correlations between the activity indices.  In Sects.~8 and~9, we investigate the closed-loop stellar activity variation and periodicities in the CARMENES data set.  Finally in Sect.~10 we develop a new technique to infer the CRX for missing phases using the Doppler-imaging technique.

\section{Stellar activity of EV~Lac}

The stellar parameters of EV~Lac are listed in Table~\ref{tab:stellarparam}.  The M3.5 dwarf EV~Lac is very active: its log(H$\alpha$/L$_{\rm bol}$) measurements can vary between --3.62 and --4.744  \citep{Jeffers2018A&A...614A..76J,Reiners2018A&A...612A..49R}. Its $\log{R'_{\rm HK}}$value of --3.75 is one of the highest values in the sample of more than 4000 stars, spanning spectral types F to M, compiled by \cite{BoroSaikia2018A&A...616A.108B} as shown in Fig.~\ref{fig:RHK_EVLac}.  EV~Lac exhibits frequent flaring events and X-ray emission in the saturated regime  \citep{wright2011ApJ...743...48W}, 
which is in agreement with its location on the $\log{R'_{\rm HK}}$versus period relation of \cite{Astudillo2017A&A...600A..13A}. 

\begin{figure*}
\mbox{
\hspace{1.65cm}
\includegraphics[angle=270,width=7cm]{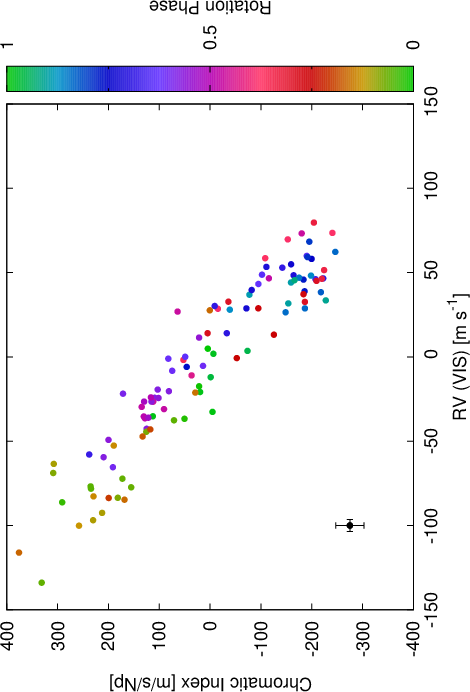}
\hspace{0.25cm}
\includegraphics[angle=270,width=7cm]{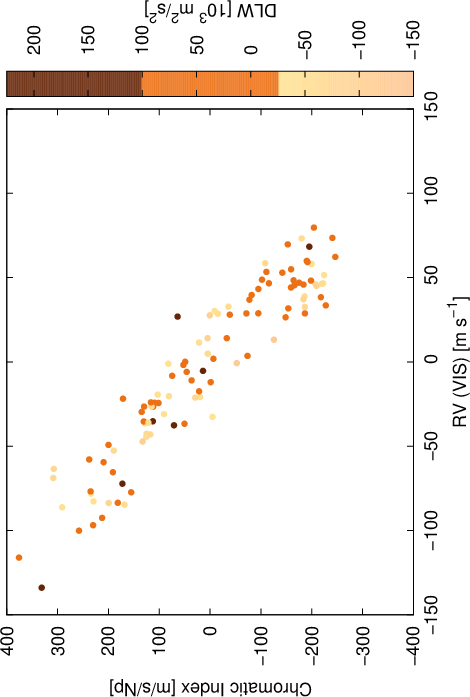}
}
\mbox{
\hspace{1.65cm}
\includegraphics[angle=270,width=7cm]{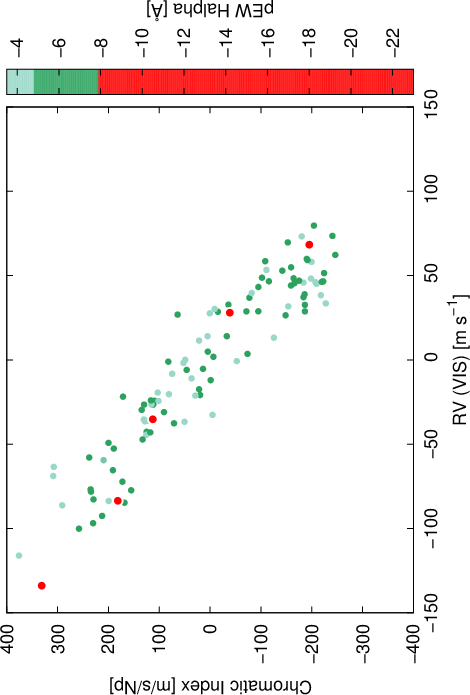}
\hspace{0.35cm}
\includegraphics[angle=270,width=7.2cm]{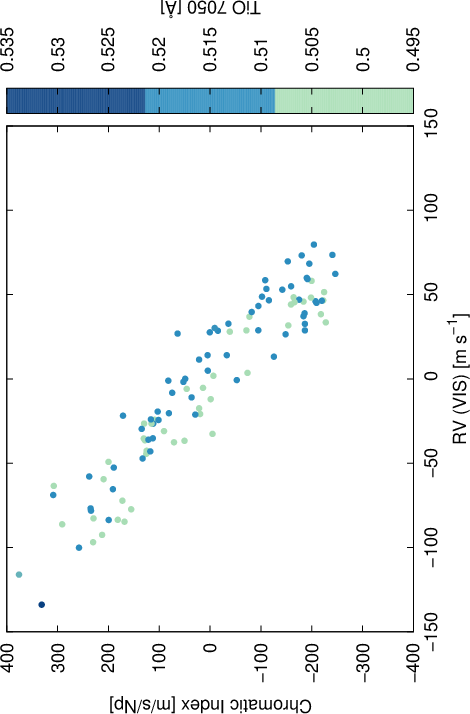}
}
\caption{CRX-RV correlation with phase for all CARMENES observations.  Points are coloured based on rotational phase (top left, with a representative error bar), dLW (top right), the equivalent width of the H$\alpha$ line (bottom left), and the TiO 7050\,\AA\ band strength (bottom right).  }
\label{fig:CRX_RV}
\end{figure*}

\subsection{Large-scale magnetic field geometry}

The large-scale magnetic field geometry of EV~Lac has previously been reconstructed from the rotationally modulated time-series of circularly polarized line profiles using the tomographic technique of Zeeman-Doppler imaging by \cite{Morin2008MNRAS.390..567M}.  The large-scale magnetic field of EV~Lac is composed of two strong concentrations of radial field of opposite polarities.  The positive-polarity spot is located near the equator, while the negative-polarity spot is located at an approximate latitude of 50\,deg.  The spectropolarimetric observations of EV~Lac were secured approximately one year apart.  Over this time-span, the large-scale magnetic field geometry of EV~Lac remained stable and only showed some slight smaller-scale variations. This is in contrast to the large-scale field reconstructed for earlier M dwarfs ($M >$0.5\,$M_\odot$) and K and G dwarfs \citep[e.g.][]{Donati2008MNRAS.390..545D,Morin2008MNRAS.384...77M,Jeffers2011MNRAS.411.1301J}, which mainly comprise a  toroidal field.  The stability of the large-scale field of  EV~Lac is in contrast to the reconstructed large-scale field for the late-F star $\tau$~Boo \citep{Mengel2016MNRAS.459.4325M, Jeffers2018A&A...614A..76J}, where the geometry and polarity of the large-scale field reverse every 120\,d when EV~Lac is located at the boundary to full convection.  These results indicate that the magnetic field generation processes in EV~Lac are different compared to these more massive late-type stars.  The magnetic flux of EV~Lac is calculated to be up to 2\,kG using individual magnetic features \citep{Morin2008MNRAS.390..567M}, 4.1$\pm$0.2\,kG for the total disk-integrated magnetic field \citep{Shulyak2019A&A...626A..86S}, and it ranges from 5.6 to 6.8\,kG using molecular lines \citep{Afram2019A&A...629A..83A}.  The difference in these values indicates that the small-scale fields, or magnetic features, comprise a significant part of the magnetic energy.

\section{Observations and data processing}
\label{section:observations}

In this section we describe observations and processing of the high-precision CARMENES spectroscopic data.  In addition, we also use photometric data from the {\em TESS} satellite.  

\subsection{CARMENES RVs}
The high-precision RV data were obtained with the CARMENES spectrograph \citep{Quirrenbach2014SPIE.9147E..1FQ,Quirrenbach2018SPIE10702E..0WQ}  located at the 3.5\,m Calar Alto telescope. The instrument has a VIS and a NIR channel covering a broad wavelength range of 550--960\,nm and 960--1700\,nm with spectral resolutions of 94\,600 and 80\,400, respectively. The average sampling per resolution element is 2.8\,pixels.   The spectra were processed by the standard pipeline {\tt caracal} \citep[CARMENES reduction and calibration software;][]{Zechmeister2014A&A...561A..59Z, Caballero2016SPIE.9910E..0EC}.  A total of 108 spectra of EV~Lac were secured with CARMENES over the time period from January 2016 to December 2017. The extracted spectra show mean S/N ratios (calculated over all extracted orders) of $16.4\,\leq$ S/N $\leq\,109$. However, only 10 observations have S/N $< 40$.  The complete data set has \hbox{<S/N> $= 67.7 \pm 21.3$}. 

The RVs were computed with {\tt serval}\footnote{\url{https://github.com/mzechmeister/serval}} \citep[Spectrum radial-velocity analyzer,][]{Zechmeister2018A&A...609A..12Z}, which also computes the chromatic index. Given that stellar activity will introduce a wavelength dependence of the RVs, both the RV and CRX are computed as weighted averages \citep[Eq 14, 15 and 21 in][]{Zechmeister2018A&A...609A..12Z}, where the weighting of the individual orders is not fixed but is taken as defined in Eq 11 of \cite{Zechmeister2018A&A...609A..12Z}.  {\tt serval} additionally computes the differential line width (dLW) for both the VIS and NIR wavelength ranges. The dLW quantifies the changes in the equivalent width at a fixed contrast level.   The VIS and NIR stellar activity indices, namely $\log(L_{{\rm H}\alpha}/L_{\rm bol})$, pEW(H$\alpha$),  HeD3, NaD, Ca IRT-a,-b,-c,	He\,10833, Pa$\beta$, CaH2, CaH3, TiO\,7050, TiO\,8430,	TiO\,8860, VO\,7436, VO\,7942, and FeH Wing-Ford, were computed following \cite{Schoefer2019A&A...623A..44S}, and the cross-correlation function (CCF) parameters, CCF-Contrast, CCF-RV, CCF-FWHM, and CCF-Bisector were computed following \cite{Lafarga2020A&A...636A..36L}.  This resulted in a total of four parameters from {\tt serval}, 16 spectral line indices, and four CCF parameters.

The CARMENES spectra of EV~Lac span 160 stellar rotation periods.  Although the large-scale magnetic field of EV~Lac has been shown to be stable over year-long timescales \citep{Morin2008MNRAS.390..567M} the spectroscopic activity indices show additional smaller-scale activity features that vary on much shorter timescales on the order of a few rotation periods.  To understand both the large- and small-scale stellar activity of EV~Lac, we obtained eight subsets of data each of which densely samples the rotation period of EV~Lac.  The data contained in each of the subsets covers not more than several rotation periods to ensure that evolution of stellar activity is minimised.

\begin{figure*}
\centering
\includegraphics[angle=270,width=0.7\textwidth]{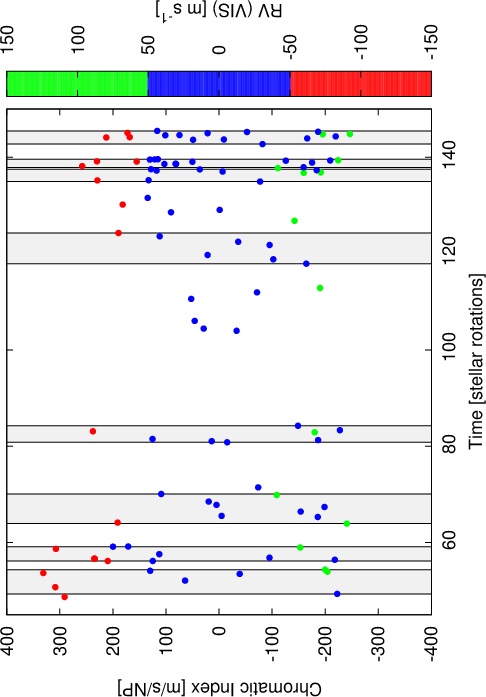}
\caption{Variation in the chromatic index as a function of time.  Colour indicates the RV value as output from {\tt serval}.  The vertical grey shaded boxes indicate the phase coverage of each of the eight subsets of data we used for the low-resolution Doppler-imaging maps.}
\label{fig:CRX_time_COL_RV}
\end{figure*}

\subsection{{{\em TESS}} light curve}

EV~Lac was observed by the Transiting Exoplanet Survey Satellite mission \citep[{\em{TESS}},][]{Ricker2015JATIS...1a4003R} in sector 16 and covers six rotational periods. We used the regression corrector in the Lightkurve package \citep{2018ascl.soft12013L} to remove instrument noise and systematics from the TESS data. The routine employs principal component analysis (PCA) on the background pixels to remove the systematics. The phase-folded light curve is shown in Figure~\ref{fig:TESSlc}.  The derived rotation period is 4.349$\pm$0.004\,d.  We used an emphemeris of 245739.35702 corresponding to the time-stamp of the first CARMENES observation of EV~Lac.  The {\em TESS} observations show that EV~Lac has a double-dip light curve over the six stellar rotations.  Previously, the light curve of EV~Lac has been a single-dip light curve \citep[][]{Alekseev2017ARep...61..221A, DiezAlonso2019A&A...621A.126D}.  The double-dip light curve indicates that the spots are distributed over the whole stellar surface \citep{Basri2020ApJ...901...14B} and are not concentrated in two active longitudes, as has been extensively and incorrectly assumed in the literature for such double-dip light curves.  In contrast, the simulations of \cite{Basri2020ApJ...901...14B} showed that if the spot coverage were confined to one hemisphere, the resulting light curve would be a single-dip light curve.   While the large-scale magnetic features remain constant over the {\em TESS} observations, there are many flares and small-scale variations, indicating rapidly evolving smaller activity features.  

\begin{figure*}
\centering
\mbox{
\includegraphics[angle=270,width=0.4\textwidth]{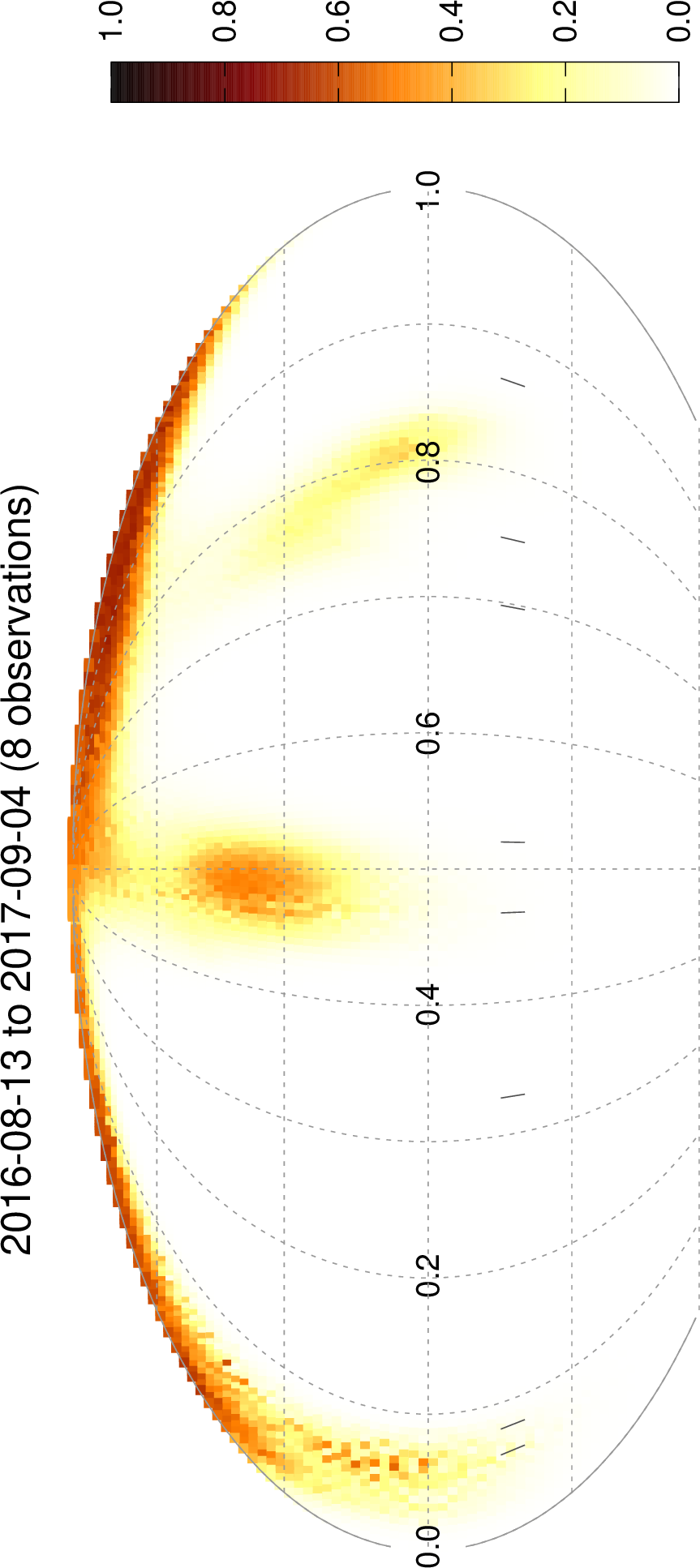}
\includegraphics[angle=270,width=0.3\textwidth]{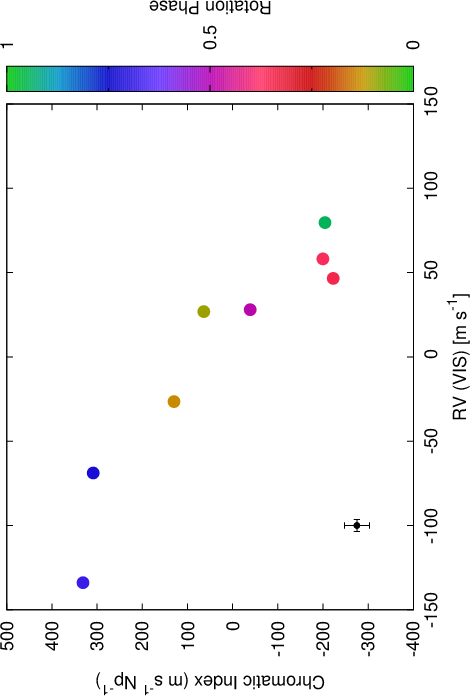}
\includegraphics[angle=270,width=0.3\textwidth]{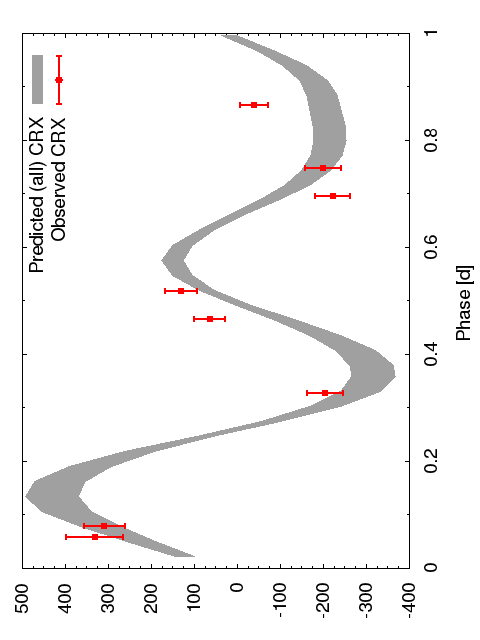}}
\mbox{
\includegraphics[angle=270,width=0.4\textwidth]{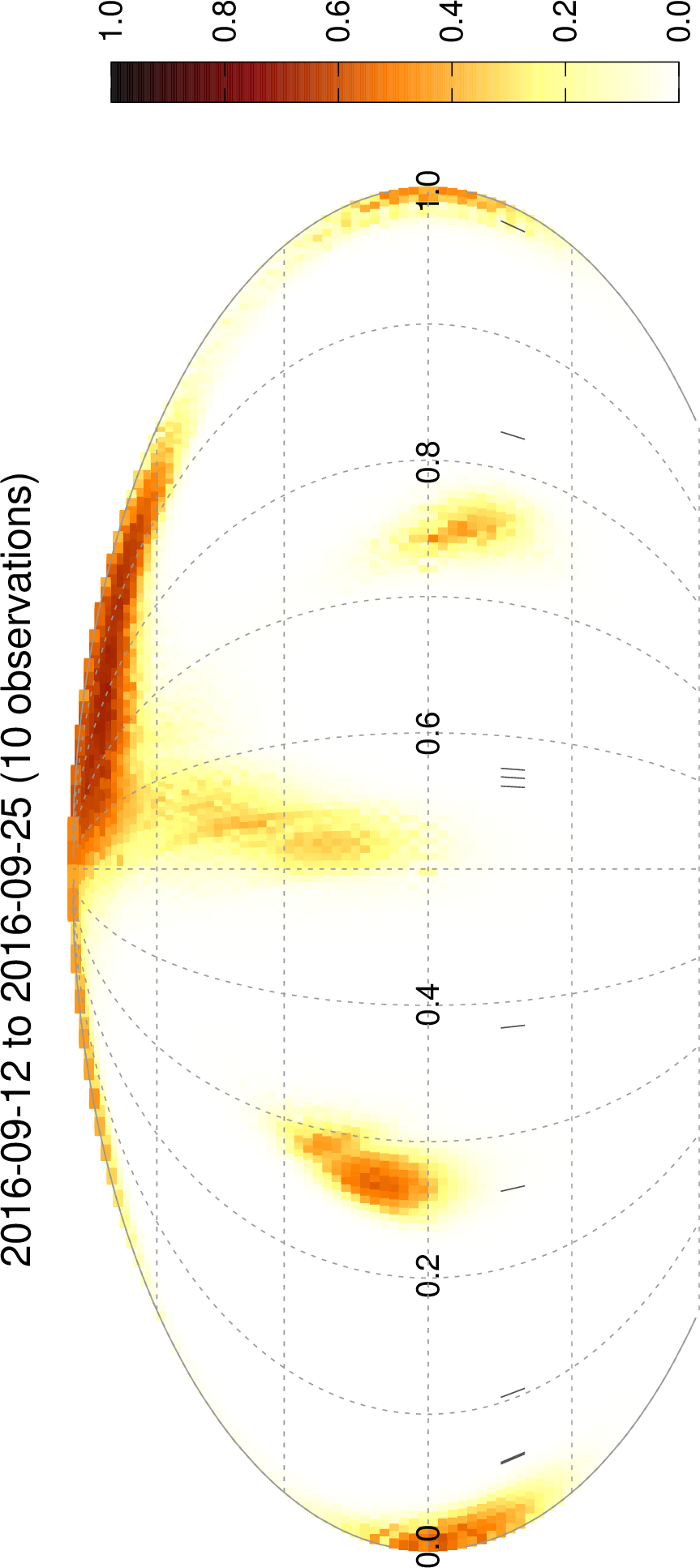}
\includegraphics[angle=270,width=0.3\textwidth]{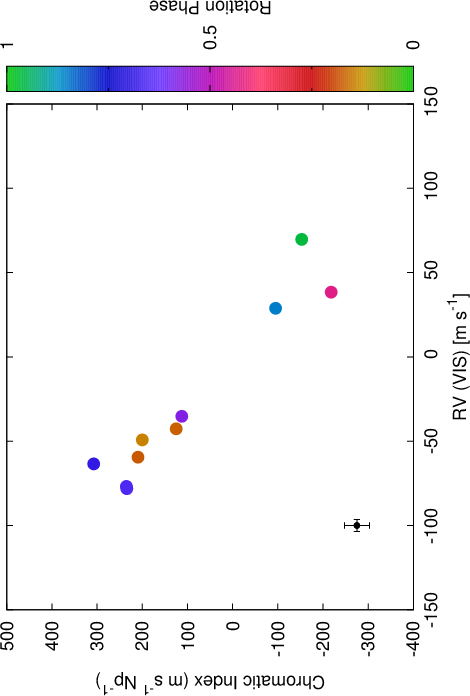}
\includegraphics[angle=270,width=0.3\textwidth]{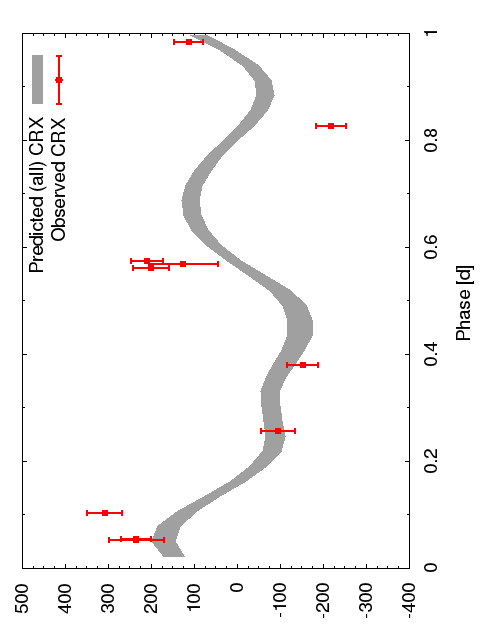}}
\\
\mbox{
\includegraphics[angle=270,width=0.4\textwidth]{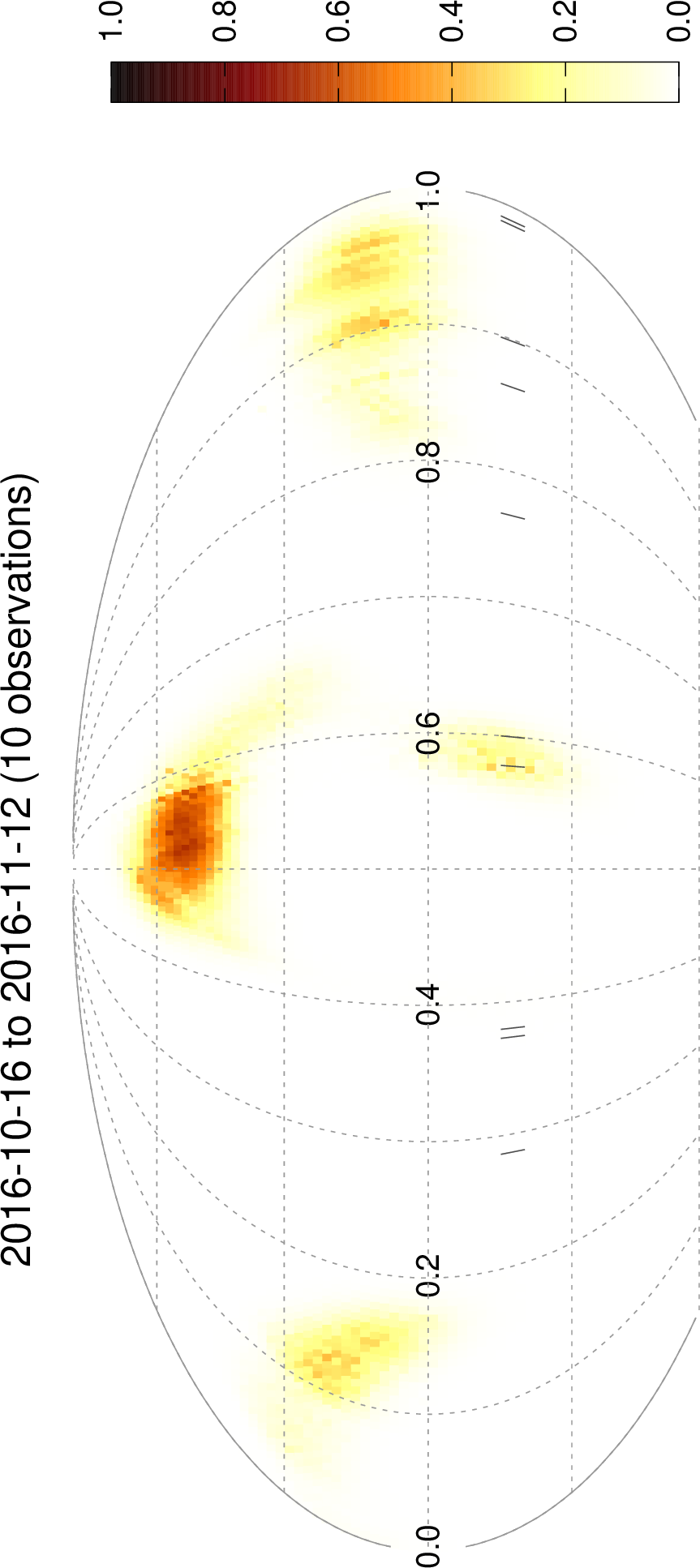}
\includegraphics[angle=270,width=0.3\textwidth]{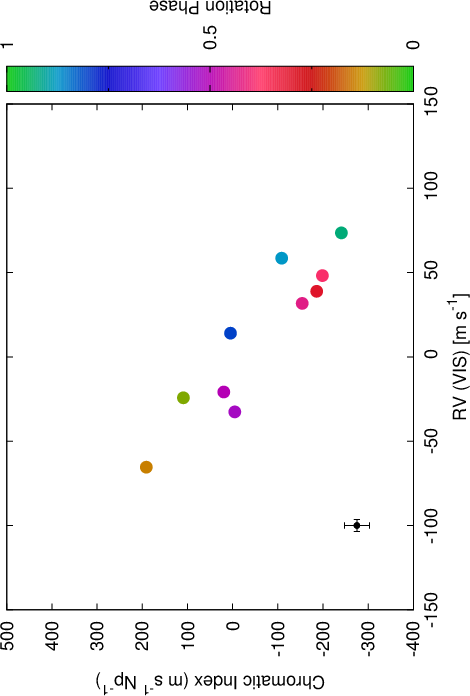}
\includegraphics[angle=270,width=0.3\textwidth]{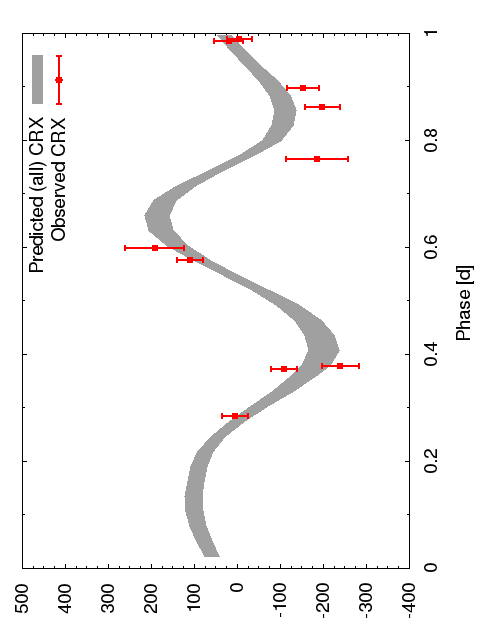}}
\\
\mbox{
\includegraphics[angle=270,width=0.4\textwidth]{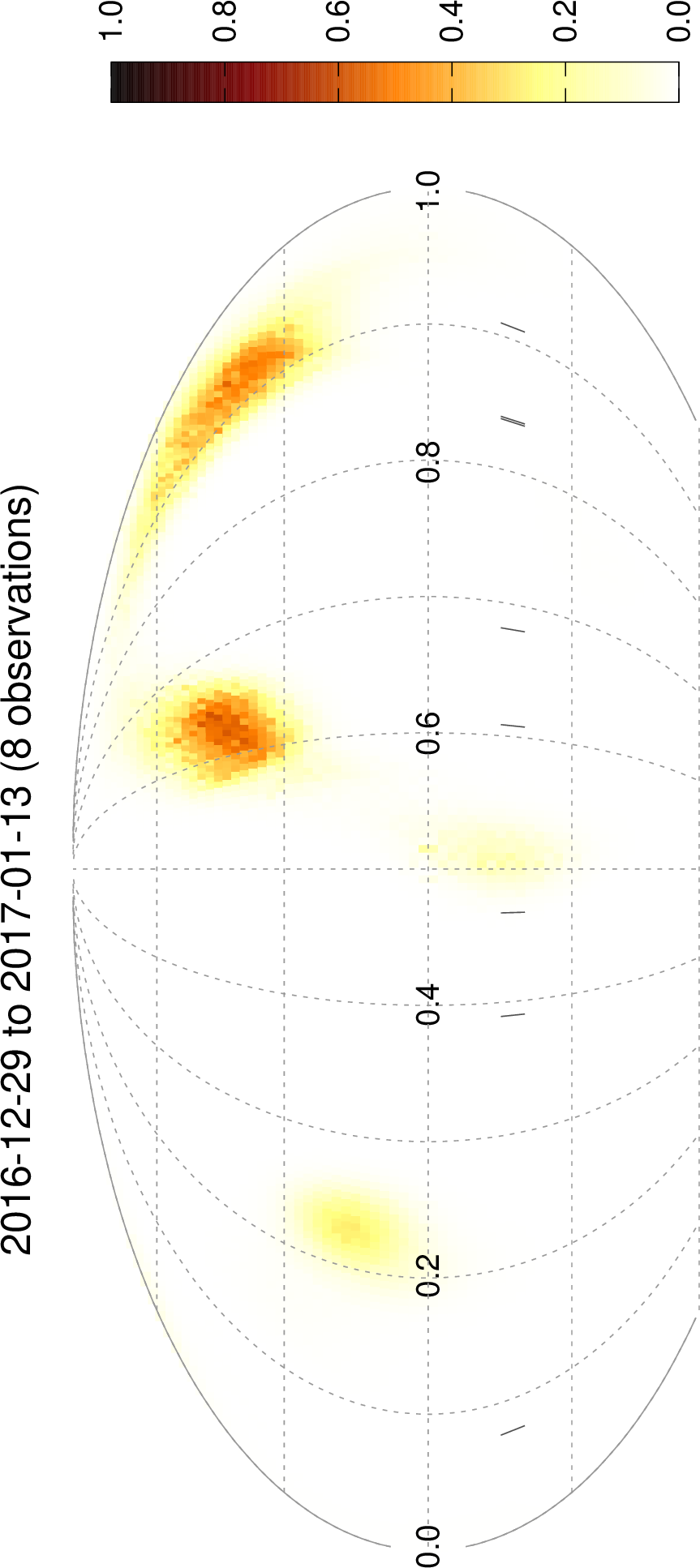}
\includegraphics[angle=270,width=0.3\textwidth]{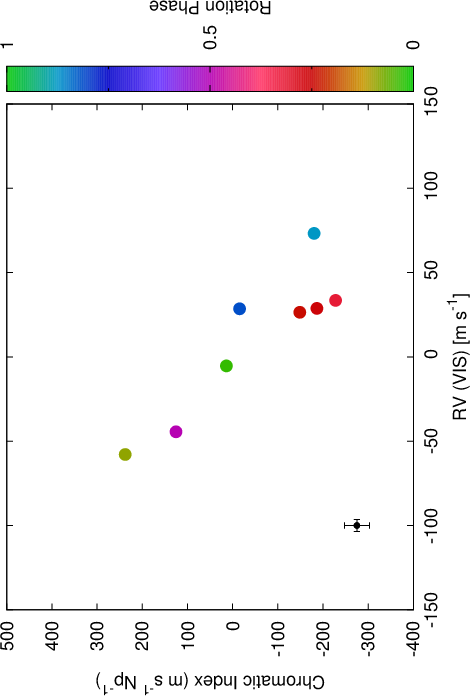}
\includegraphics[angle=270,width=0.3\textwidth]{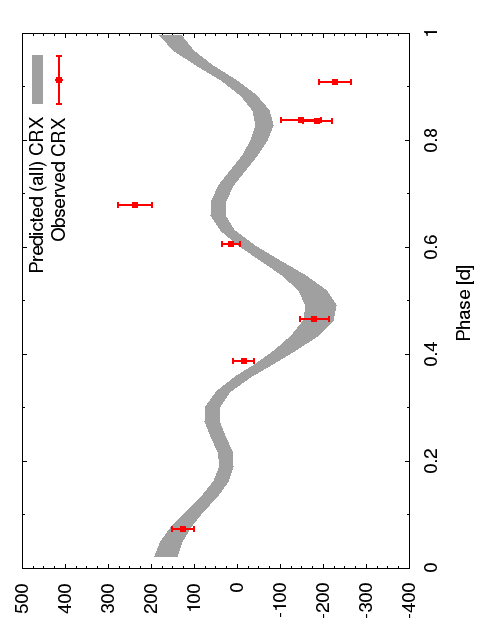}}
\\
\mbox{
\includegraphics[angle=270,width=0.4\textwidth]{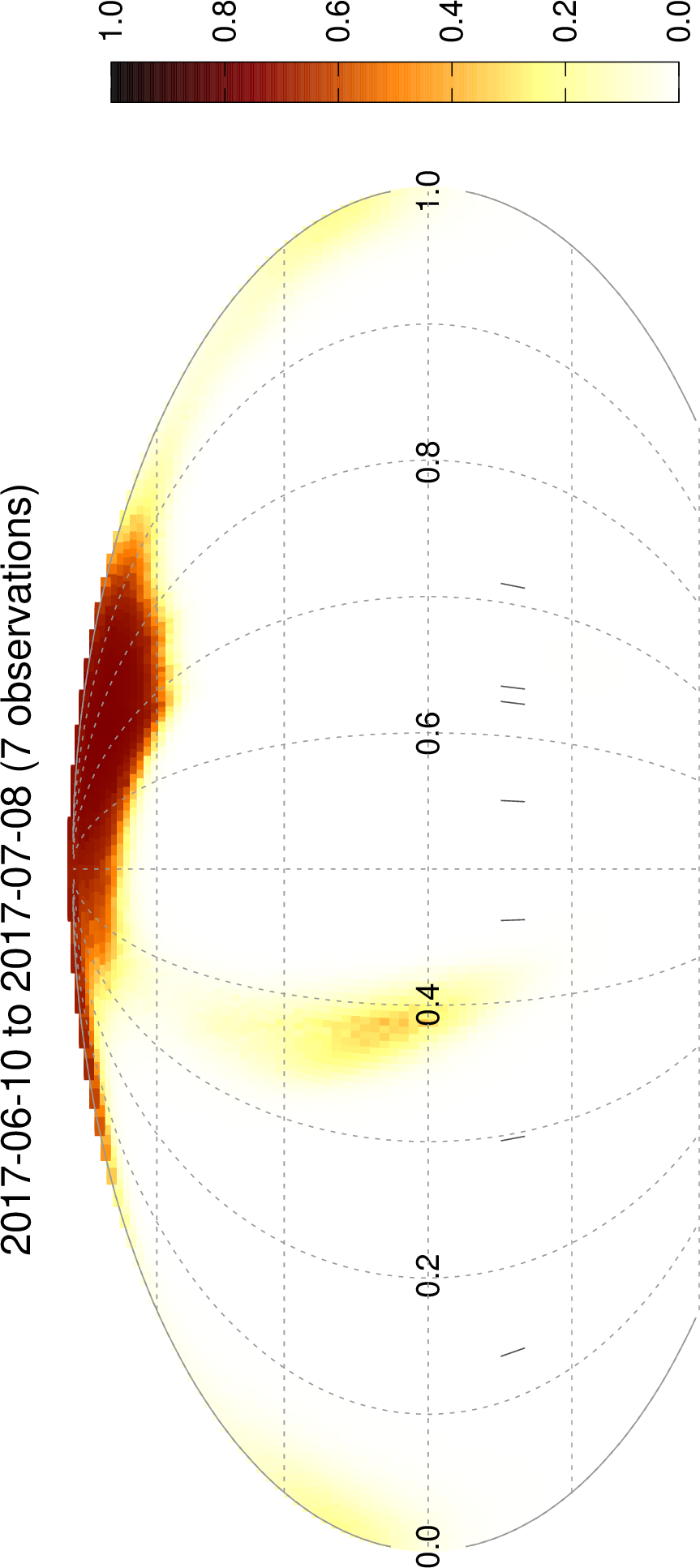}
\includegraphics[angle=270,width=0.3\textwidth]{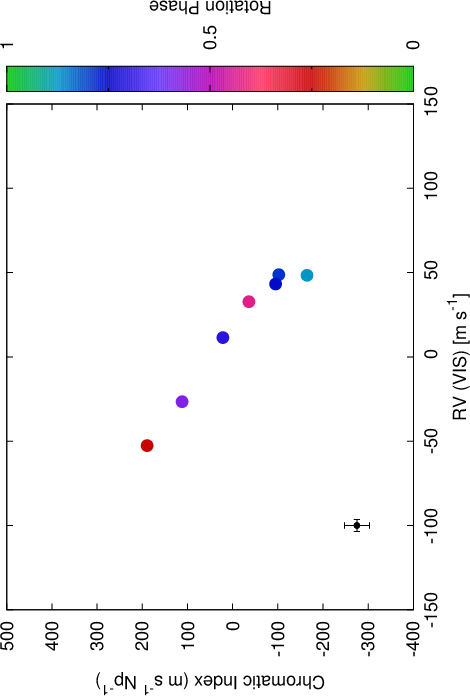}
\includegraphics[angle=270,width=0.3\textwidth]{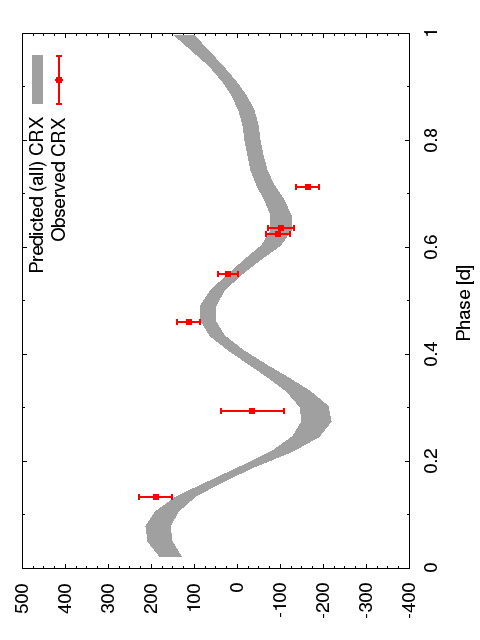}}
\\
\mbox{
\includegraphics[angle=270,width=0.4\textwidth]{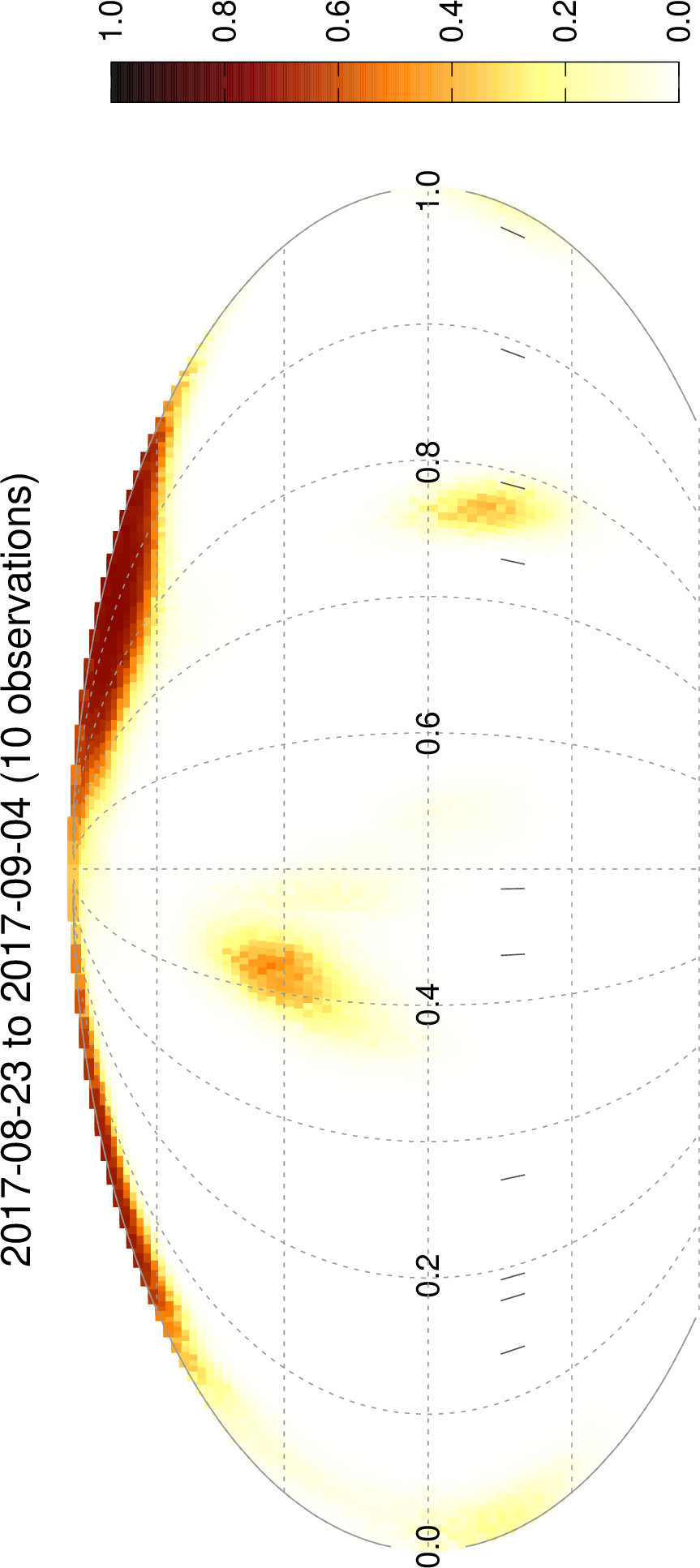}
\includegraphics[angle=270,width=0.3\textwidth]{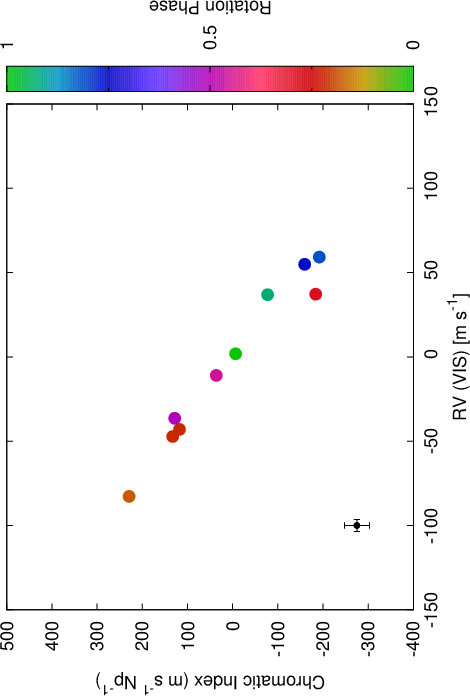}
\includegraphics[angle=270,width=0.3\textwidth]{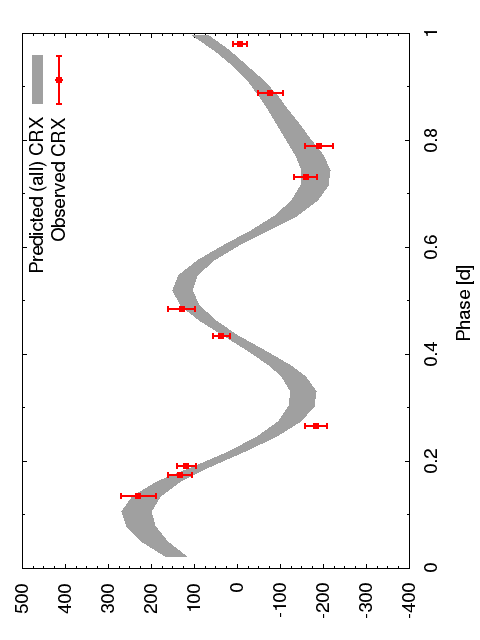}
}
\end{figure*}
\begin{figure*}[ht] 
\centering
\mbox{
\includegraphics[angle=270,width=0.4\textwidth]{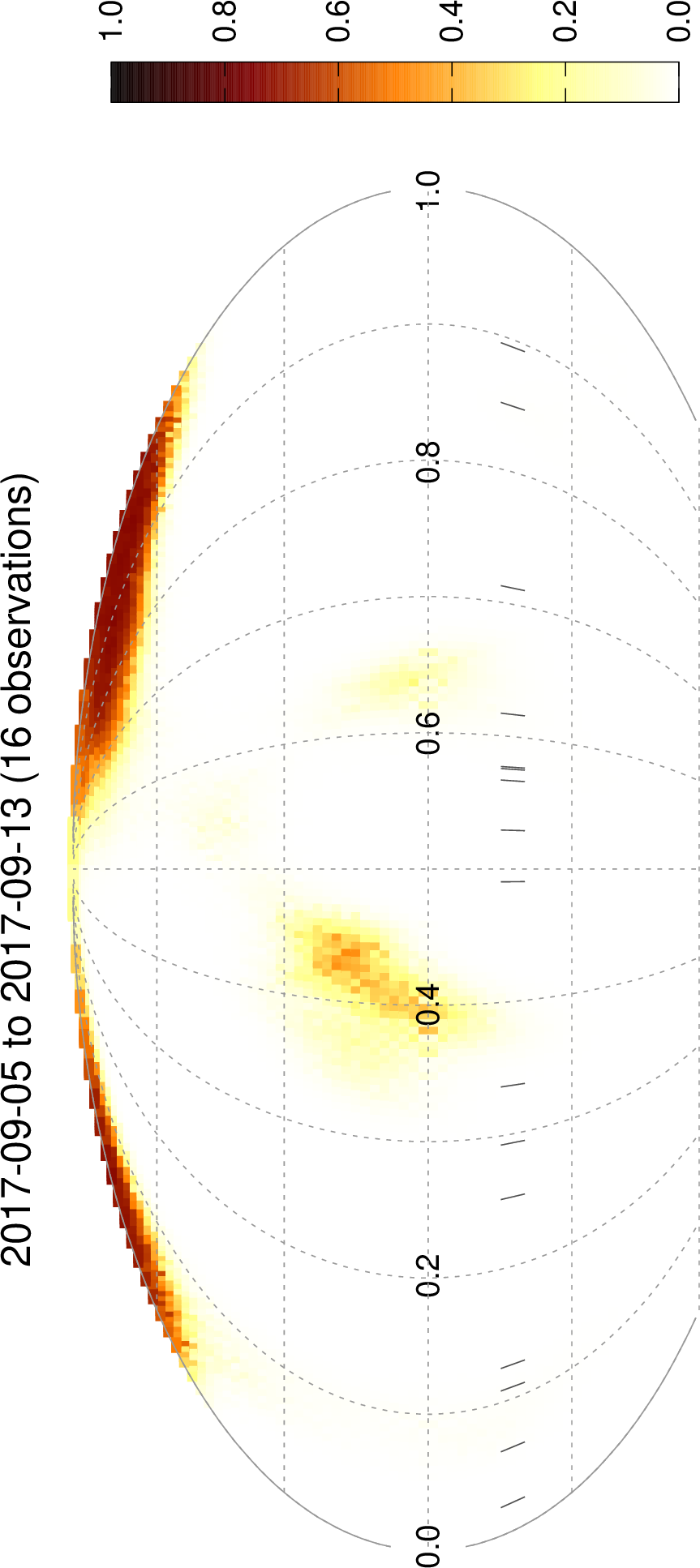}
\includegraphics[angle=270,width=0.3\textwidth]{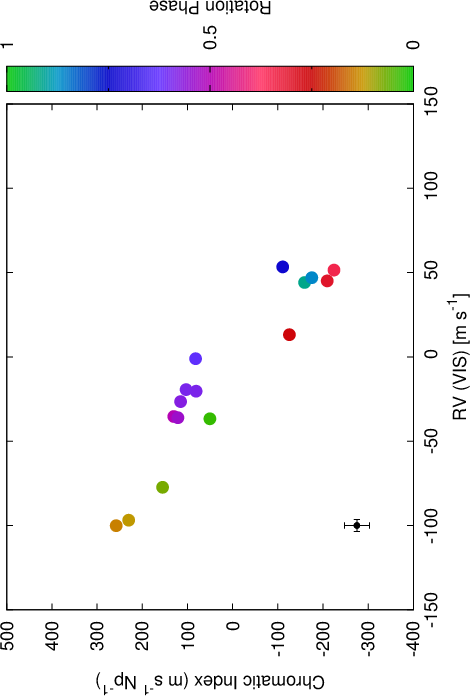}
\includegraphics[angle=270,width=0.3\textwidth]{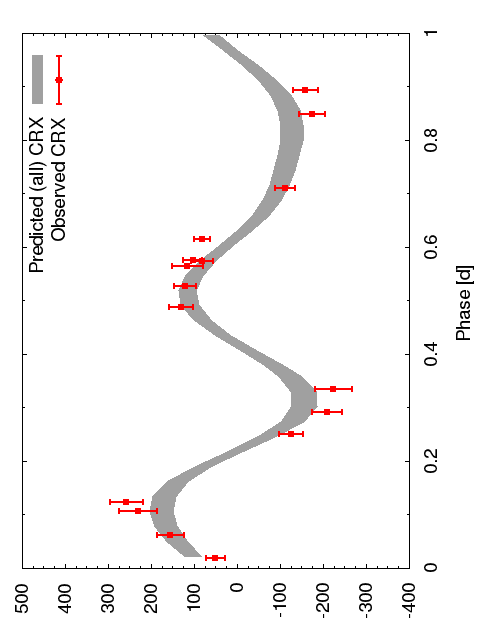}
}
\\
\mbox{
\includegraphics[angle=270,width=0.4\textwidth]{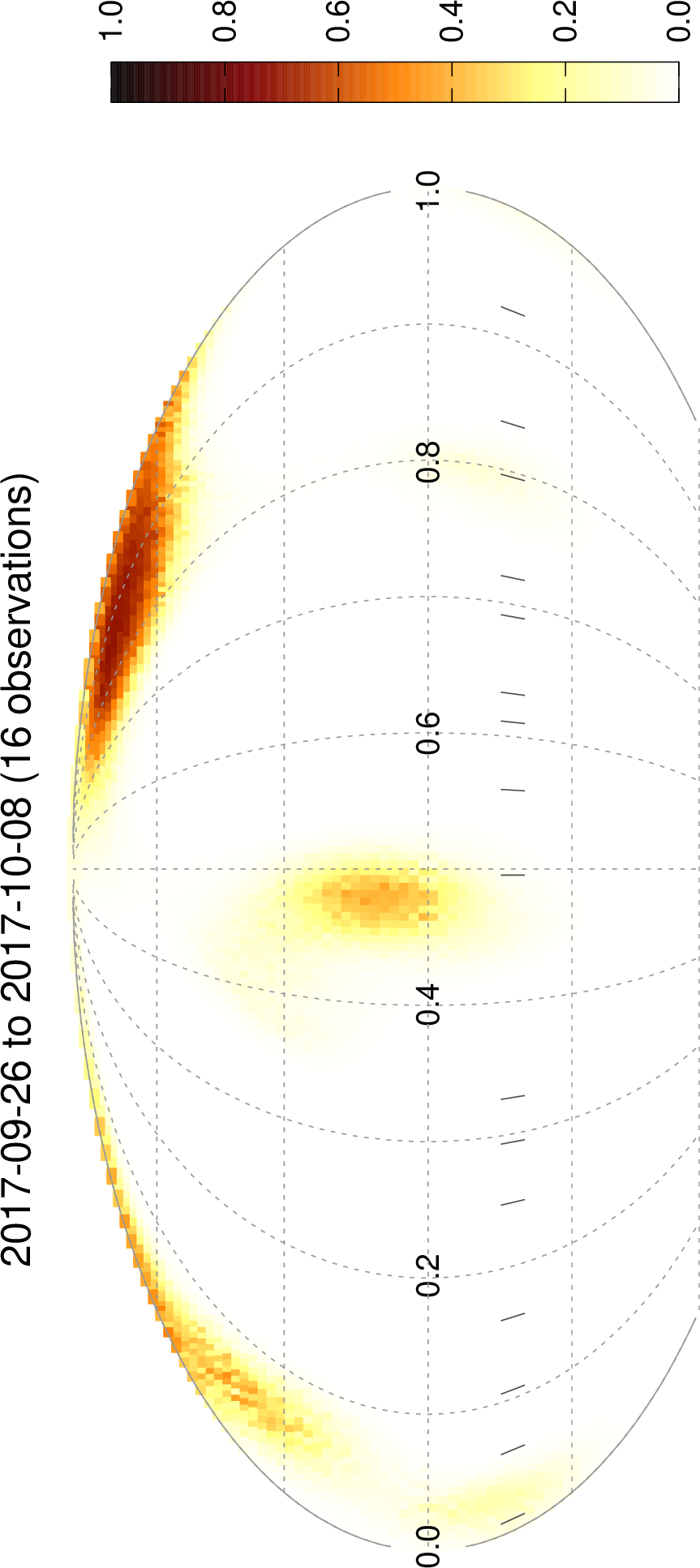}
\includegraphics[angle=270,width=0.3\textwidth]{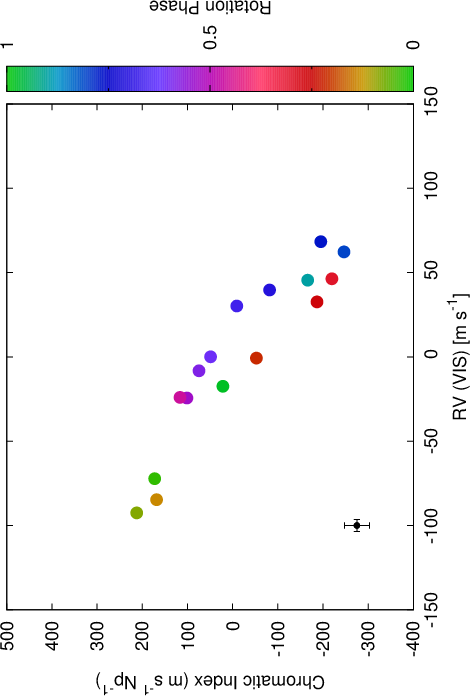}
\includegraphics[angle=270,width=0.3\textwidth]{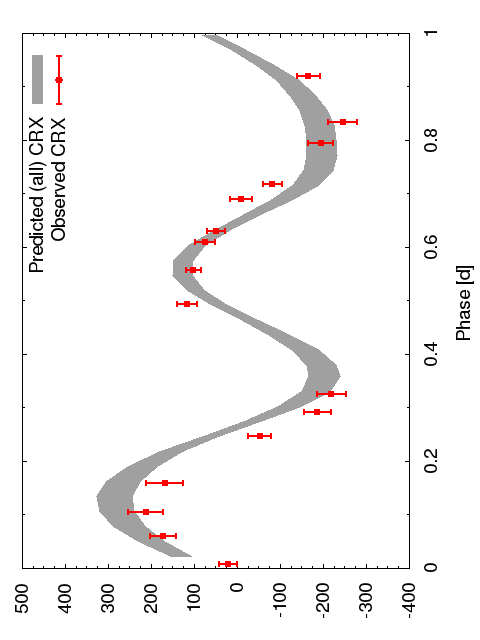}}
\caption{Low-resolution Doppler images of EV~Lac (left panel) with the observed CRX as a function of RV (middle panel) and the centre-of-light values (right panel), and the predicted CRX values (right panel) as a continuous distribution.  The CRX vs RV points are coloured by the rotational phase computed using the {\em TESS} rotation period. A representative error bar is shown in black.}
\label{fig:DImaps2}
\end{figure*}

\section{Wavelength dependence of RVs}

The CRX for EV~Lac shows a strong anti-correlation with RV, or negative chromaticity \citep{Zechmeister2018A&A...609A..12Z}, meaning that the RV measured per order decreases towards longer wavelengths.  This anti-correlation indicates that the main source of RV variation measured on EV~Lac are features that induce a smaller RV at longer wavelengths, such as dark starspots.  \\

The CRX-RV anti-correlation is shown in Figure~\ref{fig:CRX_RV}, where the data points are coloured for rotation phase, dLW and pEW (H$\alpha$), and TiO 7050 \AA\ band.  The CRX shows a slight dependence on the stellar rotation phase, that is,  green and orange points appear to be clustered in the upper left part of the anti-correlation while blue points have the tendency to be located in the lower right part (low CRX, high RV values).  The spread of colours might indicate more complex and evolving activity patterns, and is consistent with the variations in the TESS light curve. 
This is in contrast to the recent results of \cite{Baroch2020A&A...641A..69B} (see their Fig. 5) for the mid-M dwarf YZ CMi, where similar rotational phases are clustered at similar CRX / RV values.   Additionally, the CRX of EV~Lac does not show a dependence on dLW (Figure~\ref{fig:CRX_RV} top right) where high, zero, and low values of the dLW can have the full range of CRX values.  Even the presence of flares on EV~Lac, as indicated by red points in the lower left panel of Figure~\ref{fig:CRX_RV}, does not significantly impact the CRX or RV value.  This was also reported for the very active M star CN Leo \citep{Reiners2009A&A...498..853R} and the inactive M star Gl 699 \citep{Kuester2003A&A...403.1077K}, which is also known as Barnard's star . \\

The TiO bands can be used as a measure for the relative evolution of spot coverage fraction for a given star, where lower values indicate higher spot coverage fractions, and vice versa.  We investigated the changes in spot coverage for EV~Lac using the TiO band at 7050 \AA.  The results are shown in the lower right panel of Figure~\ref{fig:CRX_RV} where no clustering of high or low values of TiO 7050 with CRX is visible.  The lack of a correlation of these four proxies of stellar activity indicates that there is significant evolution of the  magnetic activity over the time-span of the CARMENES observations, and that it does not necessarily occur simultaneously in all indicators.

\begin{figure*}
\centering
\mbox{
\includegraphics[angle=270,totalheight=4.3cm,width=5.5cm]{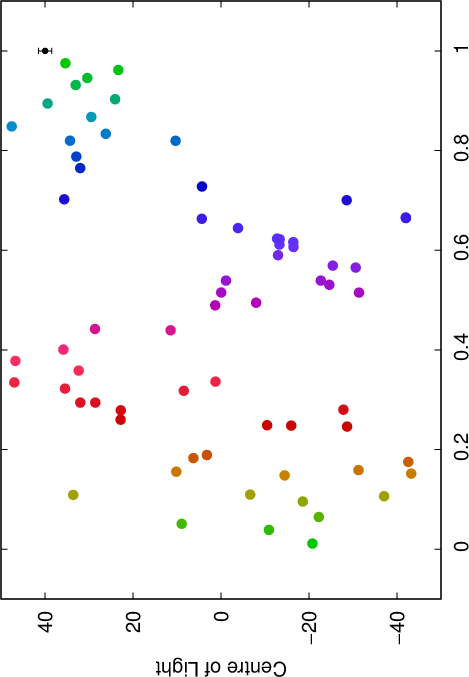}
\includegraphics[angle=270,totalheight=4.3cm,width=5.5cm]{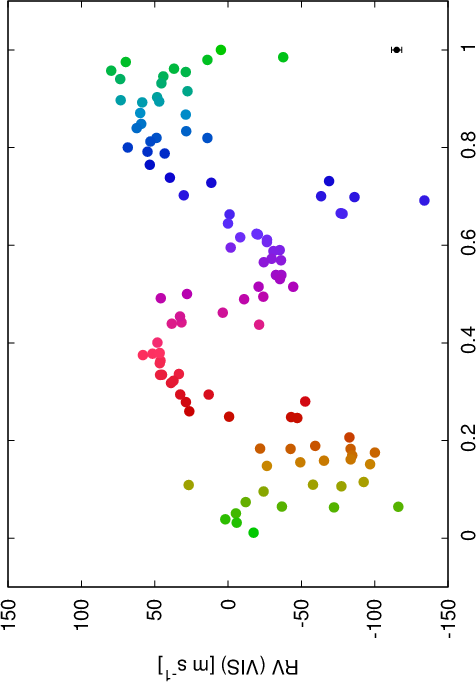}
\includegraphics[angle=270,totalheight=4.3cm,width=6.5cm]{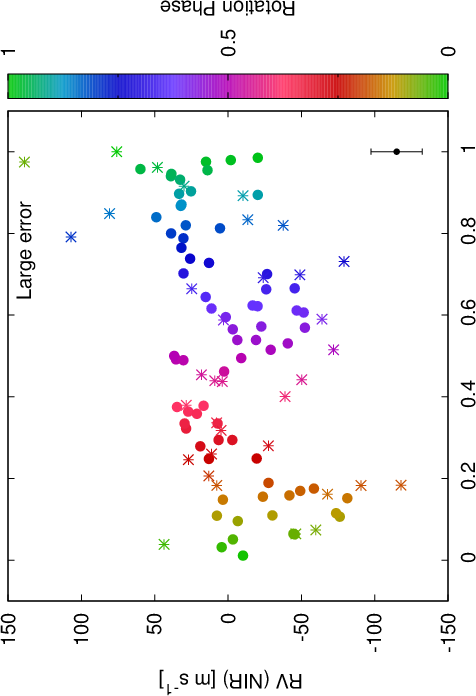}
}
\mbox{
\includegraphics[angle=270,totalheight=4.3cm,width=5.5cm]{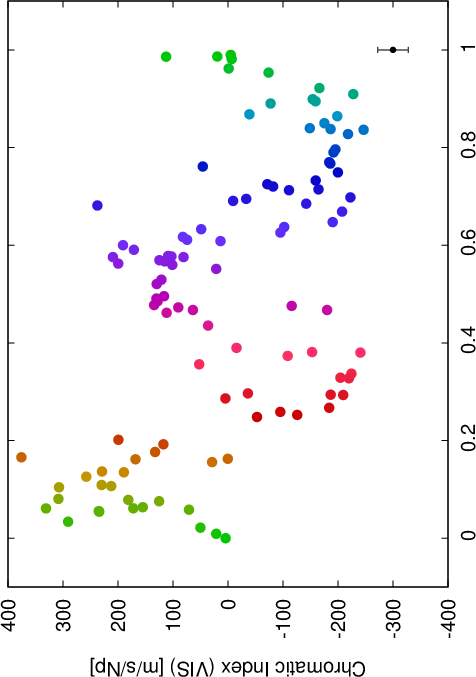}
\includegraphics[angle=270,totalheight=4.3cm,width=5.5cm]{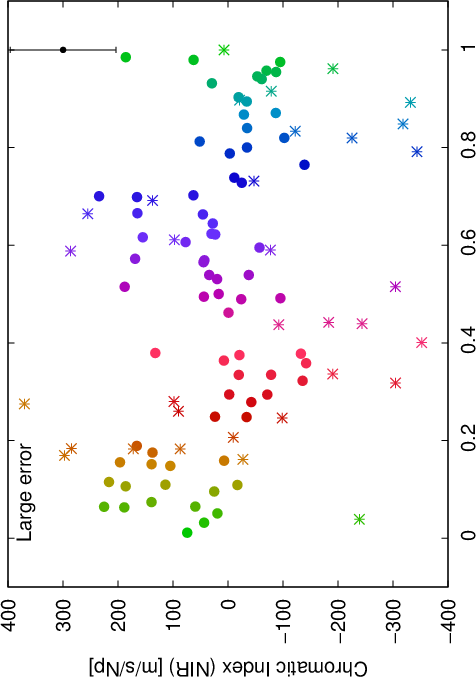}
\includegraphics[angle=270,totalheight=4.3cm,width=6.5cm]{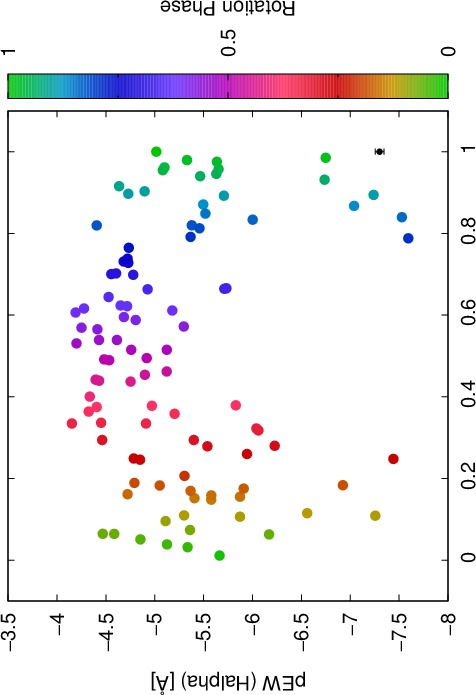}
}
\mbox{
\includegraphics[angle=270,totalheight=4.3cm,width=5.5cm]{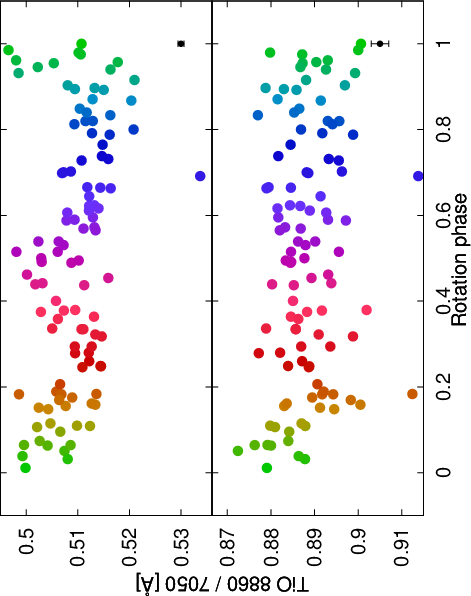}
\includegraphics[angle=270,totalheight=4.3cm,width=5.5cm]{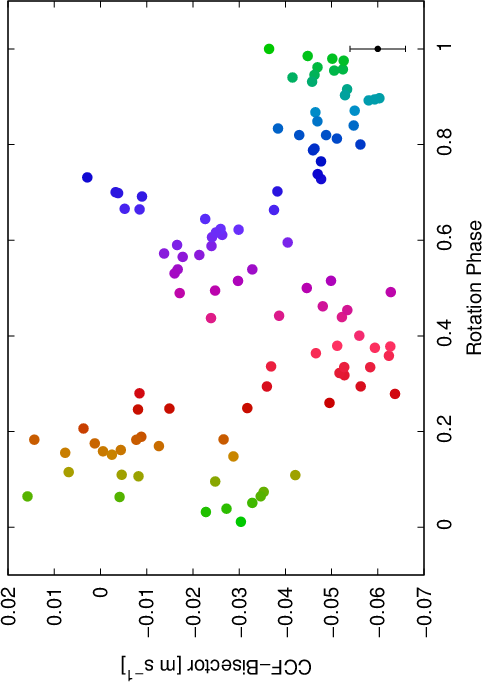}
\includegraphics[angle=270,totalheight=4.3cm,width=6.5cm]{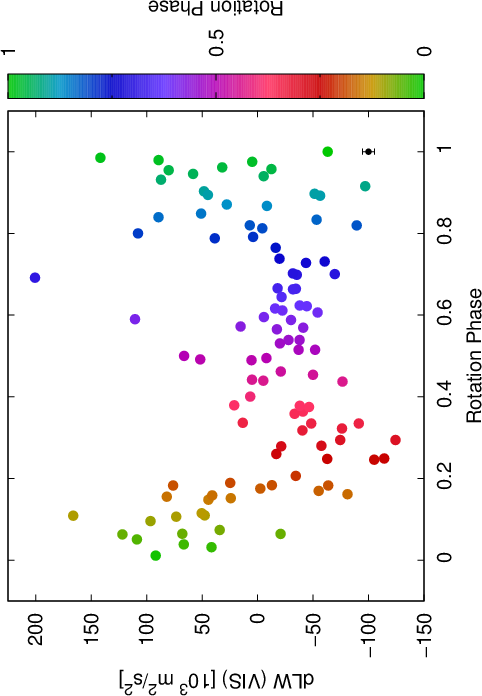}
}
\caption{ centre-of-light values for all of EV~Lac epochs used in the low-resolution Doppler images, RV (VIS) vs phase, RV (NIR) vs phase (top row), CRX (VIS) vs phase, CRX (NIR) vs phase, pEW H$\alpha$ vs phase (second row),  TiO 8860 / TiO 7050 vs phase, CCF-Bisector vs phase, and dLW (VIS)vs phase (third row).  All panels are coloured by rotation phase. For the NIR data, points with error bars that are greater than the mean error shown are indicated by an asterisk. A representative error bar is shown in black. }
\label{fig:activity}
\end{figure*}

\begin{figure*}
\def\imagetop#1{\vtop{\null\hbox{#1}}}
\centering
\mbox{
\imagetop{\includegraphics[angle=270,width=0.47\textwidth]{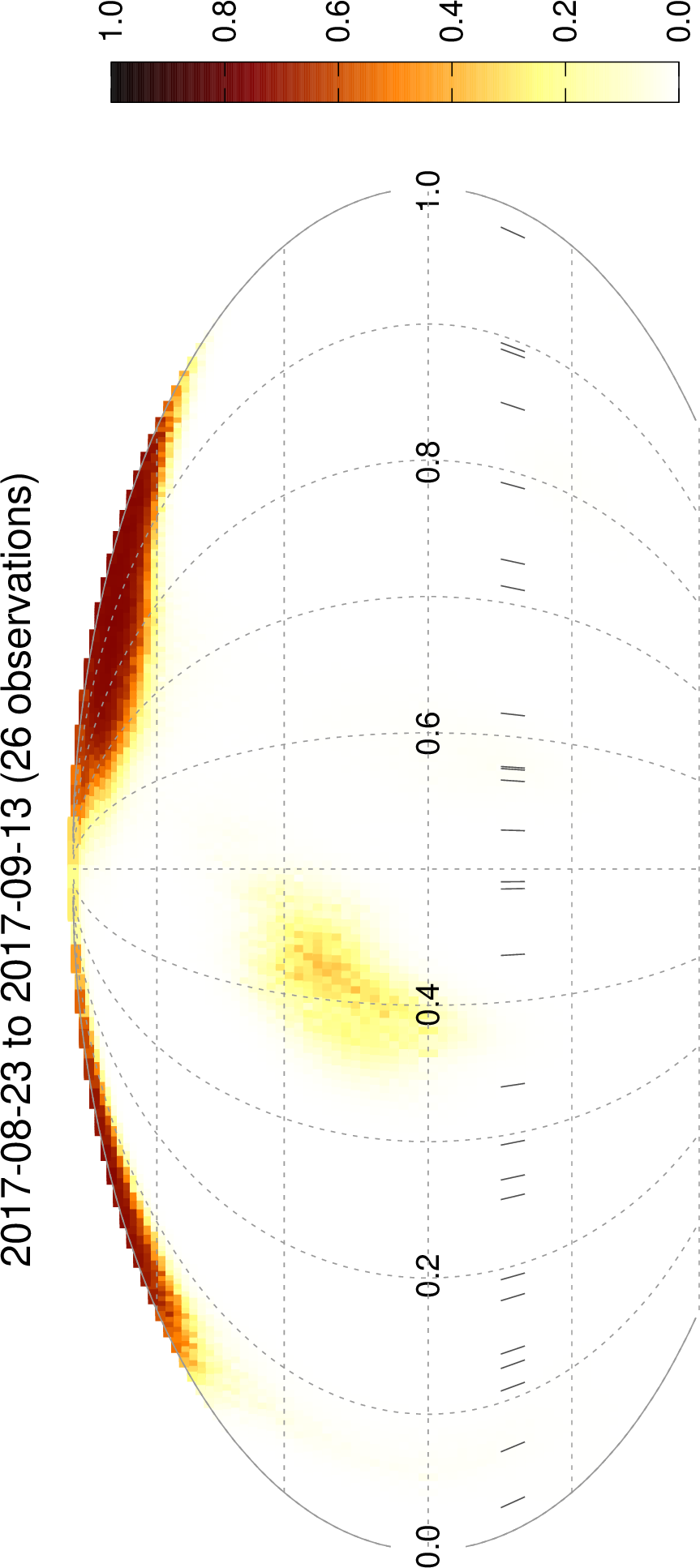}}
\imagetop{\includegraphics[angle=270,,totalheight=4.3cm,width=6.5cm]{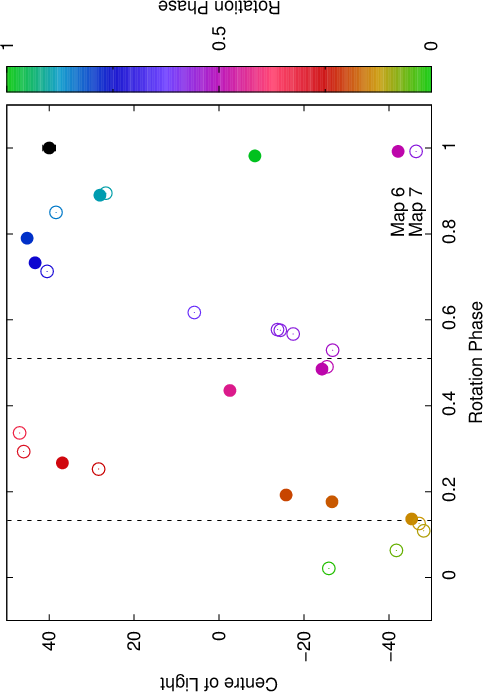}}
}
\mbox{
\includegraphics[angle=270,totalheight=4.3cm,width=5.5cm]{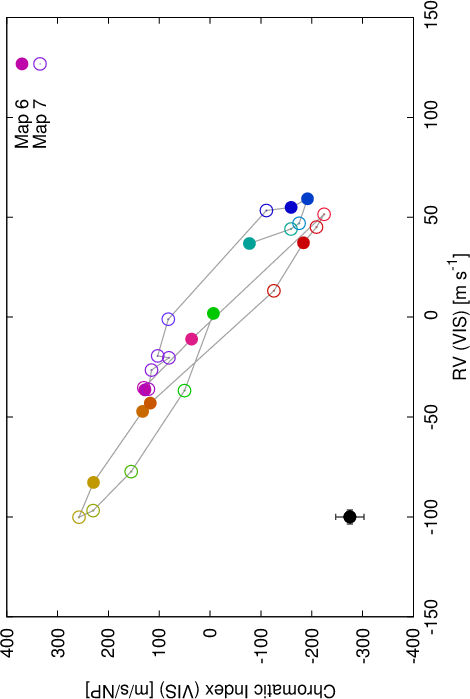}
\includegraphics[angle=270,totalheight=4.3cm,width=5.5cm]{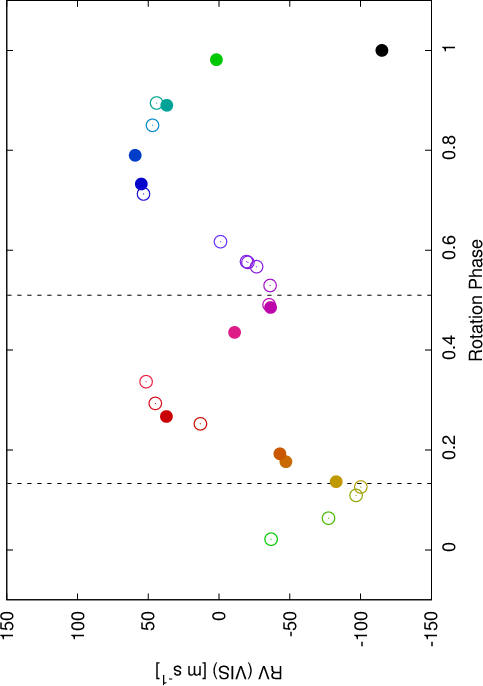}
\includegraphics[angle=270,totalheight=4.3cm,width=6.5cm]{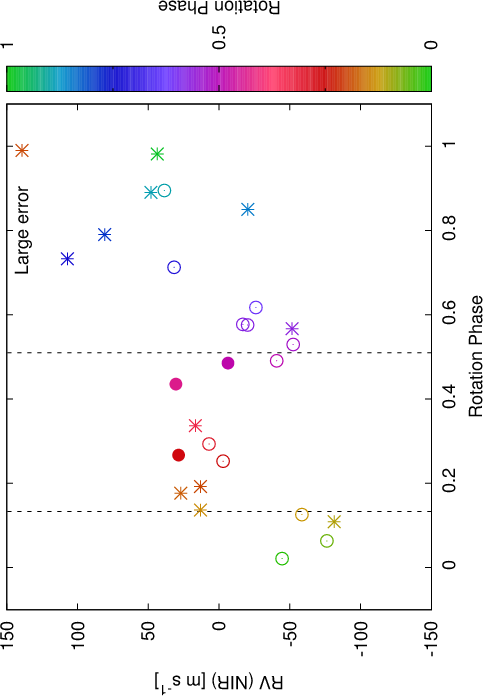}
}
\mbox{
\includegraphics[angle=270,totalheight=4.3cm,width=5.5cm]{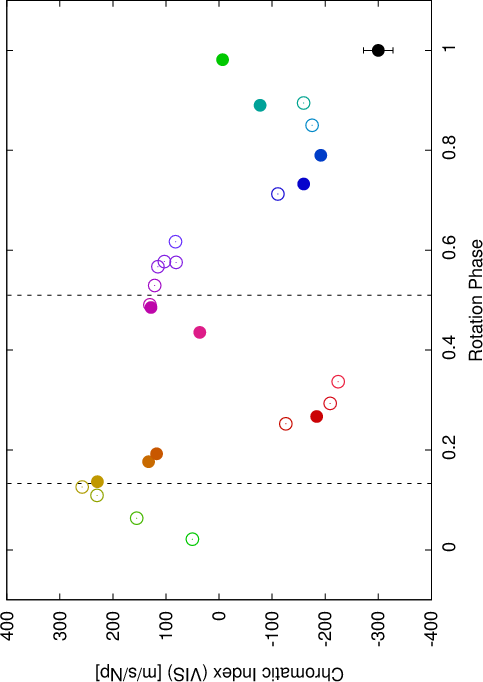}
\includegraphics[angle=270,totalheight=4.3cm,width=5.5cm]{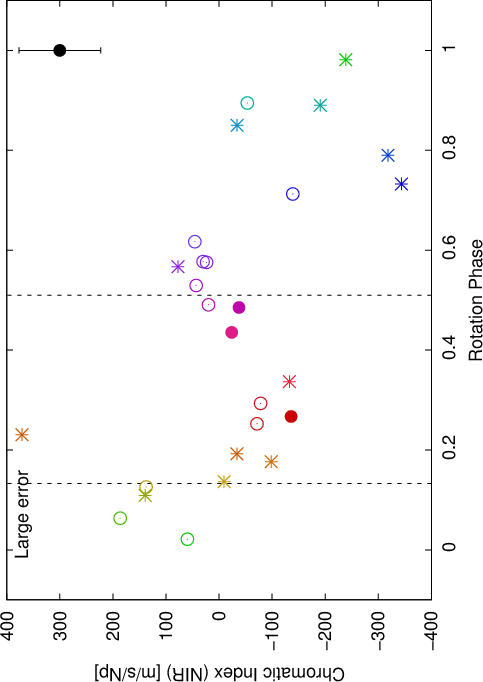}
\includegraphics[angle=270,totalheight=4.3cm,width=6.5cm]{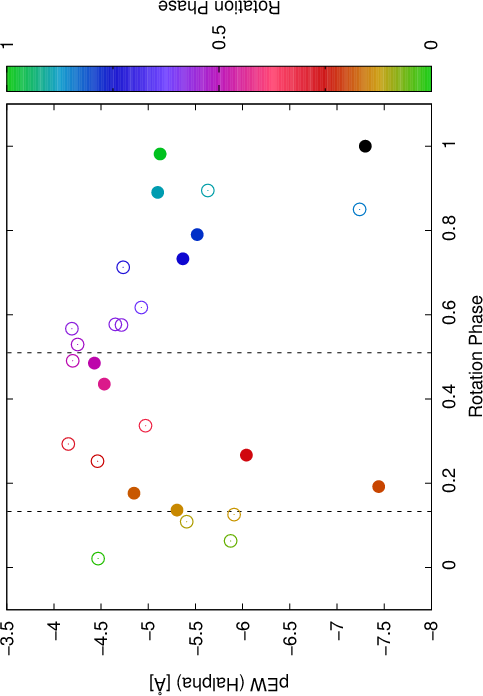}
}
\mbox{
\includegraphics[angle=270,totalheight=4.3cm,width=5.5cm]{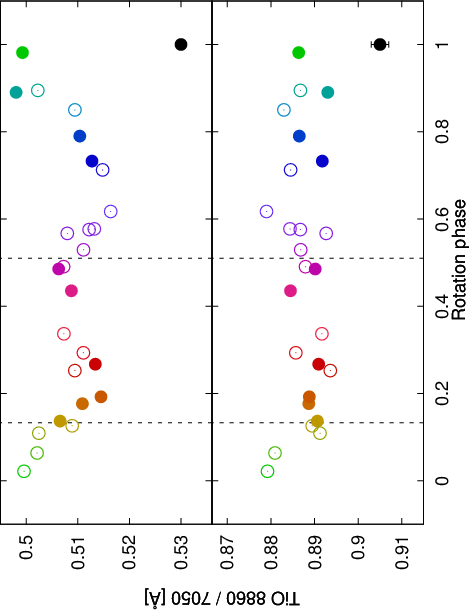}
\includegraphics[angle=270,totalheight=4.3cm,width=5.5cm]{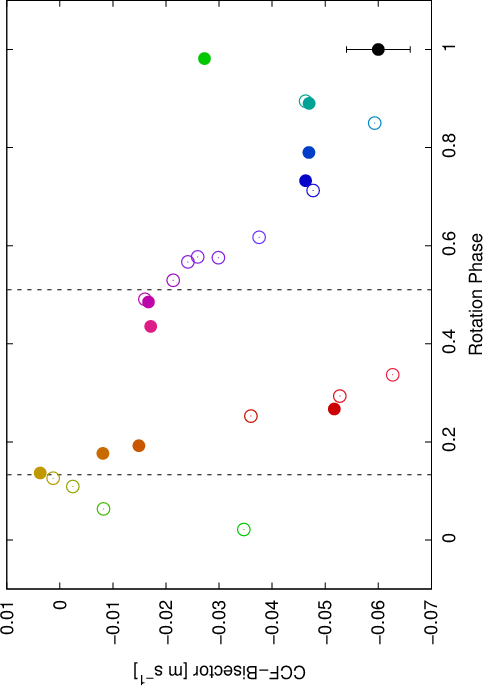}
\includegraphics[angle=270,totalheight=4.3cm,width=6.5cm]{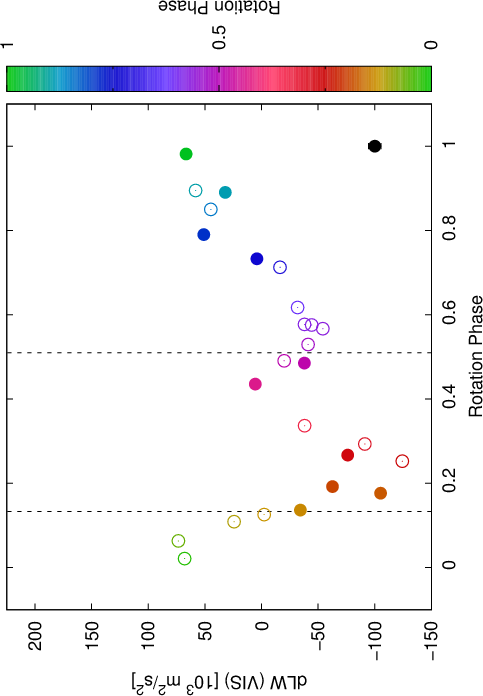}
}
\mbox{
\includegraphics[angle=270,totalheight=4.3cm,width=6.5cm]{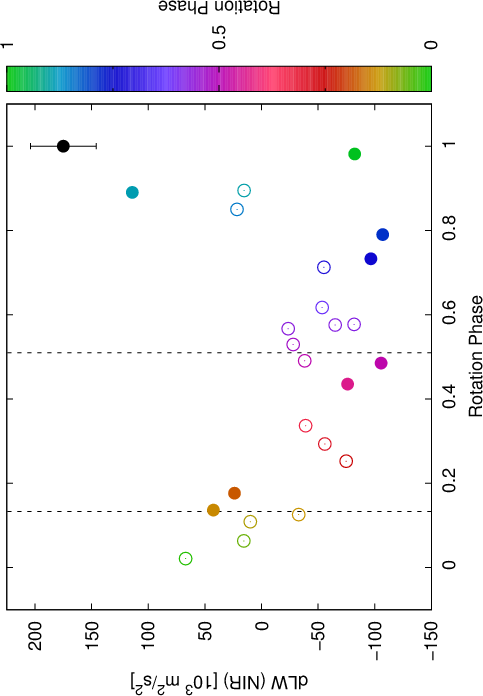}
}
\caption{As Figure 5 but considering only the observational epochs used in map 9 (= combined maps 6 and 7; top row).  Closed and open symbols are from maps 6 and 7 respectively.  (top row) Low-resolution surface brightness map (combined) using a Mollweide projection, and the resulting centre-of-light values as a function of rotational phase.  (second row) CRX as a function of RV where the rotational phases are joined by solid lines to indicate how the CRX and RV evolve as a function of phase, RV as a function of rotational phase for the VIS and NIR channels, (third row) CRX as a function of rotational phase for the VIS and NIR channels and pEW(H$\alpha$) as a function of rotation phase (fourth row) TiO 7050, 8860, CCF-Bisector and dLW (VIS) as a function of rotation phase (last row), and dLW for the NIR channel as a function of rotation phase. For the NIR data, points with error bars that are greater than the mean error shown are indicated by an asterisk. A representative error bar is shown in black for all panels.}
\label{fig:activity_maps9+10}
\end{figure*}

\section{Low-resolution surface brightness maps}

In this section we describe the activity-correction technique  by \cite{Barnes2017MNRAS.466.1733B}, which uses low-resolution Doppler imaging, and its application to EV~Lac.  While tomographic techniques are able to recover more complex images for rapidly rotating stars, low-resolution images can also be useful to identify larger active regions. Moreover, the regularised profile fits used to derive the images can also be used to model activity-induced asymmetries of the stellar absorption line profile. This is particularly useful when searching for exoplanets in the presence of astrophysical noise and has been shown to improve planet mass detection limits by up to an order of magnitude \citep{Barnes2017MNRAS.466.1733B}.

\subsection{CARMENES observations}

The EV~Lac spectra were divided up into eight subsets, each comprising at least ten observations and covering not more than a few stellar rotation periods.  This observational strategy is important to mitigate the impact of correlated noise induced by the evolution of stellar activity.  An example of this is the lifetime of starspots which typically last for several rotation periods.  

In Figure~\ref{fig:CRX_RV} (top left panel) we show the CRX-RV anti-correlation with points coloured for the rotation phase of the observation.    In Figure~\ref{fig:CRX_time_COL_RV} the CRX is plotted as a function of time (in units of rotation phase), and coloured from RV bins, to further understand the evolution of EV~Lac over the full CARMENES data set.  The eight subsets that are used for low-resolution Doppler imaging are shown as vertical light grey regions.  

While the general CRX-RV anti-correlation holds where positive CRX values correspond to negative RV values, and vice versa, similar CRX values can have correspondingly medium or very high RV values.  An example of this are the CRX values in the range of --200 to --100 where the points can be both green and blue, corresponding to high and medium RV values. This is likely a consequence of  the deviation from a simple single-spot scenario and shows the importance of using observations that are well sampled with respect to rotational phase when for investigating and quantifying the activity of active stars.  

\subsection{Method description}
\label{section:method}
The technique of Doppler imaging is typically applied to stars with \vsin\ values $>$ 20\,km\,s$^{-1}$, for which starspots or starspot groups can be resolved more precisely due to the large effective number of resolution elements across the stellar disk. For slower rotators, the spatial resolution that can be achieved is more limited, although this depends on the spectral resolution of the instrument, and ultimately on the astrophysical broadening mechanisms. \cite{Barnes2017MNRAS.466.1733B} showed that the technique of Doppler imaging can be used to create a low-resolution image that is then used to calculate the centre-of-light, or starspot-induced, RV correction.  The technique has been shown to be effective for stars with \vsin\ of 1\,-\,10 km\,s$^{-1}$. It can reduce the starspot-induced noise by an order of magnitude. For a more in-depth  description of the technique we refer to \cite{Barnes2017MNRAS.466.1733B}. 

\subsection{Least-squares deconvolution}
\label{section:method_LSD}

Since a Doppler image solution is not unique for finite data sampling and finite S/N, iterative fitting with image regularisation is usually employed. A common method is to require that the image entropy is maximised for a fixed goodness of fit. This ensures that images are recovered with the least amount of information so that noise is minimised, and the degree of spot coverage is minimised. The regularisation constraint is particularly important if the data S/N is low. A higher S/N can be achieved by optimally combining the many spectral features in a single line profile. We used a least-squares deconvolution (LSD) to obtain a single high S/N line profile for each spectrum \citep{donati97zdi,barnes98aper}. This method enables the small line distortions due to stellar temperature inhomogeneities to be seen while ensuring that subsequent iterative regularised fitting of line profiles is not too demanding computationally. The LSD procedure removes the effects of line blending when considering large numbers of lines. LSD assumes that all absorption lines have  the same shape. However, a starspot will induce a larger distortion at bluer wavelengths than at redder wavelengths, which  yields an RV shift with a wavelength-dependent magnitude, similar to the CRX. An LSD profile represents a mean unblended absoprtion line. For EV~Lac, we used a reference line list derived from observations of YZ CMi that was also used compute the CCF parameters for EV~Lac \citep{Lafarga2020A&A...636A..36L}. A total of $N_{\rm lines} = 2193$ absorption lines in the wavelength range $5362 \AA < \lambda < 8939 \AA$ were used to derive the LSD profile from each spectrum, resulting in a mean effective wavelength of $\lambda_{\rm mean} = 7252 \AA$.   We find a continuum S/N$_{\rm LSD} = 1110 \pm 393$, with an effective multiplex gain of 17 when compared with the mean input spectrum noise of $67.7 \pm 21.3$ (see \S \ref{section:observations}).  Because we performed low-resolution Doppler imaging, we needed to use the information content from all lines.   Each LSD profile was derived in the barycentric drift-corrected (CARMENES) velocity reference frame; any remaining velocity shifts were assumed to be due to stellar activity or (an) orbiting exoplanet(s).  We did not use the measured CARMENES RV in the image reconstruction process.

\subsection{Doppler images of EV~Lac}

Low-resolution brightness maps of EV~Lac were reconstructed from the LSD profiles using the  Doppler Tomography of Stars routine, DoTS \citep{cameron01mapping}. DoTS assumes a two-temperature model with limb-darkened intensities appropriate for assumed photospheric and spot temperatures. The stellar disk model has a finite number of pixels. Each pixel can take an intensity between the spot and photospheric level and is represented by a normalised spot filling value. An image is derived by iterative, maximum-entropy regularised fitting of the time-series spectra. DoTS has been used to image cool starspots on fast-rotating early-M dwarfs \citep{barnes01mdwarfs,barnes04hkaqr} and has more recently revealed starspot patterns on some of the most active known mid-M dwarf stars \citep{barnes15mdwarfs, 2017MNRAS.471..811B}.  \cite{berdyugina05starspots} and our findings for mid-M dwarfs indicate that intensities corresponding to $T_{\rm phot} - T_{\rm spot}$ of a \hbox{few-hundred K} are needed to map cool starspots. This corresponds to intensity ratios at disk centre in our model of $I_{\rm c,phot} / I_{\rm c,spot} < 10$.  By contrast, for EV~Lac, we require $T_{\rm phot} - T_{\rm spot} = 1000$ K, corresponding to $I_{\rm c,phot} / I_{\rm c,spot} = 50$ at the $\lambda_{\rm mean} = 7252 \AA$ of the LSD profiles. We were unable to account for the line distortions adequately with smaller contrasts. This is primarily due to the low \vsin\ of EV~Lac and to the limited number of resolution elements across the stellar disk.

\subsection{Low-resolution images}

As part of the low-resolution Doppler-imaging procedure, the large-scale surface brightness distributions were reconstructed. In other words, small groups of distributed spots cannot be distinguished from individual large spots and were reconstructed as such. Therefore, we emphasise that the images should not be over-interpreted. Nevertheless, using the reconstructed features, we can determine the location of the line centre and account for the distortion in the line shape resulting from the presence of spots.    

The resulting spot distributions are shown in Figure~\ref{fig:DImaps2}.   The large-scale spot distributions of EV~Lac over the time-span of our observations typically have one to two high latitude spots with several low latitude spots that are observed to evolve on quite short timescales.  Two maps from 2016 October 16 to 2017 January 13 (maps 3 and 4) do not show any high-latitude structure despite the good rotational phase coverage of the observations. 

We tested the reliability of the reconstructed surface images by also excluding lines from the LSD procedure. We removed lines that are known to be chromospherically sensitive (Lopez Gallifa et al. 2021 CS 20.5 poster).  A total of 63 lines were excluded from the original line list.  The resulting reconstructed images do not change noticeably suggesting that the removed lines have little impact on the overall LSD line profile. The degree to which these lines might impact on the brightness images and RVs requires further investigation. The effects may become more noticeable for stars that are less active than EV~Lac.  Other works that have investigated the optimisation of linelist for EV~Lac include \cite{Bellotti2021arXiv211010633B}.

\subsection{centre-of-light calculation}

For each observation, the mean intensity-weighted Doppler velocity of the visible pixels $j$ on the stellar model is given by

\begin{equation}
   <v_{\rm CofL}> =  \frac{\Sigma_j (v_j i_j)}{\Sigma_j (i_j)}
   \protect\label{eqn:v_cofl}
\end{equation}

where $v_j$ and $i_j$ are the velocity and intensity of the $j{\rm th}$ pixel. We subsequently refer to the velocity $<v_{\rm CofL}>$ as the centre-of-Light (CofL). The centre-of-light for each observation is calculated directly from the DI maps after iteratively fitting the time-series profiles. 
 Since it is effectively a weighted first moment of the stellar velocity, it is expected to yield a very similar result to the activity-induced cross-correlation velocity shifts. Because the centre-of-light is derived from the spot model, it can be estimated at any observation phase, including during small gaps in the phase coverage of the observations. It thus has the potential of identifying the full range of activity-induced RVs that poorly phase-sampled observations do not permit. A more thorough comparison of changes in activity-induced effects between observing epochs can thus be made.

\section{Tracing the activity features}

\subsection{Combined subsets of data}

The computed centre-of-light values are shown along with the activity and CCF parameters in Figure~\ref{fig:activity} for the full data set. The global anti-correlation of the CRX with RV is clearly visible as is the global double-dip shape of the RV (VIS), the CRX (VIS), CRX (NIR), and TiO 7050 activity indices. While the global shape of each activity indicator remains constant over many rotations, there is a significant evolution of smaller-scale activity, which decreases the amplitude of the dips, and in some cases fills in the dip.  In all panels, points with the same rotation phase, or colour,  can show a wide range in activity variation.  

\subsection{Map 9: Combined maps 6 and 7}

We combined the 26 observations we used to obtain maps 6 and 7 (in Fig. 13) to derive a single map 9, spanning 21 days from 2017 August 23 to 2017 September 13. This provides the best phase sampling in a relatively small (i.e. $\sim 4.8$) number of stellar rotations.  The surface brightness distribution for the map 9 reconstruction is shown in the top left panel of Figure~\ref{fig:activity_maps9+10}, which shows one large high-latitude or polar spot from phases 0.7 to 0.05, and one smaller and weaker low-latitude spot at phase 0.4.  The centre-of-light values trace these reconstructions.  

The variation in CRX as a function of RV is shown in row 2, the left panel of Figure~\ref{fig:activity_maps9+10}.  Over the rotational phase of the combined map 9, the CRX first increases and then decreases, and then this pattern is repeated, leading to two figure-of-eight shapes.  This pattern is closely mirrored by both the centre-of-Light values and the RV.  The CRX at NIR wavelengths shows a similar dependence, but it is less apparent as the photosphere to spot contrast ratio is much lower at longer wavelengths and also because the errors of the NIR RVs are larger.

The variation in pEW (H$\alpha$) does not appear to be directly correlated with the RVs, the CRX, or the centre-of-light values.  Similarly, it seems that for the TiO 7050 band, there is a phase shift in the rotation phases where the maximum spot value occurs.  The reason is that TiO is sensitive to the total spot coverage on the star.  For example, the RV will not be very strongly influenced by the large spot at phase 0.9 because of its high latitude; the TiO tracer will nevertheless indicate an increased spot coverage independent of its location.   There could be additional small spot features that are not resolved by either the RV or the low-resolution Doppler imaging.  

The CCF-Bisector (lower left panel) is very efficient at tracing the spot coverage and agrees very well with the global shape of the centre-of-light, RV and CRX variations as a function of rotational phase.   This is because a chromatic effect is also associated with the CCF-Bisector as redder lines are generally weaker.  The dLW also seems to have a double-dip structure, but again slightly shifted in phase compared to the RV and spot variations.  

\section{Correlation of activity indices}

We searched for correlations between the main parameters and activity indicators, namely the CRX, centre-of-light, CCF-Contrast, RV ({\tt serval}), CCF-FWHM, CCF-Bisector, dLW (VIS), and log ($L_{\rm H\alpha}/L_{\rm bol}$) against the full set of parameters and activity indices, as previously described in Section 3.1.  The results for each surface brightness map of EV~Lac are listed in Table~\ref{tab:activity_corr}, where a Pearson $r$ coefficient $>0.7$ or $<-0.7$ indicates strong correlation or anti-correlation. The Student t-test probability value, 
$p$,
is also used to assess the statistical significance of the Pearson $r$ value following the procedure described by \cite{Jeffers2020Sci...368.1477J} in application to the star GJ\,887.
Values of $p > 0.03$ imply no strong evidence to reject the (no correlation) null hypothesis.  

In the following text, the use of the word correlation refers to both correlations and anti-correlations.  For all maps, both the CRX and the centre-of-light have a strong correlation with the position-related parameters such as RV (VIS), CCF RV, RV (NIR), and the shape-related parameter CCF-Bisector.  The CCF-Contrast typically correlates strongly with the dLW or CCF-FWHM or with the individual chromospheric activity indicators. 

Each parameter, such as the output from {\tt serval}, activity line indicator or CCF parameter, is categorised into six subcategories.  
These can be categorised by the form of the absorption line distortion that they induce. The absorption line shift, scaling or distortions are described by the moments of the line profile as follows:

\begin{itemize}
    \item{\underline {Moment 0}}: (depth) CCF-Contrast
    \item {\underline {Moment 1}}: (mean / velocity shift) position of the line includes parameters such as RV and CofL
    \item {\underline {Moment 2}}: (variance) CCF-FWHM and also includes the dLW 
    \item {\underline {Moment 3}}: (skewness) 
 line bisector span of the CCF
    \item {\underline {Parameter 4}}:  the activity line indices
     \item {\underline {Parameter 5}}:  the CRX
\end{itemize}

\begin{figure}
\centering
\includegraphics[angle=270,width=0.45\textwidth]{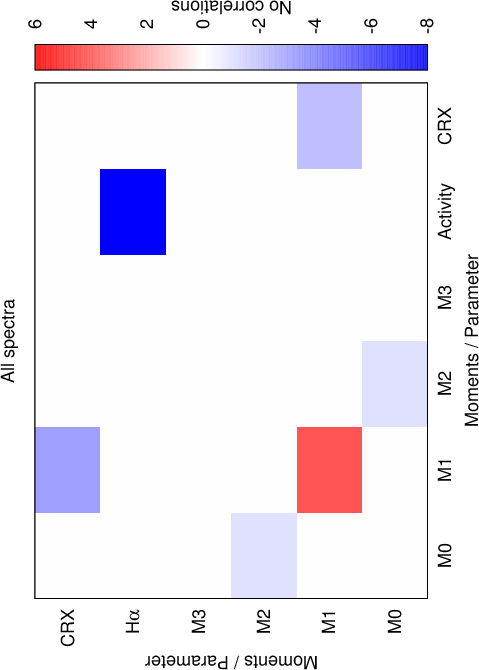}
\caption{Correlation of the moments / activity indices with each other for all CARMENES spectra of EV~Lac.  Red rectangles indicate positive correlations, and blue rectangles indicate anti-correlations.   Only strong correlations with a high significance are considered. We only consider H$\alpha$ for the Parameter 4 correlations, but we compare it to all activity lines. The lack of correlations for M0--M0 and M3--M3 moments arises because there is only one activity parameter with these moments}
\label{fig:correlations_all}
\end{figure}

\noindent
The detailed classification of the parameters for the full data set is also listed in Table~\ref{tab:activity_corr} for ($i$) the eight low-resolution Doppler imaging maps, ($ii$) map 9, which is the combination of maps 6 and 7, and ($iii$) all of the CARMENES spectra of EV~Lac. Figure~\ref{fig:correlations_all} illustrates how the activity parameters associated with each line profile moment correlate with activity parameters associated with the other moments for all CARMENES observations of EV Lac. Only strong correlations with $r > 0.7$ are considered.   For example, indicators that are effectively associated with moment 0, show strong correlations with moment 2 indicators.   Moment 1 indicators show strong correlations with other moment 1 indicators and strong correlations with the CRX.  The commonly used stellar activity indicator log ($L_{\rm H\alpha}/L_{\rm bol}$) has a strong correlation with the other activity indicators, but not with either the photospheric band or RV-related indicators. 

\begin{figure}
\centering
\includegraphics[angle=270,width=0.45\textwidth]{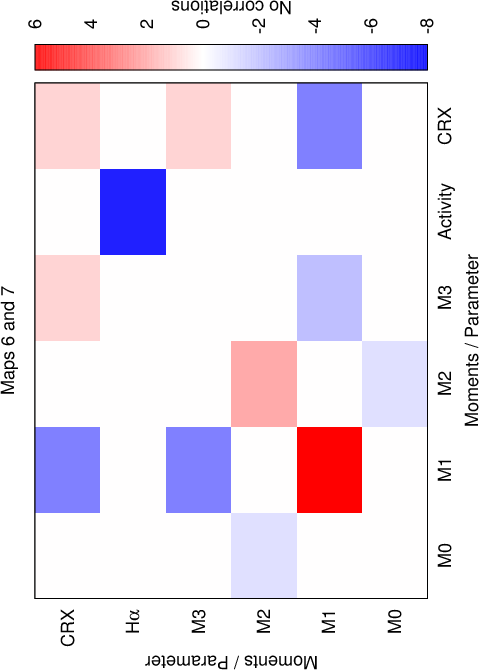}
\caption{Correlation of the moments / activity indices with each other for map 9 (combined maps 6 and 7).  Only strong correlations with a high significance are considered. }
\label{fig:correlations_maps9+10}
\end{figure}

Many more strong (anti-) correlations between the six moments / parameters are shown in Figure~\ref{fig:correlations_maps9+10} for the reduced subsample of data comprising map 9, which is the combination of maps 6 and 7.   The correlations in map 9 are particularly important for benchmarking the stellar activity of EV~Lac as the features of stellar activity will not have evolved significantly compared to the full CARMENES data set.  In particular and compared to the full CARMENES data of EV~Lac, there is now a strong correlation of odd moments with other odd moments.   For example, moment 3 has a strong correlation with moment 1 RV-related or position variations.  This means that the RV variations are mainly caused by a distortion in the shape of the line profile, which is typically caused by the presence of stellar activity.   This is expected following the results of \cite{Saar1997ApJ...485..319S}, who modelled the impact that a fixed starspot has on RV and bisector measurements.  They demonstrated that the bisector span varies as (\vsin)$^{3.3}$, making it a suitable diagnostic for stars such as EV~Lac.  Likewise, the CRX has a strong anti-correlation with moment 3 indicators.  This is because the CCF-Bisector also shows a wavelength dependence as redder lines are generally weaker.    

The remaining correlations in Figure~\ref{fig:correlations_maps9+10} show that even-numbered moments, such as CCF-contrast, CCF-FWHM, or dLW, are strongly correlated with each other.   It is not surprising that the CCF-contrast has a strong correlation with dLW given that we assigned the dLW as a moment 2 parameter for simplicity when in reality it also has a moment 0 component.  This is because the calculation of dLW assumes a fixed contrast or depth value, as described in \cite{Zechmeister2018A&A...609A..12Z} (see their section 4.3 for a comparison of dLW with FWHM). The impact of this simplified assumption is that there is a slightly increased correlation of moment 2 indices with moment 0 indices, as shown in Figure~\ref{fig:correlations_all} and Figure~\ref{fig:correlations_maps9+10}.  For the analysis in this work, the assumption that dLW is purely moment 2 is justified, but caution would be advised in applying this assumption to a more in-depth analysis.  As previously described, log ($L_{\rm H\alpha}/L_{\rm bol}$) correlates strongly with other chromospheric activity indicators.  

\section{Closed-loop relations}

For M dwarfs with stable or very low levels of stellar activity, high-precision RV observations show a stable and well-behaved evolution of the activity indices with stellar rotation phase.  This results in correlations, such as the closed-loop correlations of stellar activity with RV, as reported by \cite{Bonfils2007A&A...474..293B} for the M 2.5 dwarf GJ 674.  In addition to hosting an 11 $M_{\rm \oplus}$ planet with a period of 4.69 days, GJ 674 shows a period at 35 days that is attributed to the stellar rotation period.  This is further supported by their analysis of the Ca II H\&K lines, which show a regular sinusoidal variation with rotational phase or a closed-loop variation with RV.  The closed-loop pattern results from the Ca II H\&K value induced by a starspot or active region having a maximum value when the spot is in the meridian of the stellar disk, while the RV will have a value of zero as the line profile is neither blue- or redshifted.  Since GJ\,674 is a very low activity star with a very stable configuration of spots or active regions this results in a closed-loop pattern that is stable over several years.  

We investigated whether similar relations are present in the CARMENES data of EV~Lac over the time-span of (1) the full data set and (2) the subset of data that comprises map 9.  The results are shown in Figure~\ref{fig:closedloop}.  The centre-of-light, the CRX, and the CCF-Bisector all show a distinct correlation (or anti-correlation) with the RV for the full data set.  The corresponding plots for map 9 (see the right panel of Figure~\ref{fig:closedloop}) trace the evolution of these indices over a timescale where the evolution of stellar activity is considered to be insignificant.  The centre-of-light, the CRX, and the CCF-Bisector all show a double figure-of-eight evolution which corresponds to the double-dip and double-peak previously shown as a function of rotational phase in Figure~\ref{fig:activity}.  This shows more complex activity patterns than the single figure-of-eight shape report by \cite{Baroch2020A&A...641A..69B} for YZ CMi.  The authors also reported that the figure-of-eight shape results from a phase offset that is primarily caused by limb darkening and convective blueshift.  For M dwarfs such as EV~Lac, the convective blueshift is expected to be close to zero \citep{Liebing2021arXiv210803859L}.  We note that the centre-of-light values for ``All data'' are the centre-of-light values from the individual low-resolution Doppler-imaging maps combined and not the values resulting from one low-resolution Doppler image reconstructed using the full data set.  

The CRX-NIR as a function of RV$_{\rm NIR}$ shows a cloud of points without any significant structure or correlations with RV$_{\rm NIR}$ or rotation for the full data set.  The corresponding plot for the smaller subset comprising map 9 shows a distinct correlation with RV$_{\rm NIR}$.  The double figure-of-eight evolution is present, but with additional small-scale variations due to the increased error bars at NIR wavelengths.  Similarly, the TiO 7050 index, TiO 8430 index, and dLW (VIS) show a similar pattern with a cloud of points for the full data set and a clearer double closed-loop structure for the subset of map 9.  Surprisingly, the CCF-FWHM shows a more complicated double-loop and possibly a triple-loop structure than the dLW. This is because of the different methods that the two parameters use to monitor the changing line shape.  The dLW assumes a fixed contrast or depth, whereas the CCF-FWHM does not.  This makes the dLW more sensitive to changes in contrast, such as those caused by stellar activity variations or other instrumental changes, whereas the FWHM will be less sensitive to these parameters.  This is illustrated in Figure~\ref{fig:closedloop} where the dLW follows the double-loop structure also seen in the activity indices.  The applicability of dLW as an activity indicator is further complicated as it can also contain an additional instrumental contribution, which is minimal for the data presented in Figure~\ref{fig:closedloop} as the data were secured over only a few stellar rotation periods.  For the remaining activity indices, we do not see any significant correlations in the full data set or the smaller subset of map 9.   This is consistent with the lack of rotational evolution, as previously illustrated in Figure~\ref{fig:activity_maps9+10}.

\begin{figure*}
\centering
\begin{minipage}{10.85cm}
\includegraphics[angle=270,width=10.5cm,right]{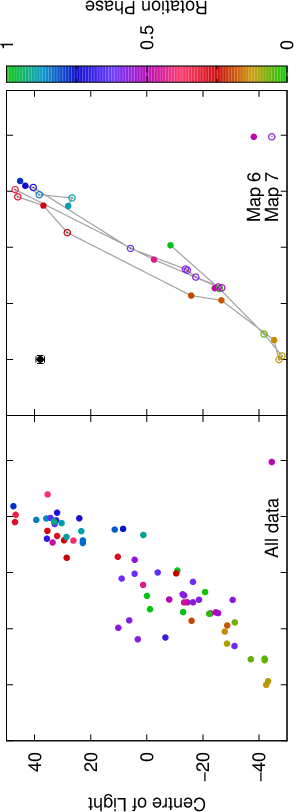}
\includegraphics[angle=270,width=10.75cm,right]{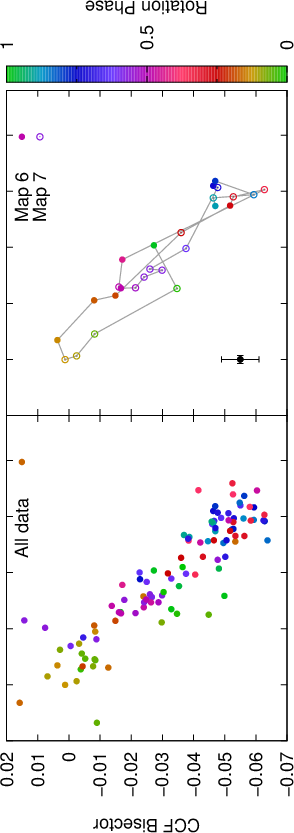}
\includegraphics[angle=270,width=10.61cm,right]{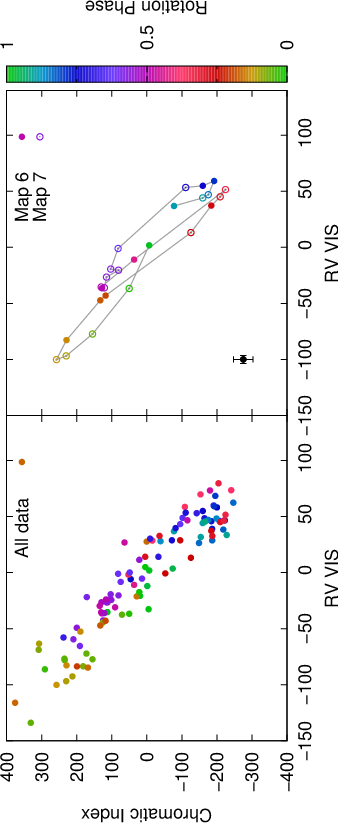}
\includegraphics[angle=270,width=10.595cm,right]{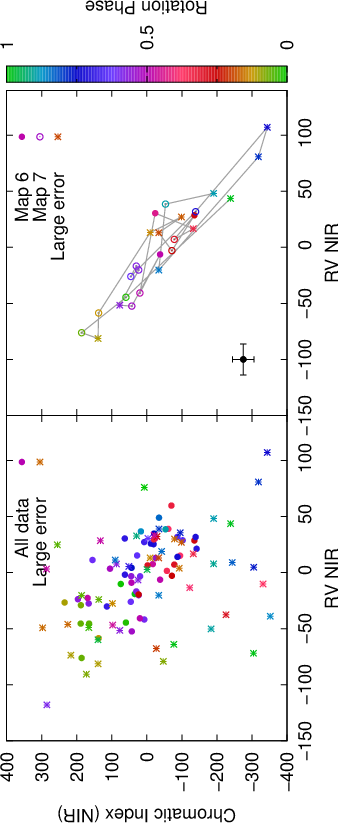}
\includegraphics[angle=270,width=10.85cm,right]{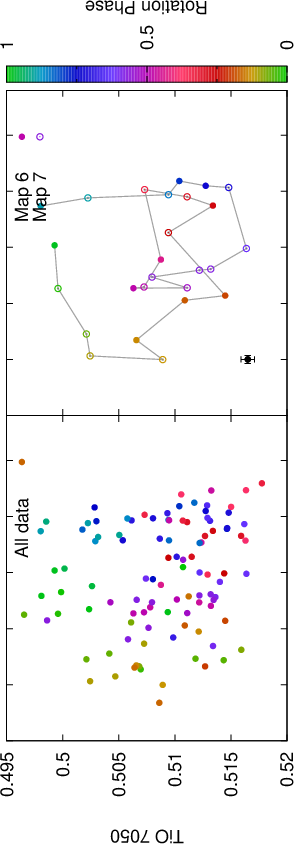}
\includegraphics[angle=270,width=10.83cm,right]{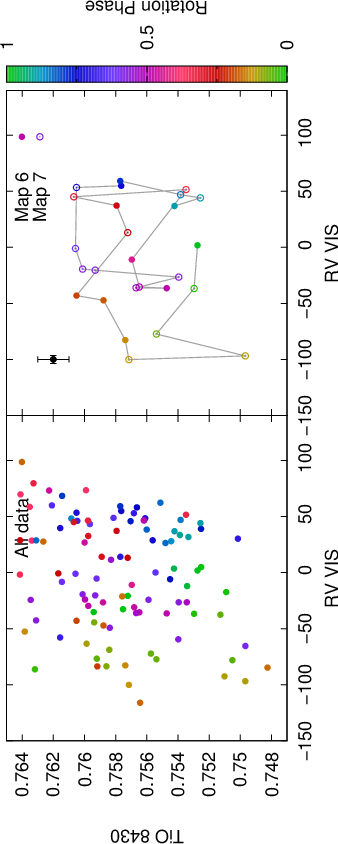}
\end{minipage}
\end{figure*}

\begin{figure*}
\centering
\begin{minipage}{10.85cm}
\includegraphics[angle=270,width=10.565cm,right]{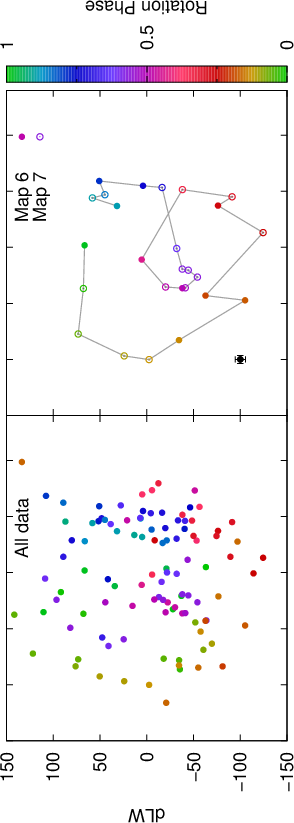}
\includegraphics[angle=270,width=10.67cm,right]{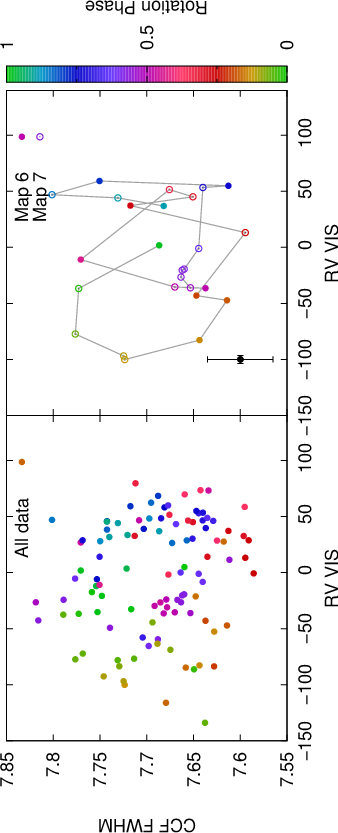}
\end{minipage}
\caption{Variation in indicators as a function of RV for the full data set (left panels) and the smaller subsample comprising map 9 (right panels).  Colour indicates rotational phase.  Closed circles are from map 6, and open circles are from map 7. Data points with successive rotational phases, from phases 0 to 1.0, are connected with the light grey solid line.  The corresponding surface brightness image for the map 9 (combined maps 6 and 7) is shown in Figure 6.   For the NIR data, points with error bars that are greater than the mean error shown are indicated by an asterisk. A representative error bar is shown in black.}
\label{fig:closedloop}
\end{figure*}

\section{Periodicities}


\begin{figure*}
\centering
\includegraphics[width=0.9\textwidth]{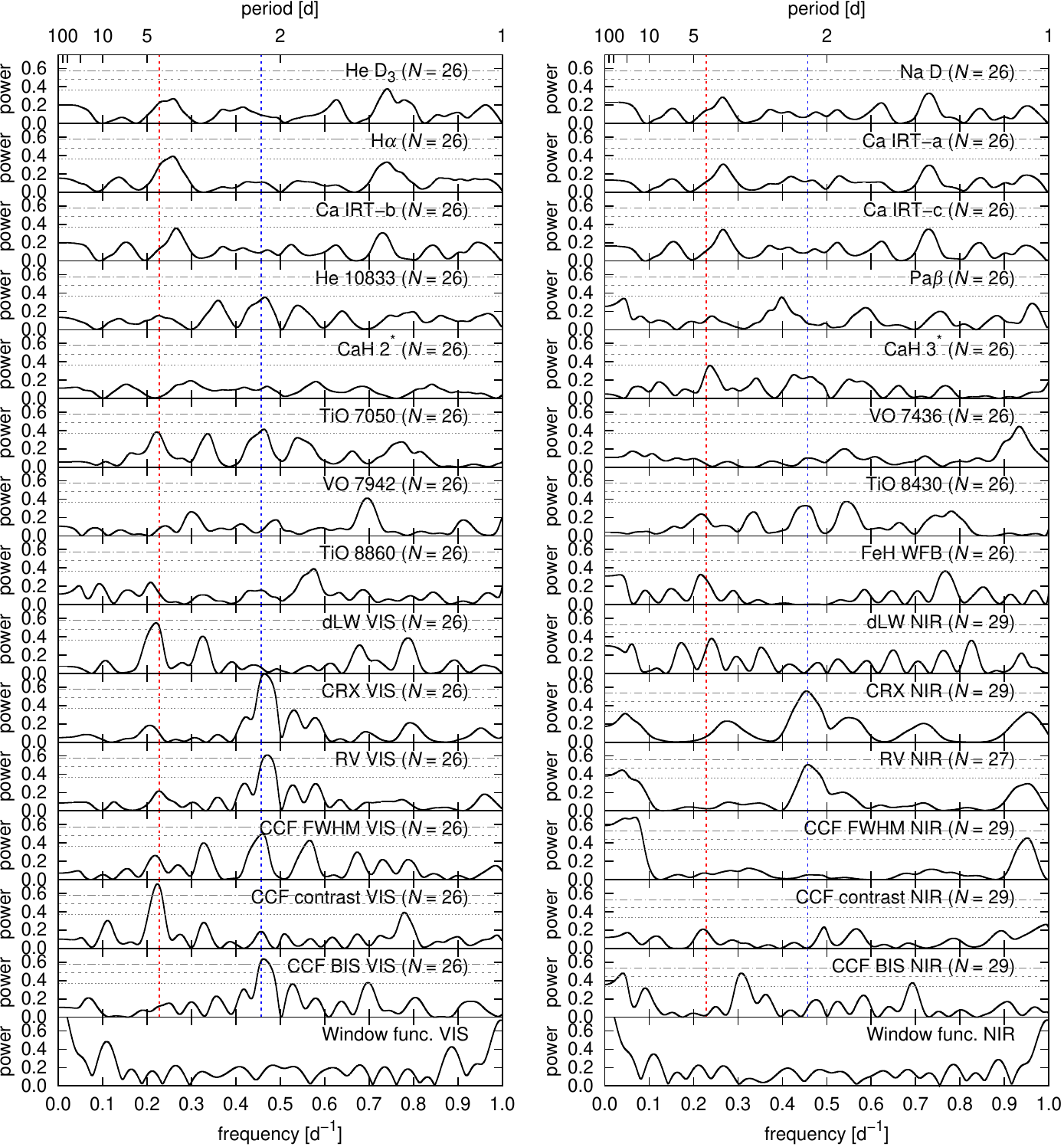}
\caption{GLS periodograms for the combined CARMENES data set of maps 6 and 7 for each of the considered indicators.  The dotted red line marks the rotation period ($P_{\rm rot}$ = 4.34 d, or frequency = 0.228 d$^{{-1}}$) and the dotted blue line indicates the second harmonic at $P_{\rm rot}$/2. The horizontal lines indicate the statistical significance of the detected periodicities shown as the 10\% (dotted), 1\% (dashed), and 0.1\% (dash-dotted) false-alarm probability levels.  The number of used CARMENES spectra is indicated by $N$ after a 2$\sigma$ clipping. }
\label{fig:GLS_Period}
\end{figure*}

\begin{figure*}
\centering
\includegraphics[width=0.9\textwidth]{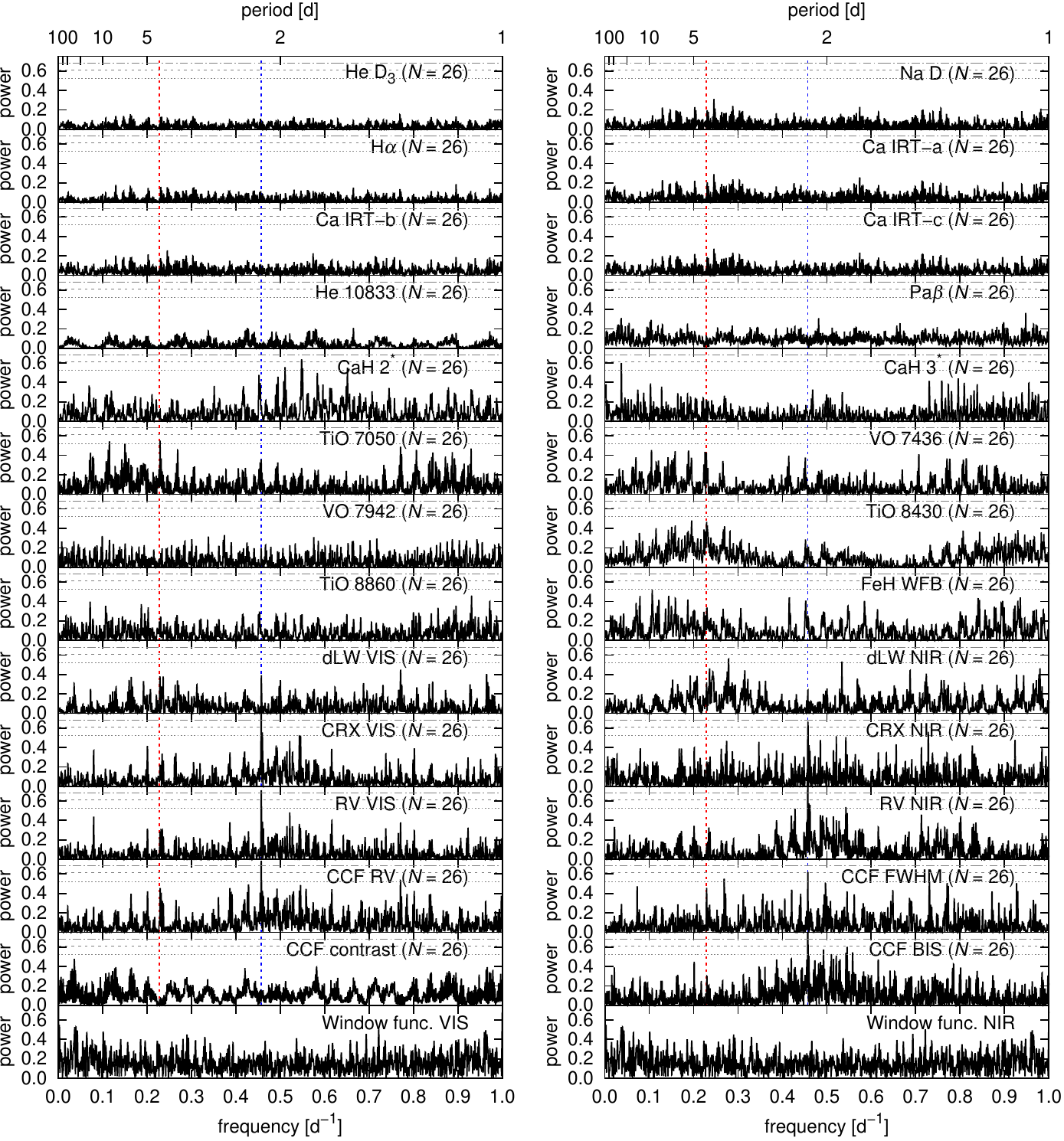}
\caption{GLS periodograms for a random selection of 26 observations from the complete CARMENES data set}
\label{fig:GLS_Period_small}
\end{figure*}

\subsection{Full data set}
The periodicities for the activity indicators and CCF parameters have previously been investigated by Sch\"ofer et al (2021 submitted) for the full CARMENES data set of EV~Lac.  They used the generalised Lomb-Scargle (GLS) periodogram analysis tool \citep{Zechmeister2009A&A...496..577Z} and reported that the chromospheric indicators, the TiO band indices, dLW, and CCF-Contrast show significant peaks, with a false-alarm probability (FAP) of 0.1\% levels, at the rotation frequency, whereas CRX, RV, CCF-FWHM, and CCF-Bisector show periodicities at its second harmonic ($P_{\rm rot}/2$).  These periodicities are  consistent with the phase-folded results that are shown in Figure~\ref{fig:activity} for the variation of the  RV$_{\rm VIS}$, RV$_{\rm NIR}$, CRX$_{\rm VIS}$, CRX$_{\rm NIR}$, pEW (H$\alpha$), TiO 8860, TiO 7050, and dLW for the full data set.  It should be noted that the panel showing TiO 7050 also shows significant structure at $P_{\rm rot}/2$.  For a discussion of the periodicities in the full data set of EV~Lac we refer to \cite{Lafarga2020A&A...636A..36L}, and Sch\"ofer et al. (2021 submitted).

\subsection{Individual maps} \label{sec:periods_maps}
We also searched for periodicities using GLS periodograms in each of the eight subsets of data as described in Section 4.  Over the time-span of each subset, which covers not more than a few rotation periods, it was not possible to detect any significant periodicities. To increase the time-base to look for periods, we combined two of the data sets for the low-resolution Doppler imaging that were taken over successive rotation periods (map 9 = combined maps 6 and 7).  The combined maps comprise a total of 26 spectra secured over a time-span of 20.34 days, or 4.64 rotation periods of EV~Lac.  The resulting GLS periodograms mostly show significant peaks with FAP of 0.1\% levels at the same frequencies as the full data-set periodograms, although the peaks are much broader.  The GLS periodograms for each activity index in this subset are shown in Figure ~\ref{fig:GLS_Period}.   We find that indicators that correspond to even line moments (i.e. CCF-Contrast -- moment 0; dLW -- moment 2; chromospheric line pEWs, photospheric band -- parameter 4) tend to show periodicities at the rotation period, whereas indicators based on odd moments such as line position (i.e. RV, CRX -- moment 1, parameter 5;  CCF-Bisector -- moment 3) favour $P_{\rm rot}$/2.  An interesting point is that the peak in the RV is just at our detection threshold, while other indices such as CRX$_{\rm VIS}$ show peaks at $P_{\rm rot}$/2 that are well above the detection threshold.

In general, these results are consistent with the phase-folded plots for map 9 as previously shown in Figure~\ref{fig:activity_maps9+10} and the closed-loop plots in Figure~\ref{fig:closedloop}.  However, the lack of periodicities at $P_{\rm rot}$/2 in indices such as dLW is somewhat surprising given the double-dip nature of the phase-folded dLW for the subset of map 9 (Figure~\ref{fig:activity_maps9+10}) and the double-loop nature of the closed-loop plots (Fig~\ref{fig:closedloop}).   An explanation could be that there is a slight phase lag in the locations of the two dips compared to the dips in indicators such as the RV-VIS and CRX (Figure~\ref{fig:activity_maps9+10}, where  the vertical dashed lines in this figure show the locations of the dips of the RV-VIS values.).  

The TiO7050 and TiO8430 bands have a slightly higher peak at $P_{\rm rot}$/2 in this subset than in the periods found in the full data set, although the rotation period peak is clearly higher in the full data set.  This can be explained by the smaller subset of data comprising map 9 showing less evolution of the stellar activity features over the time-span of the data set.  This is supported by the phase-folded and closed-loop correlations for TiO bands as shown in Figures~\ref{fig:activity_maps9+10} and ~\ref{fig:closedloop}, respectively.

\subsection{Random phases}

We randomly selected a total of 26 observations of EV~Lac from the total data set of more than 100 spectra.  This number of observations was chosen as it is the same number of spectra as in the combined map 9 investigated in the previous subsection (Section ~\ref{sec:periods_maps}), but not observed over just a few stellar rotation periods.  
The resulting periodograms show significant peaks with an FAP of 0.1\% levels at half the rotation period in the CRX (VIS and NIR), the RV (VIS and NIR), and CCF-Bisector periodograms, similar to the full data set.  It should be noted that many additional peaks fall mostly just below the detection threshold.  The TiO band shows a weaker peak with an FAP of 0.1\% at the rotation period of EV~Lac.   In addition, the CaH 2 band index shows a significant peak at the rotation period of EV~Lac that is twice the value calculated for the TiO indices, although this could result from the small sample size. Moreover, the CCF RVs show stronger signals than the {\tt serval} RVs.  The GLS periodograms for each activity index in this subset are shown in Figure ~\ref{fig:GLS_Period_small}.

\section{Inferring CRX from CofL}

Figure~\ref{fig:DImaps2} illustrates the power of Doppler imaging: with sufficient S/N and image regularisation, we can recover the phase and latitude of spots with several observations that sample the stellar rotation period well. The resulting image thus has predictive power because we can rotate our image to any phase and forward-model the corresponding absorption line profile. This is useful when trying to compare effects between epochs (such as cross-correlation shift due to starspot-induced line asymmetries or centre-of-light variation) when only a few observations were made at random phases at each epoch. In other words, we can obtain a more consistent measure of starspot-induced effects on the line profiles from one epoch to the next by predicting them at regular phase intervals from the Doppler images we derive. We first calculated the correlation between the centre-of-light and CRX for the observations all eight epochs that we considered. For the combined data, we find a Pearson $r = 0.85$ and Student probability of $p = 7.3 \times 10^{-25}$. This confirms a strong linear correlation between CRX and centre-of-light.  A linear fit yields

\begin{equation}
{\rm CRX}(v_{\rm CofL}) = ( 3.17 \pm 12.35 ) + (4.08 \pm 0.70) \times v_{\rm CofL}
\label{eqn:CRX_CofL}
\end{equation}

Using each map in Figure~\ref{fig:DImaps2}, we calculated the centre-of-light for 36 regularly spaced observation phases (i.e. at stellar rotation intervals $10\degr$). Eqn. \ref{eqn:CRX_CofL} was then used to predict the corresponding CRX values, from which we also derived the range in CRX. This procedure mitigates the bias introduced by comparing only the observed sparsely sampled and variable phases at each epoch.

\section{Discussion}

We have investigated the stellar activity of the very active mid-M dwarf EV~Lac using high-precision RV measurements secured with the CARMENES spectrograph. In this section we discuss our results in a broader context. 

\subsection{Low-resolution Doppler imaging}

We have demonstrated that Doppler imaging at low-resolution can be used to separate the spot-induced component from the measured RV.   Even though the \vsin\ of EV~Lac is below the threshold of $\approx$ 20 km\,s$^{-1}$ typically used for Doppler-imaging studies, we have shown both in the application to simulated data \citep{Barnes2017MNRAS.466.1733B} and in this study using CARMENES observations of EV~Lac, that the derived centre-of-light values can identify the spot-induced RV component.  The centre-of-light quantifies the asymmetry induced in the star's spectral lines by the presence of starspots.  The advantage of this technique is that it is also applicable to stars with low \vsin\ values.  Similar techniques based on mapping the spot distributions on the stellar surface include those of ~\cite{Baroch2020A&A...641A..69B}, where the CRX and simultaneous photometry were used to infer the spot distributions and as part of the Doppler-imaging technique itself \citep{Petit2015A&A...584A..84P}, and recently for AU~Mic \citep{Klein.AUMic.2021MNRAS.502..188K}.

The resulting low-resolution Doppler images typically show a large-scale spot feature at high latitudes and are consistent with the Zeeman Doppler-imaging reconstructions of the large-scale magnetic field of EV~Lac by \cite{Morin2008MNRAS.390..567M}.  For the two epochs without the high-latitude large-scale feature, this does not result from a lack of phase coverage but we consider that this is due to the evolution of spot features on the stellar surface.  In every epoch, additional small spots  evolve from one epoch to the next.  This is particularly noticeable in the last three maps, where the observations of EV~Lac were taken over a short time-interval.    We tested the reliability of the smaller features to systematics in the image reconstruction process, such as over-fitting the data.

The rapid evolution of stellar activity features is supported by variability in the TESS light curve, which covers six rotational periods (Figure~\ref{fig:TESSlc}).  The first stellar rotation (shown as blue dots) shows many flares with a reasonable amplitude, while the fourth rotational cycle (shown as red dots) also shows flares, but with a significantly lower amplitude than the first rotational cycle.  Additionally, in rotation cycle 4, the level of the continuum appears to increase between phases 0.5 and 0.7, which could be caused by micro-flaring or increased regions of bright plage.  A more detailed analysis of the flare activity in EV Lac has been recently conducted by \cite{Paudel2021ApJ...922...31P}.  

The computed centre-of-light value is correlated with the CRX, which is computed independently from the CARMENES spectra for each map.  This is true even though the resulting surface brightness images have a lower resolution than the Doppler-imaging studies in the literature.  Since the technique of Doppler imaging reconstructs the full surface brightness distribution on the star even when there are small phase gaps, we can use this missing information to infer the CRX values for these missing phases.  The results are shown in the right panel of Figure~\ref{fig:DImaps2}, where the measured CRX values together with the inferred CRX values are shown for each map.  This demonstrates that care is needed in interpreting apparent changes in activity levels, particularly when only a few phase samples are obtained at a given epoch as an activity minimum or maximum could be missed.   

\subsection{Regular-cadence observations} \label{sec:reg_cadence}

Regular-cadence observations are important to mitigate the impact of correlated noise induced by the lifetime of starspots, which typically last for several rotation periods. For example, the observations should be secured with a cadence much shorter than the evolution times of the activity feature or the rotation period of the star.   Even in the case of EV~Lac, where the large-scale activity patterns are stable over the time-span of the observations, there is significant variability in the smaller-scale activity features. This is evident in the increased number of strong correlations in Figure~\ref{fig:correlations_maps9+10} compared to the full CARMENES data set shown in Figure~\ref{fig:correlations_all} and the clear closed-loop plots shown for the data set comprising map 9 (combined maps 6 and 7), right panels in Figure~\ref{fig:closedloop} compared to the full data set, left panels in Figure~\ref{fig:closedloop}.    The regular-cadence observations also produce more precise periodograms than the random sample, which shows many additional peaks due to correlated noise (see below in the discussion). 

The results highlight the importance of stellar activity as a source of correlated noise, which has important consequences for recovering signals using Gaussian processes.  As investigated by \cite{Cabot2021AJ....161...26C} regular-cadence observations are required to detect low-mass planets orbiting active stars.  They reported that sparse sampling prevents GPs from learning the noise structure, which has the implication that GPs are more likely to absorb potential planetary signals.  A detailed investigation into how different activity patterns are treated by different GP kernels was performed by \cite{Perger2021A&A...645A..58P}.

Previous detections of the low-mass planets orbiting GJ887 \citep{Jeffers2020Sci...368.1477J} and Proxima Centauri \citep{Anglada2016Natur.536..437A} were only possible using regular-cadence observations of approximately one RV measurement per clear night, even though there were many previous high-precision RV measurements over a time-span of approximately 20 years.   The regular-cadence observations are very important if the planetary signal is smaller than or similar to correlated noise sources originating from stellar activity.  

The evolution of the small-scale stellar activity features that we have reconstructed using the low-resolution Doppler-imaging technique will be a solid basis for investigating the limitations and how to optimise the GP kernels used in modelling RVs in planet searches.   A similar investigation was performed by \cite{Cabot2021AJ....161...26C}, who used photometric light curves to investigate the intrinsic starspot variability of HD101501.  This approach is limited because the inversion of photometric light curves has been demonstrated to only recover two to three spots irrespective of the input spot distribution \cite{Jeffers2005MNRAS.359..729J, Jeffers2009AIPC.1094..664J,Basri2020ApJ...901...14B}.

\subsection{Closed-loop relations}

Previously, \cite{Zechmeister2018A&A...609A..12Z} presented closed-loop correlations for YZ CMi using the dLW.  Using the S-index, \cite{Bonfils2007A&A...474..293B} showed closed-loop correlations for the planet-hosting star GJ 674, and \cite{Forveille2009A&A...493..645F} reported similar circular correlations for the super-Earth host star Gl 176.  The presence of a closed-loop activity modulation indicates a non-complex and stable activity pattern as is observed in the CARMENES observations of YZ CMi \cite{Baroch2020A&A...641A..69B}. We find for the small subset of data comprising map 9 that there are closed-loop correlations with RV for the CRX, the CCF-Bisector, TiO 7050, TiO 8430, and the dLW that are not apparent in the full data set as the stellar activity features have significantly evolved.  This is consistent with the results from the correlations of the line moments and parameters as shown in Figures~\ref{fig:correlations_all} and ~\ref{fig:correlations_maps9+10} where the map 9 data set shows many more strong correlations than the full data-set.

Stellar parameters such as CCF-FWHM and dLW can be impacted by changes both in the levels of activity on the star and in the instrumental profile or sky background.  This is particularly notable for the dLW as it computes the changing equivalent width assuming a fixed contrast and is sensitive to the changing levels of stellar activity, as well as to intrinsic instrumental variations.The CCF-FWHM is less sensitive to these changes as it also includes CCF-contrast variations.  Our results are consistent with both the rapid evolution of stellar activity on EV~Lac and with the GLS periodograms for these data sets.  Further analysis of the full CARMENES GTO sample will indicate how ubiquitous these closed-loop relations are and how they depend on the stellar parameters and activity level.

\subsection{Periodicities of indicators}

Recently Sch\"ofer et al. (2019 and 2021 submitted) and \cite{Lafarga2020A&A...636A..36L} presented analyses of periodicities using activity indices in the full CARMENES GTO sample.  \cite{Schoefer2019A&A...623A..44S} showed that in 15 out of 133 stars with rotation periods longer than one day, a significant periodicity was detected in at least two activity indicators at the stellar rotation period.    The most likely lines to show this effect are H$\alpha$, the Ca~{\sc ii} IRT-b line, and the TiO 7050 and 8430 bands.  In our analysis of EV~Lac, CRX and RV show periodicities at $P_{\rm rot}$/2, while dLW, H$\alpha$, and TiO 7050 show periodicities at $P_{\rm rot}$.  EV~Lac is only one of four stars from the entire CARMENES GTO sample that shows periodicities in at least three different indicators. 

The search for periodicities in the CARMENES GTO sample was further extended by \cite{Lafarga2020A&A...636A..36L}.  They searched for periods at the stellar rotation period and harmonics in 98 stars with at least 40 observations where 71\,\% of stars are H$\alpha$ inactive.  \cite{Lafarga2020A&A...636A..36L} concluded that not all indicators of stellar activity trace exactly the same activity effects.  For example, CRX and BIS are useful for the most active stars, as we have shown in this work for EV~Lac, while the chromospheric lines $H_{\rm \alpha}$ and the Ca IRT are more suitable for tracing activity in stars with lower activity levels.  This is consistent with the lack of periodicities detected in pEW(H$\alpha$) for EV~Lac.  Even in the short subset of data that comprises map 9, there is no significant periodicity at $P_{\rm rot}$ or $P_{\rm rot}$/2 in pEW(H$\alpha$), which is consistent with the phase-folded plots  presented in Figure~\ref{fig:activity_maps9+10}.  In these plots, the pEW(H$\alpha$) indicator appears to show periodic structure but this is both out of phase for phases $>$ 0.5, and in-phase for phases $>$0.5 with the RV dips (indicated by the dashed vertical lines).  In the phase-folded plots of pEW(H$\alpha$) for the full CARMENES data set of EV~Lac, as shown in Figure~\ref{fig:activity}, there is a significant variation in pEW(H$\alpha$) over similar rotational phases, indicating the rapid evolution of chromospheric activity features.  

The TiO bands are photospheric activity indicators, and we have investigated periodicities in the TiO bands at 7050 \AA, 8430 \AA\, and 8860 \AA.  As reported by \cite{Schoefer2019A&A...623A..44S}, the full data set shows significant periodicities at $P_{\rm rot}$ for TIO 8860 \AA, and $P_{\rm rot}$/2 for TiO 7050 \AA, and at $P_{\rm rot}$ for TiO 8430 \AA.    In the smaller subset comprising map 9, the same periodicities at $P_{\rm rot}$ and $P_{\rm rot}$/2 are shown only for TiO 7050 \AA.  It should be noted that the $P_{\rm rot}$/2 detection in TiO 7050 \AA\ is much stronger than in the full data set and is consistent with the closed-loop plots for TiO 7050 \AA\ shown in Figure~\ref{fig:closedloop}.   In the `random data set', which is composed of the same number of observations as map 9, but selected randomly from the full CARMENES data set of EV~Lac, there is a peak in TiO 7050 \AA\ at $P_{\rm rot}$, but not at $P_{\rm rot}$/2.  There are also many additional peaks with an equivalent  statistical significance.  

The comparison of periodicities in the random data set to those detected in map 9 gives an important insight into the importance of sufficient phase coverage or time-span of observations to accurately detect activity signals.   The random data set shows the same significant detections in CRX (VIS + NIR), RV (VIS+NIR), CCF-FWHM, CCF-Bisector, but with many additional, and sometimes statistically significant, signals at different frequencies or periods and in different activity indicators, which are not present in the full data set, nor the subset comprising map 9. This is due to the presence of correlated noise, as previously noted in the discussion of regular cadence observations in Section~\ref{sec:reg_cadence} .  Certainly in the case of EV~Lac, stellar activity evolves over a short time-span making the detection of periodicities in the full data set quite challenging.  As a follow-up from this work, we will investigate the minimum number of observations required to reproduce the periodicities detected in the full data set, while also detecting a statistically significant signal in RV.   

\subsubsection{RV signal at $P_{\rm rot}$/2}

\begin{figure}
\centering
\includegraphics[angle=270,width=0.45\textwidth]{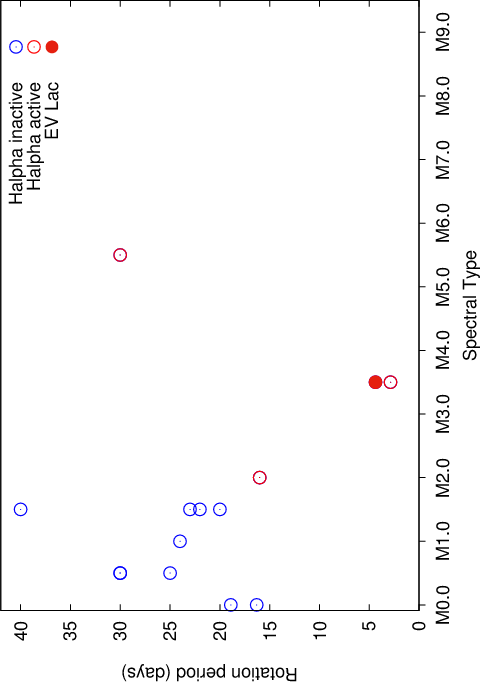}
\caption{Distribution of rotation periods (literature) as a function of M dwarf spectral type for the stars showing a significant peak in RV at $P_{\rm rot}$/2 in the CARMENES GTO sample as investigated by \cite{Lafarga2020A&A...636A..36L}.  The points are coloured based on whether the star is H$\alpha$ active (red) or inactive (blue).}
\label{fig:Prot/2:rtn_SPT}
\end{figure}

As previously mentioned,  from an original sample of 98 M dwarfs, \cite{Lafarga2020A&A...636A..36L} detected significant periodicities in 56 stars in at least one parameter.  In 44 of these stars they detected a significant peak in GLS periodograms of RV. The vast majority of periodicities in RV were detected at $P_{\rm rot}$, and in 15 stars (including EV~Lac), there is a detection at $P_{\rm rot}$/2.  The majority of these 15 stars have rotation periods greater than 16 days, with the exception of EV~Lac and the M\,3.5 dwarf J18498--238, which has $P_{\rm rot}$ =2.87$\pm$0.01 days and \vsin\ =3.0\,km\,s$^{-1}$ and is also H$\alpha$ active.   The subsample of 15 stars also shows a dependence on spectral type, where 11 of the 15 stars have spectral types earlier than M\,2.0\,V.  The remaining 4 stars are also H$\alpha$ active, as is shown in Figure~\ref{fig:Prot/2:rtn_SPT}.  There is also a decrease in rotation period with spectral type, although this could be due to the small size of the sample.  All but 5 of the stars have \vsin\ values $<$ 2\,km\,s$^{-1}$.  We also investigated whether the other 14 stars showed significant $P_{\rm rot}$/2 signals in the other indices, similar to what we have reported in this work for EV~Lac.  In addition to EV~Lac, the 2 stars J04588+498 and J11511+352, have periods only at $P_{\rm rot}$/2 in the CCF-Bisector.  Furthermore, in the full GTO sample, only EV~Lac has periods at $P_{\rm rot}$/2 in both RV$_{\rm VIS}$ and CRX$_{\rm VIS}$.  The fact that many of these stars are H$\alpha$ inactive and have low \vsin\ values indicates that it is not the intrinsic high activity of EV~Lac that causes the double-dip periodicities in many of the indicators we investigated.    EV~Lac is certainly unique in that it shows the double-dip features simultaneously in several indices.  These CARMENES data sets will be used to test methods for characterising and correcting stellar activity.

\subsubsection{Removal of stellar activity features}

\begin{figure}
\centering
\includegraphics[angle=270,width=0.47\textwidth]{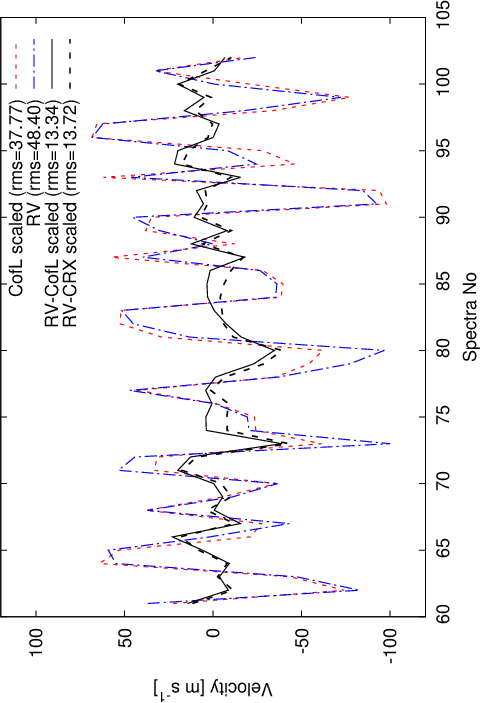}
\caption{Ability of the centre-of-light method to remove activity-induced RVs from the measured CARMENES (VIS) RVs.  The dot-dashed blue line plots the RVs measured by CARMENES and processed by {\tt serval}. The dashed red line shows the corresponding scaled activity-induced RVs derived using the low-resolution Doppler imaging process (centre-of-light) using Eqn. \ref{eqn:v_cofl}. A linear scaling to optimise the amplitude of the model centre-of-light RVs with the  observed {\tt serval} RVs is applied prior to calculating the residuals shown by the solid black line. The scaling accounts the for mismatch between observed and model RVs and also for any difference in effective wavelength at which these are measured. The data points shown are for the latter three low-resolution Doppler images combined.  The CRX subtracted from the RVs is shown by the dashed black line.}
\label{fig:RV-RVXCORR_ALL60}
\end{figure}

We have used the low resolution Doppler-imaging technique to reconstruct the large-scale starspot images of EV~Lac over eight epochs. From this reduction, we have derived the centre-of-light, and also quantified the impact that the stellar activity of EV~Lac induces on its measured RV values.  We subtracted both the centre-of-light value and the CRX from the measured CARMENES RV values.  For the centre-of-light, we achieved a decrease by a factor of 3.6 in the root-mean-squared (rms) values. The results are shown in Figure~\ref{fig:RV-RVXCORR_ALL60} for the data contained in the last three Doppler images (maps 6 to 8) as they have the best S/N and rotational phase coverage.  Our results are within the expected performance of the technique as reported by  ~\cite{Barnes2017MNRAS.466.1733B} for simulated data of the Proxima Centauri exoplanet system.  The removal of the CRX from the measured RVs results in a decrease of rms by a factor of 3.52 for maps 6 to 8 (rms = 13.72), and 2.84 for the full CARMENES data set (rms = 17.71).  These rms values translate, for both methods, to an equivalent improvement in planet-mass sensitivity. 

An important result of this work is that we have confirmed that the spot-induced centre-of-light reconstructed using Doppler-imaging techniques and the CRX are correlated with each other, which was previously only inferred.   Both the centre-of-light and the CRX are important for understanding and quantifying stellar activity in active stars.  The benefit of the CRX is that it is easily accessible given that it is an output of the {\tt{serval}} pipeline and is available for all CARMENES GTO stars.  However, the subtraction of the CRX relies on a strong correlation, defined as Pearson $r$ coefficient $>0.7$ or $<-0.7$,  with RV, which is only true for a small fraction ($<$ 10\%) of the CARMENES GTO sample.  How it performs for stars without such a strong correlation has yet to be quantified and will be investigated  (Cardona Guillén et al. 2022, submitted.).  The advantage of the low-resolution Doppler imaging technique is that it allows us to obtain maps of the actual large-scale starspot distribution it is simultaneously able to fill in gaps in phase coverage and can determine the full activity variability in any given rotation cycle.  It also applicable to a larger number of stars as it does not rely on a strong correlation with RV.  The disadvantage of the centre-of-light technique is that it is not applicable to very slowly rotating stars (with \vsin $\leq$ 2.0 km s$^{-1}$) and that it requires regular-cadence sampling over not more than several stellar rotation periods.   Further works will focus on applying these techniques to the full CARMENES GTO sample, currently comprising more than 20,000 spectra of more than 350 stars, to understand the optimal stellar parameter range for each method. 

\section{Conclusions}

We have investigated the impact of stellar activity on high-precision RV measurements of the mid-M dwarf EV~Lac. By applying the newly developed technique of low-resolution Doppler-imaging we were able to quantify the activity-induced RV shift, or centre-of-light velocity shift.  We report ($i$) a strong correlation between the centre-of-light and the chromatic index, which is a measure of the wavelength variation of RVs, confirming that the CRX originates from stellar activity, and ($ii$) by subtracting the centre-of-light component, we obtain an improvement of at least a factor of three in planet mass sensitivity.  Additionally, we find closed-loop relations for a subset of data that covers only several rotation periods which are not visible in the full CARMENES data set.  We additionally evaluate the impact of large phase gaps in the periodograms of activity indicators and conclude that  a regular-cadence observing strategy is the most efficient way to identify and eliminate sources of correlated noise caused by stellar activity.


\begin{acknowledgements}
We thank the referee for their very constructive comments that helped to improve the paper.
SVJ acknowledges the support of the German Science Foundation (DFG) priority program SPP 1992 ``Exploring the Diversity of Extrasolar Planets'' (JE 701/5-1). 
CARMENES is an instrument for the Centro Astron\'omico Hispano-Alem\'an de Calar Alto (CAHA, Almer\'{\i}a, Spain). 
CARMENES was funded by the German Max-Planck-Gesellschaft (MPG), 
  the Spanish Consejo Superior de Investigaciones Cient\'{\i}ficas (CSIC),
  the European Union through FEDER/ERF FICTS-2011-02 funds, 
  and the members of the CARMENES Consortium 
  (Max-Planck-Institut f\"ur Astronomie,
  Instituto de Astrof\'{\i}sica de Andaluc\'{\i}a,
  Landessternwarte K\"onigstuhl,
  Institut de Ci\`encies de l'Espai,
  Institut f\"ur Astrophysik G\"ottingen,
  Universidad Complutense de Madrid,
  Th\"uringer Landessternwarte Tautenburg,
  Instituto de Astrof\'{\i}sica de Canarias,
  Hamburger Sternwarte,
  Centro de Astrobiolog\'{\i}a and
  Centro Astron\'omico Hispano-Alem\'an), 
  with additional contributions by the Spanish Ministry of Economy, the German Science Foundation through the Major Research Instrumentation 
    Programme and DFG Research Unit FOR2544 ``Blue Planets around Red Stars'', 
  the Klaus Tschira Stiftung, 
  the states of Baden-W\"urttemberg and Niedersachsen, 
  and by the Junta de Andaluc\'{\i}a.
  Based on data from the CARMENES data archive at CAB (INTA-CSIC).
  We acknowledge financial support from the Agencia Estatal de Investigaci\'on of the Ministerio de Ciencia, Innovaci\'on y Universidades and the ERDF through projects
   PID2019-109522GB-C5[1:4]/AEI/10.13039/501100011033,  
   PID2019-107061GB-C64, 
   PGC2018-098153-B-C33, 
   AYA2016-79425-C3-1/2/3-P, 
   BES-2017-080769, 
and the Centre of Excellence ``Severo Ochoa'' and ``Mar\'ia de Maeztu'' awards to the Instituto de Astrof\'isica de Canarias 
(CEX2019-000920-S, SEV-2015-0548-16-3), Instituto de Astrof\'isica de Andaluc\'ia 
(SEV-2017-0709), and Centro de Astrobiolog\'ia (MDM-2017-0737), and the Generalitat de Catalunya/CERCA programme.  
This paper includes data collected by the TESS mission. Funding for the TESS mission is provided by the NASA's Science Mission Directorate.
\end{acknowledgements}

\bibliographystyle{bibtex/aa.bst} 
\bibliography{bibtex/refs.bib} 

\begin{appendix}

\section{Correlations of activity indices}
In this section we list all of the correlation parameters from the activity correlations and CCF correlation.

\begin{table*}
 \caption{\label{tab:activity_corr} Correlations with the stellar activity indices. Listed are the Pearson '$r$-coefficients and the Student t-test $stp$-values. }
 \centering
 \begin{tabular}{lccc}
 \hline
 \hline
Category & Pairs of activity indices & Pearson ($r$) & Student t-test ($stp$) \\
 \hline
 \multicolumn{4}{c}{Map 1 : phase  49.35--54.36  Good } \\
  \multicolumn{4}{c}{high lat. / polar spot + 1 weak low lat. spot } \\
\hline
CRX-RV$_{\rm VIS}$ & 5-1 & --0.9354 & 0.00060 \\ 
CRX-CCF-RV & 5-1 & --0.9179 & 0.00130 \\ 
CRX-CofL & 5-1 & --0.9046 & 0.00200 \\ 
CRX-RV$_{\rm NIR}$ & 5-1 & ---0.8528 & 0.00710 \\ 
CRX-TiO8860 & 5-4 & 0.7110 & 0.04800 \\ 
CRX-CCF-Bisector & 5-3 & 0.9434 & 0.00040 \\ 
\\
CofL-CRX$_{\rm VIS}$ & 1-5 & --0.9046 & 0.00200 \\ 
CofL-CCF-Bisector & 1-3 & --0.8006 & 0.01700 \\ 
CofL-CCF-RV & 1-1 & 0.9214 & 0.00110 \\ 
CofL-RV$_{\rm VIS}$ & 1-1 & 0.9268 & 0.00090 \\ 
CofL-RV$_{\rm NIR}$ & 1-1 & 0.9473 & 0.00030 \\ 
\\
CCF-Cont-(LHa) & 0-4 & --0.9283 & 0.00080 \\ 
CCF-Cont-dLW$_{\rm VIS}$ & 0-2 & --0.9283 & 0.00080 \\ 
CCF-Cont-TiO8860 & 0-4 & --0.8087 & 0.01510 \\ 
CCF-Cont-TiO8430 & 0-4 & --0.7733 & 0.02440 \\ 
CCF-Cont-CaH2 & 0-4 & --0.7692 & 0.02570 \\ 
CCF-Cont-TiO7050 & 0-4 & --0.7493 & 0.03240 \\ 
CCF-Cont-VO7942 & 0-4 & --0.7149 & 0.04620 \\ 
CCF-Cont-Pabeta & 0-4 & 0.9020 & 0.00220 \\ 
CCF-Cont-HeD3 & 0-4 & 0.9258 & 0.00090 \\ 
CCF-Cont-Halpha & 0-4 & 0.9316 & 0.00070 \\ 
CCF-Cont-NaD & 0-4 & 0.9334 & 0.00070 \\ 
CCF-Cont-He10833 & 0-4 & 0.9395 & 0.00050 \\ 
CCF-Cont-CaIRT-c & 0-4 & 0.9531 & 0.00020 \\ 
CCF-Cont-CaIRT-b & 0-4 & 0.9547 & 0.00020 \\ 
CCF-Cont-CaIRT-a & 0-4 & 0.9583 & 0.00010 \\ 
\\
RV$_{\rm VIS}$-CRX$_{\rm VIS}$ & 1-5 & --0.9354 & 0.00060 \\ 
RV$_{\rm VIS}$-CCF-Bisector & 1-3 & --0.8890 & 0.00310 \\ 
RV$_{\rm VIS}$-TiO8860 & 1-4 & --0.8618 & 0.00600 \\ 
RV$_{\rm VIS}$-CaIRT-c & 1-4 & 0.7062 & 0.05000 \\ 
RV$_{\rm VIS}$-CaIRT-a & 1-4 & 0.7073 & 0.04980 \\ 
RV$_{\rm VIS}$-He10833 & 1-4 & 0.7108 & 0.04810 \\ 
RV$_{\rm VIS}$-CaIRT-b & 1-4 & 0.7124 & 0.04740 \\ 
RV$_{\rm VIS}$-HeD3 & 1-4 & 0.7177 & 0.04500 \\ 
RV$_{\rm VIS}$-Pabeta & 1-4 & 0.7353 & 0.03770 \\ 
RV$_{\rm VIS}$-RV$_{\rm NIR}$ & 1-1 & 0.7994 & 0.01730 \\ 
RV$_{\rm VIS}$-CofL & 1-1 & 0.9268 & 0.00090 \\ 
RV$_{\rm VIS}$-CCF-RV & 1-1 & 0.9979 & 2.31e-08 \\ 
\\
RV$_{\rm NIR}$-CRX$_{\rm VIS}$  &1-5 &  -0.8528  &  0.00710  \\  
RV$_{\rm NIR}$-CCF-Bisector & 1-3 &  -0.7101  &  0.04840 \\   
RV$_{\rm NIR}$-CCF-RV & 1-1 &  0.7884  &  0.02010  \\ 
RV$_{\rm NIR}$-RV$_{\rm VIS}$  & 1-1 &  0.7994  &  0.01730  \\ 
RV$_{\rm NIR}$-CofL   & 1-1 &  0.9473  &  0.00040  \\ 
\\
CCF-Bisector-RV$_{\rm VIS}$ & 3-1 & --0.8890 & 0.00310 \\ 
CCF-Bisector-CCF-RV & 3-1 & --0.8769 & 0.00430 \\ 
CCF-Bisector-CofL & 3-1 & --0.8006 & 0.01700 \\ 
CCF-Bisector-RV$_{\rm NIR}$ & 3-1 & --0.7101 & 0.04840 \\ 
CCF-Bisector-TiO8860 & 3-4 & 0.7574 & 0.02950 \\ 
CCF-Bisector-CRX$_{\rm VIS}$ & 3-5 & 0.9434 & 0.00040 \\ 
\\
dLW$_{\rm VIS}$-CCF-Cont & 2-0 & --0.9283 & 0.00080 \\ 
dLW$_{\rm VIS}$-CaIRT-a & 2-4 & --0.8164 & 0.01340 \\ 
dLW$_{\rm VIS}$-CaIRT-c & 2-4 & --0.8016 & 0.01670 \\ 
dLW$_{\rm VIS}$-CaIRT-b & 2-4 & --0.7988 & 0.01740 \\
 \hline
    \end{tabular}
\end{table*}

\begin{table*} 
 \centering
 \begin{tabular}{lccc}
  \hline
 \hline
Category & Pairs of activity indices & Pearson ($r$) & Student t-test ($stp$) \\
 \hline
dLW$_{\rm VIS}$-NaD & 2-4 & --0.7889 & 0.01990 \\ 
dLW$_{\rm VIS}$-Halpha & 2-4 & --0.7822 & 0.02180 \\ 
dLW$_{\rm VIS}$-He10833 & 2-4 & --0.7818 & 0.02190 \\ 
dLW$_{\rm VIS}$-HeD3 & 2-4 & --0.7474 & 0.03310 \\ 
dLW$_{\rm VIS}$-Pabeta & 2-4 & --0.7069 & 0.04990 \\ 
dLW$_{\rm VIS}$-(L$_{\rm Ha}$/L$_{\rm \rm bol}$) & 2-4 & 0.8448 & 0.00830 \\ 
\\
log(Ha)-Halpha & 4-4 & --0.9793 & 2.20e-05 \\ 
log(Ha)-He10833 & 4-4 & --0.9711 & 5.90e-05 \\ 
log(Ha)-NaD & 4-4 & --0.9701 & 6.53e-05 \\ 
log(Ha)-CaIRT-a & 4-4 & --0.9678 & 8.16e-05 \\ 
log(Ha)-CaIRT-c & 4-4 & --0.9607 & 0.00010 \\ 
log(Ha)-CaIRT-b & 4-4 & --0.9590 & 0.00010 \\ 
log(Ha)-HeD3 & 4-4 & --0.9529 & 0.00020 \\ 
log(Ha)-Pabeta & 4-4 & --0.9413 & 0.00050 \\ 
log(Ha)-CCF-ContRAST & 4-0 & --0.9283 & 0.00090 \\ 
log(Ha)-TiO8860 & 4-4 & 0.7759 & 0.02360 \\ 
log(Ha)-dLW$_{\rm VIS}$ & 4-2 & 0.8448 & 0.00820 \\ 
log(Ha)-CaH2 & 4-4 & 0.8556 & 0.00670 \\ 
 \hline
 \multicolumn{4}{c}{Map 2 : phase 56.17--59.14 :   Good } \\
 \multicolumn{4}{c}{Small high lat. + 1 mid-lat. + 3 strong low lat. spots } \\
 \hline
CRX-RV$_{\rm VIS}$ & 5-1 & --0.9562 & 1.52e-05 \\ 
CRX-CofL & 5-3 & --0.7525 & 0.01200 \\ 
CRX-CRX$_{\rm NIR}$ & 5-5 & 0.9153 & 0.00020 \\ 
\\
CofL-CRX$_{\rm VIS}$ & 1-5 & --0.7525 & 0.01200 \\ 
CofL-RV$_{\rm VIS}$ & 1-1 & 0.8247 & 0.00330 \\ 
CofL-RVxcor & 1-1 & 0.9946 & 3.73e-09 \\ 
\\
CCF-Cont-CCF-Bisector & 0-3 & 0.7735 & 0.00860 \\ 
\\
RV$_{\rm VIS}$-CRX$_{\rm VIS}$ & 1-5 & --0.9562 & 1.52e-05 \\ 
RV$_{\rm VIS}$-CRX$_{\rm NIR}$ & 1-5 & --0.8442 & 0.00210 \\ 
RV$_{\rm VIS}$-RVxcor & 1-1 & 0.7727 & 0.00870 \\ 
RV$_{\rm VIS}$-CofL & 1-1 & 0.8247 & 0.00330 \\ 
\\
CCF-FWHM-CCF-Bisector & 2-3 & 0.7716 & 0.00890 \\ 
\\
CCF-Bisector-CCF-FWHM & 3-2 & 0.7716 & 0.00890 \\ 
CCF-Bisector-CCF-Cont & 3-0 & 0.7735 & 0.00860 \\ 
\\
dLW$_{\rm VIS}$-CaIRT-b & 2-4 & --0.7344 & 0.01560 \\ 
dLW$_{\rm VIS}$-CaIRT-a & 2-4 & --0.7093 & 0.02160 \\ 
dLW$_{\rm VIS}$-NaD & 2-4 & --0.7078 & 0.02200 \\ 
\\
log(Ha)-CaIRT-b & 4-4 & --0.9918 & 1.96e-08 \\ 
log(Ha)-CaIRT-a & 4-4 & --0.9908 & 3.12e-08 \\ 
log(Ha)-CaIRT-c & 4-4 & --0.9904 & 3.64e-08 \\ 
log(Ha)-Halpha & 4-4 & --0.9892 & 5.92e-08 \\ 
log(Ha)-NaD & 4-4 & --0.9325 & 8.35e-05 \\ 
log(Ha)-HeD3 & 4-4 & --0.9202 & 0.00010 \\ 
log(Ha)-He10833 & 4-4 & --0.8657 & 0.00120 \\ 
 \hline
 \multicolumn{4}{c}{Map 3 : phase  63.93--70.08 : Good } \\
 \multicolumn{4}{c}{1 strong mid lat + 4 weak + 1 strong low lat. } \\
\hline
CRX-RV$_{\rm VIS}$ & 5-1 & --0.9217 & 0.00010 \\ 
CRX-CCF-RV & 5-1 & --0.9101 & 0.00020 \\ 
CRX-CofL & 5-1 & --0.8640 & 0.00120 \\ 
CRX-RVxcor & 5-1 & --0.8476 & 0.00190 \\ 
CRX-CCF-Bisector & 5-3 & 0.9029 & 0.00030 \\ 
\\
CofL-CRX$_{\rm VIS}$ & 5-1 & --0.864 & 0.0012 \\ 
CofL-CCF-Bisector & 1-3 & --0.7616 & 0.01050 \\ 
CofL-TiO8430 & 1-4 & 0.7156 & 0.02000 \\ 
 \hline
    \end{tabular}
\end{table*}

\begin{table*} 
 \centering
 \begin{tabular}{lccc}
  \hline
 \hline
Category & Pairs of activity indices & Pearson ($r$) & Student t-test ($stp$) \\
 \hline
CofL-RV$_{\rm VIS}$ & 1-1 & 0.9357 & 6.93e-05 \\ 
CofL-CCF-RV & 1-1 & 0.9419 & 4.65e-05 \\ 
CofL-RVxcor & 1-1 & 0.9988 & 7.94e-12 \\ 
\\
CCF-Cont-FeHWF & 0-4 & 0.7870 & 0.00690 \\ 
CCF-Cont-VO7942 & 0-4 & 0.8167 & 0.00394 \\ 
CCF-Cont-NaD & 0-4 & 0.8283 & 0.00308 \\ 
\\
 RV$_{\rm VIS}$-CRX$_{\rm VIS}$ & 1-5 & --0.9217 & 0.00015 \\ 
RV$_{\rm VIS}$-CCF-Bisector & 1-3 & --0.8430 & 0.00219 \\ 
RV$_{\rm VIS}$-RVxcor & 1-1 & 0.9269 & 0.00011 \\ 
RV$_{\rm VIS}$-CofL & 1-1 & 0.9357 & 6.93e-05 \\ 
RV$_{\rm VIS}$-CCF-RV & 1-1 & 0.9935 & 7.7e-09 \\ 
\\
CCF-FWHM-TiO7050 & 2-4 & --0.7250 & 0.01770 \\ 
CCF-FWHM-CaH3 & 2-4 & --0.7143 & 0.02030 \\ 
\\
CCF-Bisector-RV$_{\rm VIS}$ & 3-1 & --0.8430 & 0.00219 \\ 
CCF-Bisector-CCFRV & 3-1 & --0.7888 & 0.00669 \\ 
CCF-Bisector-CofL & 3-1 & --0.7616 & 0.01050 \\ 
CCF-Bisector-RVxcor & 3-1 & --0.7390 & 0.01460 \\ 
CCF-Bisector-CRX$_{\rm NIR}$ & 3-5 & 0.7110 & 0.02120 \\ 
CCF-Bisector-CRX$_{\rm VIS}$ & 3-5 & 0.9029 & 0.00035 \\ 
\\
dLW$_{\rm VIS}$-dLWNIR & 2-2 & 0.7988 & 0.00557 \\ 
\\
log(Ha)-Halpha & 4-4 & --0.9990 & 3.75e-12 \\ 
log(Ha)-CaIRT-b & 4-4 & --0.9596 & 1.11e-05 \\ 
log(Ha)-CaIRT-a & 4-4 & --0.9380 & 5.98e-05 \\ 
log(Ha)-HeD3 & 4-4 & --0.9305 & 9.37e-05 \\ 
log(Ha)-CaIRT-c & 4-4 & --0.9153 & 0.00020 \\ 
log(Ha)-NaD & 4-4 & --0.8641 & 0.00126 \\ 
 \hline
 \multicolumn{4}{c}{Map 4 : phase  80.82--84.20 : Good } \\
 \multicolumn{4}{c}{1 weak mid lat + 2 weak + 1 strong low lat. } \\
 \hline
 CRX-RV$_{\rm VIS}$ & 5-1 & --0.9013 & 0.00223 \\ 
CRX-RV$_{\rm NIR}$ & 5-1 & --0.8991 & 0.00238 \\ 
CRX-CCF-RV & 5-1 & --0.8754 & 0.00439 \\ 
CRX-CCF-Bisector & 5-3 & 0.9681 & 7.94e-05 \\ 
\\
CofL-CCF-Bisector & 1-3 & --0.7544 & 0.03050 \\ 
CofL-RV$_{\rm NIR}$ & 1-1 & 0.8143 & 0.01390 \\ 
CofL-RV$_{\rm VIS}$ & 1-1 & 0.8625 & 0.00585 \\ 
CofL-CCF-RV & 1-1 & 0.8883 & 0.00320 \\ 
CofL-RVxcor & 1-1 & 0.9908 & 1.93e-06 \\ 
\\
CCF-Cont-NaD & 0-4 & 0.7712 & 0.02500 \\ 
\\
RV$_{\rm VIS}$-CCF-Bisector & 1-3 & --0.9657 & 9.84e-05 \\ 
RV$_{\rm VIS}$-CRX$_{\rm VIS}$ & 1-5 & --0.9013 & 0.00223 \\ 
RV$_{\rm VIS}$-CofL & 1-1 & 0.8625 & 0.00585 \\ 
RV$_{\rm VIS}$-RVxcor & 1-1 & 0.8932 & 0.00281 \\ 
RV$_{\rm VIS}$-RV$_{\rm NIR}$ & 1-1 & 0.9044 & 0.00203 \\ 
RV$_{\rm VIS}$-CCF-RV & 1-1 & 0.9973 & 5.08e-08 \\ 
\\
RV$_{\rm NIR}$-CCF-Bisector & 1-3  &  -0.9387  &  0.00050 \\ 
RV$_{\rm NIR}$-CRX$_{\rm VIS}$ & 1-5  &  -0.8990  &  0.00230  \\
RV$_{\rm NIR}$-CofL  & 1-1 &  0.8142  &  0.01386  \\
RV$_{\rm NIR}$-RVxcor & 1-1  &  0.8758  &  0.00430  \\
RV$_{\rm NIR}$-CCF-RV & 1-1  &  0.8945  &  0.00270  \\ 
RV$_{\rm NIR}$-RV$_{\rm VIS}$ & 1-1  &  0.9043  &  0.00200  \\ 
\\
CCF-FWHM-dLW$_{\rm VIS}$ & 2-2 & 0.8807 & 0.00387 \\ 
 \hline
    \end{tabular}
\end{table*}

\begin{table*} 
 \centering
 \begin{tabular}{lccc}
  \hline
 \hline
Category & Pairs of activity indices & Pearson ($r$) & Student t-test ($stp$) \\
 \hline
CCF-Bisector-RV$_{\rm VIS}$ & 1-1 & --0.9657 & 9.84e-05 \\ 
CCF-Bisector-CCF-RV & 1-1 & --0.9486 & 0.00033 \\ 
CCF-Bisector-RV$_{\rm NIR}$ & 1-1 & --0.9387 & 0.00055 \\ 
CCF-Bisector-RVxcor & 1-1 & --0.8094 & 0.01490 \\ 
CCF-Bisector-CofL & 1-1 & --0.7544 & 0.03050 \\ 
CCF-Bisector-CRX$_{\rm VIS}$ & 1-5 & 0.9681 & 7.94e-05 \\ 
\\
dLW$_{\rm VIS}$-CCF-FWHM & 2-2 & 0.8807 & 0.00387 \\ 
\\
log(Ha)-Halpha & 4-4 & --0.9993 & 1e-09 \\ 
log(Ha)-HeD3 & 4-4 & --0.9308 & 0.00078 \\ 
log(Ha)-CaIRT-a & 4-4 & --0.9019 & 0.00219 \\
log(Ha)-CaIRT-c & 4-4 & --0.8914 & 0.00295 \\ 
log(Ha)-CaIRT-b & 4-4 & --0.8567 & 0.00659 \\ 
log(Ha)-He10833 & 4-4 & --0.7445 & 0.03410 \\ 
log(Ha)-TiO7050 & 4-4 & 0.7160 & 0.04580 \\ 
 \hline
 \multicolumn{4}{c}{Map 5 : phase  117.87--124.25 : Good } \\
\multicolumn{4}{c}{1 weak mid lat. + 2 weak + 1 strong low lat.  } \\
\hline 
CRX-RV$_{\rm VIS}$ & 5-1 & --0.9781 & 0.00014 \\ 
CRX-CCF-RV & 5-1 & --0.9676 & 0.00036 \\ 
CRX-RV$_{\rm NIR}$ & 5-1 & --0.8796 & 0.00905 \\ 
CRX-TiO8860 & 5-4 & --0.8253 & 0.02230 \\ 
CRX-CofL & 5-1 & --0.8112 & 0.02680 \\ 
CRX-dLW$_{\rm VIS}$ & 5-2 & --0.7859 & 0.03620 \\ 
CRX-CCF-Bisector & 5-3 & 0.9412 & 0.00156 \\ 
\\
CofL-CCF-Bisector & 1-3 & --0.9065 & 0.00488 \\ 
CofL-CRX$_{\rm NIR}$ & 1-5 & --0.8258 & 0.02210 \\ 
CofL-CRX$_{\rm VIS}$ & 1-5 & --0.8112 & 0.02680 \\ 
CofL-CCF-RV & 1-1 & 0.8116 & 0.02670 \\ 
CofL-RV$_{\rm NIR}$ & 1-1 & 0.8171 & 0.02480 \\ 
CofL-RV$_{\rm VIS}$ & 1-1 & 0.8884 & 0.00751 \\ 
\\
CCF-Cont-CofL & 0-1 & --0.7218 & 0.06710 \\ 
\\
RV$_{\rm VIS}$-CRX$_{\rm VIS}$ & 1-5 & --0.9781 & 0.00014 \\ 
RV$_{\rm VIS}$-CCF-Bisector & 1-3 & --0.9653 & 0.00042 \\ 
RV$_{\rm VIS}$-TiO8860 & 1-4 & 0.8132 & 0.02610 \\ 
RV$_{\rm VIS}$-CofL & 1-1 & 0.8884 & 0.00751 \\ 
RV$_{\rm VIS}$-RV$_{\rm NIR}$ & 1-1 & 0.9178 & 0.00356 \\ 
RV$_{\rm VIS}$-CCF-RV & 1-1 & 0.9853 & 4.96e-05 \\ 
\\
RV$_{\rm NIR}$-CCF-Bisector & 1-1 &  -0.9558  &  0.00070 \\
RV$_{\rm NIR}$-CRX$_{\rm VIS}$  & 1-5 &  -0.8795  &  0.00900 \\
RV$_{\rm NIR}$-CaH2  & 1-4 &  -0.7972  &  0.03177 \\
RV$_{\rm NIR}$-TiO8860 & 1-4  &  0.7320  &  0.06140 \\
RV$_{\rm NIR}$-CofL  & 1-1 &  0.8170  &  0.02480 \\
RV$_{\rm NIR}$-RV$_{\rm VIS}$  & 1-1 &  0.9177  &  0.00350 \\
RV$_{\rm NIR}$-CCF-RV & 1-1  &  0.9222  &  0.00310\\
\\
CCF-FWHM-FeHWing-Ford & 2-4 & --0.8315 & 0.02040 \\ 
CCF-FWHM-TiO8430 & 2-4 & --0.8222 & 0.02320 \\ 
CCF-FWHM-CaH3 & 2-4 & 0.7718 & 0.04210 \\ 
CCF-FWHM-dLWNIR & 2-2 & 0.8124 & 0.02640 \\ 
CCF-FWHM-dLW$_{\rm VIS}$ & 2-2 & 0.9201 & 0.00332 \\ 
\\
CCF-Bisector-RV$_{\rm VIS}$ & 3-1 & --0.9653 & 0.00042 \\ 
CCF-Bisector-RV$_{\rm NIR}$ & 3-1 & --0.9559 & 0.00077 \\ 
CCF-Bisector-CCF-RV & 3-1 & --0.9362 & 0.00191 \\ 
CCF-Bisector-CofL & 3-1 & --0.9065 & 0.00488 \\ 
CCF-Bisector-CRX$_{\rm VIS}$ & 3-5 & 0.9412 & 0.00156 \\ 
 \hline
    \end{tabular}
\end{table*}

\begin{table*} 
 \centering
 \begin{tabular}{lccc}
  \hline
 \hline
Category & Pairs of activity indices & Pearson ($r$) & Student t-test ($stp$) \\
 \hline
  \multicolumn{4}{c}{Map 6 : phase  134.93--137.51 : Good } \\ 
\multicolumn{4}{c}{Weak high lat. + 1 mid lat. + 3 small low lat.}  \\
\hline
dLW$_{\rm VIS}$-CRX$_{\rm VIS}$ & 2-5 & --0.7859 & 0.03620 \\ 
dLW$_{\rm VIS}$-dLWNIR & 2-2 & 0.7540 & 0.05000 \\ 
dLW$_{\rm VIS}$-TiO8860 & 2-4 & 0.7604 & 0.04720 \\ 
dLW$_{\rm VIS}$-CCF-FWHM & 2-2 & 0.9201 & 0.00332 \\ 
\\
log(Ha)-Halpha & 4-4 & --0.9980 & 3.28e-07 \\ 
log(Ha)-NaD & 4-4 & --0.9936 & 6.23e-06 \\ 
log(Ha)-CaIRT-b & 4-4 & --0.9799 & 0.00011 \\ 
log(Ha)-CaIRT-c & 4-4 & --0.9761 & 0.00017 \\ 
log(Ha)-CaIRT-a & 4-4 & --0.9683 & 0.00034 \\ 
log(Ha)-HeD3 & 4-4 & --0.9164 & 0.00370 \\ 
\\
CRX-RV$_{\rm VIS}$ & 5-1 & --0.9825 & 4.01e-07 \\ 
CRX-CCF-RV & 5-1 & --0.9739 & 1.96e-06 \\ 
CRX-CofL & 5-1 & --0.9676 & 4.64e-06 \\ 
CRX-RVxcor & 5-1 & --0.9652 & 6.15e-06 \\ 
CRX-RV$_{\rm NIR}$ & 5-1 & --0.7538 & 0.01180 \\ 
CRX-CRX$_{\rm NIR}$ & 5-5 & 0.8053 & 0.00493 \\ 
CRX-CCF-Bisector & 5-3 & 0.9672 & 4.84e-06 \\ 
\\
CofL-CRX$_{\rm VIS}$ & 1-5 & --0.9676 & 4.64e-06 \\ 
CofL-CCF-Bisector & 1-3 & --0.9470 & 3.25e-05 \\ 
CofL-CRX$_{\rm NIR}$ & 1-5 & --0.7880 & 0.00678 \\ 
CofL-RV$_{\rm NIR}$ & 1-1 & 0.8050 & 0.004960 \\ 
CofL-CCF-RV & 1-1 & 0.9665 & 5.32e-06 \\ 
CofL-RV$_{\rm VIS}$ & 1-1 & 0.9723 & 2.49e-06 \\ 
CofL-RVxcor & 1-1 & 0.9975 & 1.71e-10 \\ 
\\
CCF-Cont-dLW$_{\rm VIS}$ & 0-2 & --0.9028 & 0.00035 \\ 
\\
RV$_{\rm VIS}$-CRX$_{\rm VIS}$ & 1-5 & --0.9825 & 4.01e-07 \\ 
RV$_{\rm VIS}$-CCF-Bisector & 1-1 & --0.9576 & 1.34e-05 \\ 
RV$_{\rm VIS}$-CRX$_{\rm NIR}$ & 1-5 & --0.8399 & 0.00236 \\ 
RV$_{\rm VIS}$-RV$_{\rm NIR}$ & 1-1 & 0.7905 & 0.00648 \\ 
RV$_{\rm VIS}$-CofL & 1-1 & 0.9723 & 2.49e-06 \\ 
RV$_{\rm VIS}$-RVxcor & 1-1 & 0.9787 & 8.83e-07 \\ 
RV$_{\rm VIS}$-CCF-RV & 1-1 & 0.9970 & 3.67e-10 \\ 
\\
RV$_{\rm NIR}$-CRX$_{\rm NIR}$ & 1-5  &  -0.9156  &  0.00020 \\ 
RV$_{\rm NIR}$-CCF-Bisector & 1-3  &  -0.7860  &  0.00700 \\ 
RV$_{\rm NIR}$-CRX$_{\rm VIS}$ & 1-5  &  -0.7537  &  0.01180 \\ 
RV$_{\rm NIR}$-RV$_{\rm VIS}$  & 1-1 &  0.7905  &  0.00648 \\ 
RV$_{\rm NIR}$-CCF-RV & 1-1  &  0.7932  &  0.00610 \\ 
RV$_{\rm NIR}$-CofL  & 1-1 &  0.8049  &  0.00496 \\ 
RV$_{\rm NIR}$-RVxcor & 1-1  &  0.8077  &  0.00470 \\ 
\\
CCF-FWHM-dLW$_{\rm VIS}$ & 2-2 & 0.7681 & 0.00946 \\ 
dLW$_{\rm VIS}$-CCF-Cont & 2-0 & --0.9028 & 0.00035 \\ 
dLW$_{\rm VIS}$-CCF-FWHM & 2-2 & 0.7681 & 0.00946 \\ 
\\
CCF-Bisector-RV$_{\rm VIS}$ & 3-1 & --0.9576 & 1.34e-05 \\ 
CCF-Bisector-CCF-RV & 3-1 & --0.9548 & 1.73e-05 \\ 
CCF-Bisector-RVxcor & 3-1 & --0.9486 & 2.87e-05 \\ 
CCF-Bisector-CofL & 3-1 & --0.9470 & 3.25e-05 \\ 
CCF-Bisector-RV$_{\rm NIR}$ & 3-1 & --0.7860 & 0.00702 \\ 
CCF-Bisector-CRX$_{\rm NIR}$ & 3-5 & 0.8045 & 0.00501 \\ 
CCF-Bisector-CRX$_{\rm VIS}$ & 3-5 & 0. & 4.84e-06 \\ 
\\
log(Ha)-Halpha & 4-4 & --0.9960 & 1.09e-09 \\ 
log(Ha)-CaIRT-c & 4-4 & --0.9558 & 1.59e-05 \\ 
log(Ha)-He10833 & 4-4 & --0.9556 & 1.61e-05 \\ 
    \hline
    \end{tabular}
\end{table*}

\begin{table*} 
 \centering
 \begin{tabular}{lccc}
  \hline
 \hline
Category & Pairs of activity indices & Pearson ($r$) & Student t-test ($stp$) \\
 \hline
log(Ha)-CaIRT-a & 4-4 & --0.9546 & 1.76e-05 \\ 
log(Ha)-NaD & 4-4 & --0.9524 & 2.12e-05 \\ 
log(Ha)-CaIRT-b & 4-4 & --0.9462 & 3.43e-05 \\ 
log(Ha)-HeD3 & 4-4 & --0.9267 & 0.00012 \\ 
log(Ha)-CaH2 & 4-4 & 0.9267 & 0.00012 \\ 
 \hline
 \multicolumn{4}{c}{Map 7 : phase  137.73--139.57 : Excellent } \\
 \multicolumn{4}{c}{Strong high lat. spots + 3 weak low lat spots} \\
 \hline 
 CRX-CofL & 5-1 & --0.9752 & 1.46e-10 \\ 
CRX-RVxcor & 5-1 & --0.9644 & 1.79e-09 \\ 
CRX-RV$_{\rm VIS}$ & 5-1 & --0.9426 & 4.74e-08 \\ 
CRX-CCF-RV & 5-1 & --0.9372 & 8.70e-08 \\ 
CRX-RV$_{\rm NIR}$ & 5-1 & --0.8485 & 3.26e-05 \\ 
CRX-CRX$_{\rm NIR}$ & 5-5 & 0.8912 & 3.60e-06 \\ 
CRX-CCF-Bisector & 5-3 & 0.9479 & 2.44e-08 \\ 
\\
CofL-CRX$_{\rm VIS}$ & 1-5 & --0.9752 & 1.46e-10 \\ 
CofL-CCF-Bisector & 1-3 & --0.9382 & 7.87e-08 \\ 
CofL-CRX$_{\rm NIR}$ & 1-5 & --0.9087 & 1.11e-06 \\ 
CofL-RV$_{\rm NIR}$ & 1-1 & 0.8744 & 9.41e-06 \\ 
CofL-CCF-RV & 1-1 & 0.9560 & 7.56e-09 \\ 
CofL-RV$_{\rm VIS}$ & 1-1 & 0.9637 & 2.00e-09 \\ 
CofL-RVxcor & 1-1 & 0.9949 & 2.30e-15 \\ 
\\
CCF-Cont-dLW$_{\rm VIS}$ & 0-2 & --0.8648 & 1.54e-05 \\ 
CCF-Cont-CCF-FWHM & 0-2 & --0.8062 & 0.00016 \\ 
CCF-Cont-log(Ha) & 0-4 & --0.7660 & 0.00054 \\ 
CCF-Cont-CaH3 & 0-4 & 0.7113 & 0.00200 \\ 
CCF-Cont-CaIRT-c & 0-4 & 0.7167 & 0.00178 \\
CCF-Cont-HeD3 & 0-4 & 0.7355 & 0.00117 \\ 
CCF-Cont-Halpha & 0-4 & 0.7667 & 0.00053 \\ 
CCF-Cont-CaIRT-b & 0-4 & 0.7847 & 0.00032 \\ 
CCF-Cont-NaD & 0-4 & 0.7855 & 0.00031 \\ 
\\
RV$_{\rm VIS}$-CCF-Bisector & 1-3 & --0.9498 & 1.90e-08 \\ 
RV$_{\rm VIS}$-CRX$_{\rm VIS}$ & 1-5 & --0.9426 & 4.74e-08 \\ 
RV$_{\rm VIS}$-CRX$_{\rm NIR}$ & 1-5 & --0.9288 & 2.06e-07 \\ 
RV$_{\rm VIS}$-RV$_{\rm NIR}$ & 1-1 & 0.9054 & 1.41e-06 \\ 
RV$_{\rm VIS}$-CofL & 1-1 & 0.9637 & 2.00e-09 \\ 
RV$_{\rm VIS}$-RVxcor & 1-1 & 0.9783 & 5.67e-11 \\ 
RV$_{\rm VIS}$-CCF-RV & 1-1 & 0.9948 & 2.71e-15 \\ 
\\
RV$_{\rm NIR}$-CRX$_{\rm NIR}$ & 1-5    &  -0.9173  &  5.68e-07\\ 
RV$_{\rm NIR}$-CRX$_{\rm VIS}$ & 1-5   &  -0.8485  &  3.25e-05 \\
RV$_{\rm NIR}$-CCF-Bisector & 1-3  &  -0.8180  &  0.00010 \\
RV$_{\rm NIR}$-CCF-RV & 1-1   &  0.8659  &  1.45e-05 \\
RV$_{\rm NIR}$-CofL & 1-1   &  0.8744  &  9.56e-06 \\
RV$_{\rm NIR}$-RVxcor & 1-1   &  0.8815  &  6.37e-06 \\
RV$_{\rm NIR}$-RV$_{\rm VIS}$ & 1-1   &  0.9054  &  0.00010 \\
\\
CCF-FWHM-CCF-Cont & 2-0 & --0.8062 & 0.00016 \\ 
CCF-FWHM-TiO7050 & 2-4 & --0.7148 & 0.00186 \\ 
CCF-FWHM-dLWNIR & 2-2 & 0.8554 & 2.39e-05 \\ 
CCF-FWHM-dLW$_{\rm VIS}$ & 2-2 & 0.9557 & 8e-09 \\ 
dLW$_{\rm VIS}$-CCF-CON & 2-0 & --0.8648 & 1.54e-05 \\ 
dLW$_{\rm VIS}$-HeD3 & 2-4 & --0.7151 & 0.00185 \\ 
dLW$_{\rm VIS}$-NaD & 2-4 & --0.7111 & 0.00201 \\ 
dLW$_{\rm VIS}$-dLWNIR & 2-2 & 0.8454 & 3.73e-05 \\ 
dLW$_{\rm VIS}$-CCF-FWHM & 2-2 & 0.9557 & 8.00e-09 \\ 
\\
CCF-Bisector-RVxcor & 3-1 & --0.9541 & 1.02e-08 \\ 
CCF-Bisector-RV$_{\rm VIS}$ & 3-1 & --0.9498 & 1.90e-08 \\ 
CCF-Bisector-CCF-RV & 3-1 & --0.9416 & 5.32e-08 \\ 
CCF-Bisector-CofL & 3-1 & --0.9382 & 7.87e-08 \\ 
  \hline
    \end{tabular}
\end{table*}

\begin{table*} 
 \centering
 \begin{tabular}{lccc}
  \hline
 \hline
Category & Pairs of activity indices & Pearson ($r$) & Student t-test ($stp$) \\
 \hline

CCF-Bisector-RV$_{\rm NIR}$ & 3-1 & --0.8181 & 0.00011 \\ 
CCF-Bisector-CRX$_{\rm NIR}$ & 3-5 & 0.8684 & 1.28e-05 \\ 
CCF-Bisector-CRX$_{\rm VIS}$ & 3-5 & 0.9479 & 2.44e-08 \\ 
\\
log(Ha)-Halpha & 4-4 & --0.9960 & 4.14e-16 \\ 
log(Ha)-CaIRT-b & 4-4 & --0.9743 & 1.84e-10 \\ 
log(Ha)-CaIRT-a & 4-4 & --0.9633 & 2.17e-09 \\ 
log(Ha)-CaIRT-c & 4-4 & --0.9482 & 2.33e-08 \\ 
log(Ha)-NaD & 4-4 & --0.9428 & 4.59e-08 \\ 
log(Ha)-HeD3 & 4-4 & --0.8754 & 8.96e-06 \\ 
log(Ha)-CCF-Cont & 4-4 & --0.7660 & 0.00054 \\ 
log(Ha)-He10833 & 4-4 & --0.7617 & 0.00061 \\ 
 \hline
  \multicolumn{4}{c}{Map 8 : phase 142.70--145.46 : Very good } \\
 \multicolumn{4}{c}{1 weak high lat + 1 weak mid lat. + 2 weak low lat.} \\
 \hline 
CRX-RV$_{\rm VIS}$ & 5-1 & --0.9308 & 1.70e-07 \\ 
CRX-CofL & 5-1 & --0.9173 & 5.68e-07 \\ 
CRX-CCF-RV & 5-1 & --0.9104 & 9.79e-07 \\ 
CRX-RVxcor & 5-1 & --0.9056 & 1.39e-06 \\ 
CRX-RV$_{\rm NIR}$ & 5-1 & --0.8179 & 0.00011 \\ 
CRX-CRX$_{\rm NIR}$ & 5-5 & 0.8169 & 0.00011 \\ 
CRX-CCF-Bisector & 1-3 & 0.9686 & 7.36e-10 \\ 
\\
CofL-CCF-Bisector & 1-3 & --0.9273 & 2.38e-07 \\ 
CofL-CRX$_{\rm VIS}$ & 1-5 & --0.9173 & 5.68e-07 \\ 
CofL-CRX$_{\rm NIR}$ & 1-5 & --0.8719 & 1.07e-05 \\ 
CofL-RV$_{\rm NIR}$ & 1-1 & 0.9270 & 2.43e-07 \\ 
CofL-CCF-RV & 1-1 & 0.9742 & 1.92e-10 \\ 
CofL-RV$_{\rm VIS}$ & 1-1 & 0.9802 & 3.01e-11 \\ 
CofL-RVxcor & 1-1 & 0.9986 & 3.24e-19 \\ 
\\
CCF-Cont-dLW$_{\rm VIS}$ & 0-2 & --0.9326 & 1.41e-07 \\ 
CCF-Cont-log(Ha) & 0-4 & --0.8265 & 7.91e-05 \\ 
CCF-Cont-dLWNIR & 0-2 & --0.8187 & 0.00011 \\ 
CCF-Cont-HeD3 & 0-4 & 0.7657 & 0.00055 \\ 
CCF-Cont-Halpha & 0-4 & 0.8169 & 0.00010 \\ 
CCF-Cont-CaIRT-a & 0-4 & 0.8174 & 0.00011 \\ 
CCF-Cont-NaD & 0-4 & 0.8187 & 0.00011 \\ 
CCF-Cont-CaIRT-c & 0-4 & 0.8215 & 9.53e-05 \\ 
CCF-Cont-CaIRT-b & 0-4 & 0.8428 & 4.15e-05 \\ 
\\
RV$_{\rm VIS}$-CRX$_{\rm VIS}$ & 1-5 & --0.9308 & 1.7e-07 \\ 
RV$_{\rm VIS}$-CCF-Bisector & 1-3 & --0.9302 & 1.81e-07 \\ 
RV$_{\rm VIS}$-CRX$_{\rm NIR}$ & 1-5 & --0.8905 & 3.76e-06 \\ 
RV$_{\rm VIS}$-RV$_{\rm NIR}$ & 1-1 & 0.9527 & 1.25e-08 \\ 
RV$_{\rm VIS}$-CofL & 1-1 & 0.9802 & 3.01e-11 \\ 
RV$_{\rm VIS}$-RVxcor & 1-1 & 0.9830 & 1.07e-11 \\ 
RV$_{\rm VIS}$-CCF-RV & 1-1 & 0.9920 & 5.69e-14 \\
\\
RV$_{\rm NIR}$-CCF-Bisector & 1-3  &  -0.8295  &  0.00010 \\
RV$_{\rm NIR}$-CRX$_{\rm VIS}$ & 1-5  &  -0.8179  &  0.00010 \\
RV$_{\rm NIR}$-CRX$_{\rm NIR}$ & 1-5  &  -0.8162  &  0.00010 \\
RV$_{\rm NIR}$-CofL  & 1-1 &  0.9270  &  2.42e-07 \\
RV$_{\rm NIR}$-CCF-RV & 1-1  &  0.9372  &  8.72e-08 \\
RV$_{\rm NIR}$-RVxcor & 1-1  &  0.9408  &  0.00010 \\
RV$_{\rm NIR}$-RV$_{\rm VIS}$ & 1-1  &  0.9527  &  1.24e-08\\ 
\\
CCF-FWHM-dLW$_{\rm VIS}$ & 2-2 & 0.8394 & 4.78e-05 \\ 
dLW$_{\rm VIS}$-CCF-Cont & 2-0 & --0.9326 & 1.41e-07 \\ 
dLW$_{\rm VIS}$-CaIRT-b & 2-4 & --0.7112 & 0.00201 \\ 
dLW$_{\rm VIS}$-NaD & 2-4 & --0.7046 & 0.00231 \\ 
dLW$_{\rm VIS}$-log(LHa) & 2-4 & 0.7096 & 0.00208 \\ 
dLW$_{\rm VIS}$-dLWNIR & 2-2 & 0.8209 & 9.74e-05 \\ 
dLW$_{\rm VIS}$-CCF-FWHM & 2-2 & 0.8394 & 4.78e-05 \\ 
\hline
    \end{tabular}
\end{table*}

\begin{table*} 
 \centering
 \begin{tabular}{lccc}
  \hline
 \hline
Category & Pairs of activity indices & Pearson ($r$) & Student t-test ($stp$) \\
 \hline
CCF-Bisector-RV$_{\rm VIS}$ & 3-1 & --0.9302 & 1.81e-07 \\ 
CCF-Bisector-CofL & 3-1 & --0.9273 & 2.38e-07 \\ 
CCF-Bisector-RVxcor & 3-1 & --0.9161 & 6.25e-07 \\ 
CCF-Bisector-CCF-RV & 3-1 & --0.8972 & 2.46e-06 \\ 
CCF-Bisector-RV$_{\rm NIR}$ & 3-1 & --0.8295 & 7.06e-05 \\ 
CCF-Bisector-CRX$_{\rm NIR}$ & 3-5 & 0.8341 & 5.92e-05 \\ 
CCF-Bisector-CRX$_{\rm VIS}$ & 3-5 & 0.9686 & 7.36e-10 \\ 
\\
log(Ha)-Halpha & 4-4 & --0.9873 & 1.35e-12 \\ 
log(Ha)-CaIRT-c & 4-4 & --0.9771 & 8.25e-11 \\ 
log(Ha)-CaIRT-b & 4-4 & --0.9748 & 1.60e-10 \\ 
log(Ha)-CaIRT-a & 4-4 & --0.9739 & 2.04e-10 \\ 
log(Ha)-NaD & 4-4 & --0.9436 & 4.17e-08 \\ 
log(Ha)-HeD3 & 4-4 & --0.9400 & 6.41e-08 \\ 
log(Ha)-He10833 & 4-4 & --0.8865 & 4.78e-06 \\ 
log(Ha)-CCF-Cont & 4-0 & --0.8265 & 7.91e-05 \\ 
log(Ha)-Pabeta & 4-4 & --0.7182 & 0.00173 \\ 
log(Ha)-dLW$_{\rm VIS}$ & 4-2 & 0.7096 & 0.00208 \\ 
\hline
    \end{tabular}
\end{table*}

\begin{table*} 
 \centering
 \begin{tabular}{lccc}
  \hline
 \hline
Category & Pairs of activity indices & Pearson ($r$) & Student t-test ($stp$) \\
 \hline
   \multicolumn{4}{c}{Map 9 (comb. maps 6 and 7): phase  134.93--139.57 : Excellent } \\
 \multicolumn{4}{c}{Strong high lat. spots +  weak low lat spots} \\
 \hline 
CRX-CCF-Bisector  & 5-3 & 0.9492 & 1.47e-13 \\
CRX-CCF-RV  & 5-1 & --0.9440 & 4.65e-13 \\
CRX-CofL  & 5-1 & --0.9700 & 2.92e-16 \\
CRX-CRX$_{\rm NIR}$  & 5-5 & 0.7070 & 5.27e-05 \\
CRX-RV$_{\rm VIS}$  & 5-1 & --0.9560 & 2.72e-14 \\
CRX-RVxcor  & 5-1 & --0.9610 & 6.52e-15 \\
\\
CofL-CCF-Bisector  & 1-3 & --0.9369 & 1.88e-12 \\
CofL-CCF-RV  & 1-1 & 0.9603 & 8.10e-15 \\
CofL-CRX$_{\rm NIR}$  & 1-5 & --0.7234 & 2.95e-05 \\
CofL-CRX$_{\rm VIS}$  & 1-5 & --0.9700 & 2.92e-16 \\
CofL-RV$_{\rm VIS}$  & 1-1 & 0.9741 & 4.67e-17 \\
CofL-RVxcor  & 1-1 & 0.9958 & 1.66e-26 \\
\\
CCF-Cont-dLW$_{\rm VIS}$  & 0-2 & --0.8792 & 3.39e-09 \\
\\
RV$_{\rm VIS}$-CCF-Bisector  & 1-3 & --0.9436 & 5.10e-13 \\
RV$_{\rm VIS}$-CCF-RV  & 1-1 & 0.9848 & 8.71e-20 \\
RV$_{\rm VIS}$-CofL  & 1-1 & 0.9741 & 4.67e-17 \\
RV$_{\rm VIS}$-CRX$_{\rm NIR}$  & 1-5 & --0.7555 & 8.09e-06 \\
RV$_{\rm VIS}$-CRX$_{\rm VIS}$  & 1-5 & --0.9560 & 2.72e-14 \\
RV$_{\rm VIS}$-RVxcor  & 1-1 & 0.9827 & 4.30e-19 \\
\\
RV$_{\rm NIR}$-CRX$_{\rm NIR}$ & 1-5   &  -0.9382  &  1.36e-12 \\
\\
CCF-Bisector-CCF-RV  & 3-1 & --0.9452 & 3.65e-13\\
CCF-Bisector-CofL  & 3-1 & --0.9369 & 1.88e-12 \\
CCF-Bisector-CRX$_{\rm VIS}$  & 3-5 & 0.9492 & 1.47e-13 \\
CCF-Bisector-RV$_{\rm VIS}$  & 3-1 & --0.9436 & 5.10e-13 \\
CCF-Bisector-RVxcor  & 3-1 & --0.9469 & 2.50e-13 \\
\\
dLW$_{\rm VIS}$-CCF-Cont  & 2-0 & --0.8792 & 3.39e-09 \\
dLW$_{\rm VIS}$-CCF-FWHM  & 2-2 & 0.8618 & 1.55e-08 \\
\\
log(Ha)-CaIRT-a  & 4-4 & --0.8887 & 1.33e-09 \\
log(Ha)-CaIRT-b  & 4-4 & --0.9023 & 2.98e-10 \\
log(Ha)-CaIRT-c  & 4-4 & --0.8858 & 1.78e-09 \\
log(Ha)-Halpha  & 4-4 & --0.9954 & 5.63e-26 \\
log(Ha)-He10833  & 4-4 & --0.8335 & 1.24e-07 \\
log(Ha)-HeD3  & 4-4 & --0.8985 & 4.62e-10 \\
log(Ha)-NaD  & 4-4 & --0.9017 & 3.21e-10 \\
\hline
    \end{tabular}
\end{table*}

\begin{table*} 
 \centering
 \begin{tabular}{lccc}
  \hline
 \hline
Category & Pairs of activity indices & Pearson ($r$) & Student t-test ($stp$) \\
 \hline
   \multicolumn{4}{c}{All CARMENES spectra } \\
 \hline
CRX-RV$_{\rm VIS}$ & 5-1 &  -0.9368 & 1.28e-39 \\
CRX-CofL & 5-1  & -0.8503 & 7.29e-25 \\
CRX-CCF-RV & 5-1  & -0.7363 & 9.73e-16 \\
CRX-CCF-Bisector & 5-1 & 0.5255 & 2.41e-07 \\
CRX-CRX$_{\rm NIR}$ & 5-5  & 0.6277 & 1.27e-10 \\
\\
CofL-CRX$_{\rm VIS}$ & 1-5  & -0.8503 & 7.29e-25 \\
CofL-CCF-RV & 1-1  & 0.8051 & 1.58e-20 \\
CofL-RV$_{\rm VIS}$ & 1-1  & 0.9196 & 1.93e-35 \\
\\
CCF-Cont-dLW$_{\rm VIS}$ &  0-2 & -0.7012 & 7.76e-14 \\
\\
RV$_{\rm VIS}$-CRX$_{\rm VIS}$ & 1-5  & -0.9368 & 1.28e-39 \\
RV$_{\rm VIS}$-CCF-RV & 1-1  & 0.8239 & 3.50e-22 \\
RV$_{\rm VIS}$-CofL & 1-1  & 0.9196 & 1.93e-35 \\
\\
dLW$_{\rm VIS}$-CCF-Cont & 2-0  & -0.7012 & 7.76e-14 \\
\\
log(Ha)-CaIRT-a & 4-4  & -0.9493 & 1.73e-43 \\
log(Ha)-CaIRT-c & 4-4  & -0.9448 & 5.38e-42 \\
log(Ha)-CaIRT-b & 4-4  & -0.9441 & 9.02e-42 \\
log(Ha)-NaD &  4-4 & -0.9320 & 2.42e-38 \\
log(Ha)-HeD3 & 4-4  & -0.8864 & 1.65e-29 \\
log(Ha)-He10833 & 4-4  & -0.8801 & 1.38e-28 \\
log(Ha)-Pabeta & 4-4  & -0.7203 & 7.76e-15 \\
\hline
    \end{tabular}
\end{table*}

\end{appendix}

\end{document}